\documentclass[twocolumn]{aastex63}
\usepackage{amsmath}
\usepackage[T1]{fontenc}
\usepackage{lineno}

\accepted{March 2, 2023}

\shorttitle{Flat spectrum radio transient VT 1137-0337}
\shortauthors{Dong et al.}

\graphicspath{{./}{figures/}}

\begin{document}

\title{A Flat-Spectrum Radio Transient at 122 Mpc consistent with an Emerging Pulsar Wind Nebula}

\correspondingauthor{Dillon Z. Dong}
\email{ddong@caltech.edu}

\author[0000-0001-9584-2531]{Dillon Z. Dong}
\affiliation{Cahill Center for Astronomy and Astrophysics, MC\,249-17 California Institute of Technology, Pasadena CA 91125, USA.}

\author[0000-0002-7083-4049]{Gregg Hallinan}
\affiliation{Cahill Center for Astronomy and Astrophysics, MC\,249-17 California Institute of Technology, Pasadena CA 91125, USA.}

\begin{abstract}

We report the discovery and follow-up observations of VT 1137-0337: an unusual radio transient found in our systematic search for extragalactic explosions in the VLA Sky Survey (VLASS). VT 1137-0337 is located in the brightest region of a dwarf starburst galaxy at a luminosity distance of 121.6 Mpc. Its 3 GHz luminosity is comparable to luminous radio supernovae associated with dense circumstellar interaction and relativistic outflows. However, its broadband radio spectrum - proportional to $\nu^{-0.35}$ over a range of $\gtrsim 10\times$ in frequency and fading at a rate of 5\% per year - cannot be directly explained by the shock of a stellar explosion. Jets launched by various classes of accreting black holes also struggle to account for VT 1137-0337's combination of observational properties. Instead, we propose that VT 1137-0337 is a decades old pulsar wind nebula that has recently emerged from within the free-free opacity of its surrounding supernova ejecta. If the nebula is powered by spindown, the central neutron star should have a surface dipole field of $\sim 10^{13} - 10^{14}$ G and a present-day spin period of $\sim 10 - 100$ ms. Alternatively, the nebula may be powered by the release of magnetic energy from a magnetar. Magnetar nebulae have been proposed to explain the persistent radio sources associated with the repeating fast radio bursts FRB 121102 and FRB 190520B. These FRB persistent sources have not previously been observed as transients, but do bear a striking resemblance to VT 1137-0337 in their radio luminosity, spectral index, and host galaxy properties.
\end{abstract}


\section{Introduction} 
\label{sec:intro}

Recent advances in hardware \citep[][]{Perley11-EVLA} and observing techniques \citep{Mooley19-CNSS2-OTF-methodology} used at the Jansky Very Large Array (VLA) have made it possible to efficiently survey large areas of the GHz radio sky at arcsecond resolution. These technologies are key components of the VLA Sky Survey (VLASS), which is observing the full sky north of declination $\delta = -40^{\circ}$ over multiple epochs \citep[][]{Lacy20-VLASS}. VLASS's survey speed and wide field coverage ($\sim$34,000~deg$^{2}$) has led to a rapid uptick in the detection rate of slow radio transients, which typically peak on timescales of months to decades post explosion in the 2-4~GHz VLASS band. Previously, slow radio transients were primarily detected via follow-up observations of events identified at other wavelengths. With VLASS, new transients can be directly identified at radio frequencies at scale, enabling an exploration of the radio transient phase space unbiased by the detectability of emission at other wavelengths. This has facilitated the systematic discovery of representatives from both known and previously hypothetical transient classes \citep[e.g.,][]{Metzger15-VLASS-predictions, Chevalier2012-CE-SNe}. Examples include a diversity of nuclear radio flares including tidal disruption events \citep[][]{Anderson-CNSS_TDE, Ravi21-reverse-TDE, Somalwar21-TDE-AGN} and a population of newly launched jets from quasars \citep[][]{Nyland2020-VLASS-Quasars}, a compelling candidate for an off-axis long gamma ray burst \citep[][]{Law18-FIRSTJ1419}, and the first known supernova set off by the merger of a compact object with a massive star \citep[][]{Dong21-VT1210}.
\\
\\Broadband radio spectra of a transient's synchrotron emission are powerful diagnostic tools of the transient's physical properties. Thus far, these spectra have primarily revealed spectral peaks, and/or steep power law tails where  $S_{\nu} \propto \nu^{-\alpha}$ with $\alpha > 0.5$. This is in line with predictions for synchrotron emission from electrons accelerated with diffusive shock acceleration \citep[][]{Blandford-Eichler-shocks, Jones-Ellison-1991-shock-theory, Pelletier-2017-relativistic-shocks} modified by free-free/synchrotron self absorption  \citep[][]{Essential-Radio-Astronomy}. The consistency is unsurprising, given that previously identified radio transient classes are associated with fast outflows that drive strong shocks in the surrounding gas. 
\\
\\In this paper, we present VLASS Transient J113706.19-033737.3 (hereafter VT 1137-0337), a new radio transient with a spectral index ($\propto \nu^{-0.35}$ measured over more than a factor of 10$\times$ in frequency) that is inconsistent with diffusive shock acceleration. In Section \ref{sec:discovery}, we discuss the discovery of VT 1137-0337 and its association with its host galaxy, a starbursting dwarf. We additionally estimate the rate of transients with similar observational properties. In Section \ref{sec:followup}, we present radio and optical follow up spectra at the location of the transient and discuss our procedure for fitting power-law and emission line models to these spectra. In Section \ref{sec:analysis}, we use these observations to derive constraints on the emitting region and its surroundings. In Section \ref{sec:comparisons}, we assess the consistency of VT 1137-0337 with a wide range of transient, variable, and persistent radio source classes. In Section \ref{sec:discussion}, we summarize our conclusions. Throughout this paper, we assume a flat $\Lambda$CDM cosmology where $H_0 = 69.6$~km~s$^{-1}$~Mpc$^{-1}$ \citep[][]{Bennett14-concordence-cosmology}, implying a luminosity distance to the source of 121.6~Mpc and an angular diameter distance of 115.2~Mpc \citep[][]{Wright06-cosmology-calc}. 

\section{Discovery of VT 1137-0337}
\label{sec:discovery}

\begin{figure*}
    \centering
    \includegraphics[clip=true, trim=0.15cm 0.3cm 0.2cm 0cm, width=\textwidth]{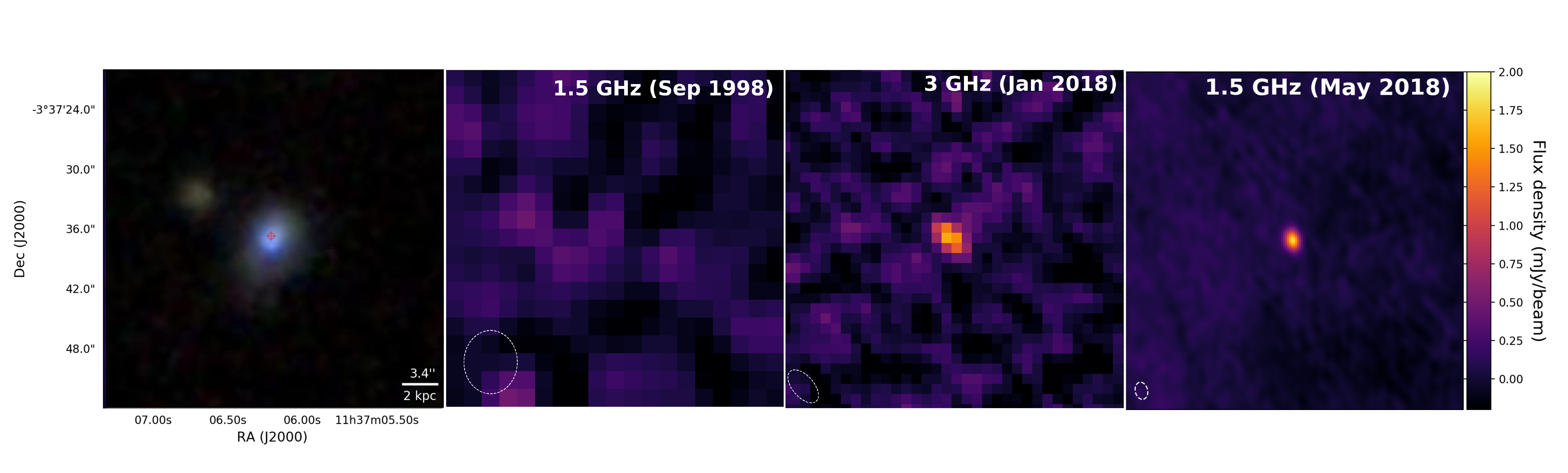}
    \caption{The detection of VT 1137-0337. Left to right: (1) The host galaxy, located at a luminosity distance of 121.6~Mpc. The radio transient position is marked with a red cross. (2) Non-detection at 1.5~GHz in the FIRST survey. (3) First detection at 3~GHz in VLASS quicklook imaging (observed in the VLA's B configuration; \citep[][]{Lacy20-VLASS}). (4) A follow-up image at 1.5~GHz taken with the VLA in the A configuration. All images are at the same size scale, and all radio images use the same colorscale shown by the colorbar. The synthesized beam of each radio image is shown by the ellipse in the bottom left.}     \label{fig:host-transient-discovery}
\end{figure*}

 VT 1137-0337, located at RA = 11:37:06.19, Dec = -03:37:37.3, was discovered the source through a blind search of VLASS Epoch 1.1 (the first half-epoch of VLASS) for transients associated with galaxies within a distance of 200~Mpc. For this search, we chose the Faint Images of the Radio Sky at Twenty-cm (FIRST) survey \citep{Becker95} as our reference epoch for its comparable resolution and sensitivity over a large area of sky (6195 deg$^2$ overlapping with Epoch 1.1). Our transient search methodology is summarized below.
\\
\\We used the source extractor PyBDSF \citep{Mohan_Rafferty_2015_pybdsf} to assemble a catalog of all sources detected at $\geq$5 times the local RMS noise the Epoch 1.1 quicklook images\footnote{VLASS quicklook images are available at \url{https://archive-new.nrao.edu/vlass/quicklook/}} \citep{Lacy20-VLASS}. We used methods implemented in the \texttt{astropy.skycoord} class \citep{Astropy-paper1-2013,Astropy-paper2-2018} to cross match these sources against the FIRST catalog \citep{Becker95}, identifying all VLASS sources that are separated by $>$10'' from the nearest known FIRST source as transient candidates. Through visual inspection, we rejected image artifacts, spatially resolved VLASS sources, and faint FIRST sources that fell under the 0.7mJy FIRST catalog threshold. After this process, we are left with $\sim$3000 point sources that are $\gtrsim$ 0.7mJy in VLASS and $<$ 0.3-0.4 mJy in FIRST. Due to the slight frequency mismatch between the reference and detection epochs, some of these transient candidates are due to non varying or slowly varying active galactic nuclei (AGN) in background galaxies which have rising spectral indices between the 1.5~GHz FIRST band and the 3~GHz VLASS band. Because our search was focused on transients within the local universe, we cross matched the transient candidates against spectroscopically verified galaxies within 200 Mpc from the Sloan Digital Sky Survey Data Release 15 \citep[SDSS DR15;][]{SDSS_DR15}, the Census of the Local Universe \citep[CLU;][]{Cook19-CLU} galaxy catalog, and the NASA Extragalactic Database (NED)\footnote{The NASA/IPAC Extragalactic Database (NED) is operated by the Jet Propulsion Laboratory, California Institute of Technology, under contract with the National Aeronautics and Space Administration.}. This process revealed 20 likely transients with statistically significant local universe galaxy associations. None of these likely transients show indications in multiwavelength data of being background sources.
\\
\\Among the 20 local universe sources, VT 1137-0337 stood out for its starbursting host galaxy (Section \ref{sec:Derived properties from optical features}) and unusual flat spectrum (Sections \ref{subsec:radio}, \ref{sec:flat-spectrum-emission-mechanisms}). It was detected in VLASS Epoch 1.1 with a 3~GHz flux of 1.7 $\pm$ 0.4 mJy in January 2018, but is absent from FIRST with a 3$\sigma$ upper limit at 1.5~GHz of $<$ 0.35 mJy in September 1998. Three VLA followup observations taken in May 2018, March 2019, and February 2022 confirmed the presence of a new 1.5~GHz source that would have been easily detectable in FIRST (Section \ref{sec:followup}, Table \ref{tab:radio_fluxes}). Figure \ref{fig:host-transient-discovery} shows the reference, discovery, and a follow-up image of this source, which is located at a separation of 0.4 arcseconds from the cataloged position of the dwarf starburst galaxy SDSS J113706.18-033737.1's nucleus \citep[][]{SDSS_DR15}. Based on our highest resolution followup observation (taken at 10~GHz in the VLA's A configuration; Section \ref{sec:followup}), we conservatively estimate that the localization uncertainity is $< 0.1$ arcsec (half the synthesized beam FWHM). This suggests that the source is slightly off-nuclear, though we note that the concept of a nucleus is somewhat ambiguous in this case since the host is a dwarf irregular galaxy with a nuclear position determined by the peak of a large star cluster.
\\
\\In our searched catalogs, there are $\sim$70,000 galaxies within the 6195 deg$^2$ VLASS/FIRST overlap. The fraction of the overlap area that falls within 0.4 arcseconds of one of their nuclei is 1.4 $\times$ 10$^{-7}$. There are $\sim$3000 VLASS transient candidates that could have been located within this area. Thus, the expected number of coincidental associations within this sample is $\sim$4 $\times$ 10$^{-4}$. Among the 20 galaxy-associated transients, we observed 12 that are within 0.4'' of their host galaxy's nucleus. The probability of one or more of these being falsely associated is $\approx$ 4 $\times$ 10$^{-4}$ (formally, 1 - Poisson($\lambda$ = 4 $\times$ 10$^{-4}$, $k$ = 0), where $\lambda$ is the mean and $k$ is the number of false positive occurrences). Given the small false alarm probability, we conclude that VT 1137-0337 is not a foreground or background source, and is indeed associated with SDSS J113706.18-033737.1. 
\\
\\The $> 500\%$ variability at 1.5~GHz is inconsistent with refractive scintillation which, given the estimated critical frequency of $\sim$8.5~GHz at the location of the transient \citep[][]{Cordes-Lazio02-ne2001}, would imply a typical point source modulation of $\sim$40\% \citep[][]{Walker1998-scintillation}. If the source diameter $d$ is resolved on the scale of the scattering disk (corresponding to a diameter of $d_{rs} \sim 2 \times 10^{17}$~cm at the distance to the host galaxy), the RMS modulation is further reduced by a factor of $(d_{rs}/d)^{7/6}$. Diffractive scintillation is suppressed at even smaller scales, being reduced by a factor of $(d_{ds}/d)$, where for our source $d_{ds} \sim 10^{15}$~cm. Additionally, the modulations would be decorrelated on a bandwidth of $\sim$6~MHz and would be substantially reduced by averaging over a single VLA spectral window. This implies that the variability of VT 1137-0337 between FIRST and VLASS is not primarily due to propagation effects in the Milky Way, and is instead intrinsic to the source itself or its immediate surroundings.

\subsection{Implied detection rate}
\label{subsec:detection-rate}
In our search of $\sim$6195 deg$^2$ out to $\sim$200~Mpc, we detected no other transients similar to VT 1137-0337, which we define as having an optically thin spectral index flatter than $\alpha = 0.5$ and being located in a galaxy at or above the star forming main sequence. This suggests that (1) the phenomenon is rare, or (2) we have detected one of the easiest ones to detect, or (3) both. Here we present a rate estimate assuming that VT 1137-0337 is a representative example of the population from which it is drawn (option 1). If instead there are less luminous analogs or if they are predominantly located in faint galaxies, this estimate should be treated as a lower limit.
\\
\\Based on VT 1137-0337's 1.7~mJy flux in the detection epoch, we estimate that the transient would have been detectable in our search out to $\sim$190~Mpc. The host galaxy, which has a g band magnitude of $\sim$17.5, would have easily been detectable by SDSS out to that distance. The effective searched volume is $V_{\textrm{search}} \approx$ $\frac{4}{3} \pi$ (190~Mpc)$^3$ $(6195~ \textrm{deg}^2/41253~ \textrm{deg}^2) \approx 4 \times 10^6$ Mpc$^3$. Within this volume, \citet[][]{Kulkarni18-galaxy-catalog-completeness} estimate that the overall redshift completeness $f_z$ of all NED galaxies is $\sim$60\%. This fraction is lower for low mass galaxies but higher for star forming galaxies. If we use the host galaxy's K band absolute magnitude (a proxy for mass), $f_z \approx 10\%$, but if we use its NUV flux (a proxy for star formation), $f_z \approx 75\%$. Given that the host galaxy is undergoing a starburst (Section \ref{sec:Derived properties from optical features}), we adopt an intermediate value of $f_z \approx 50\%$ for the redshift completeness, which is closer to the star-formation only value. Finally, VT 1137-0337 appeared in the radio sky on a timescale of $t_{\textrm{span}} <$ 19.3 years, and persisted for $t_{\textrm{span}} \gtrsim$ 4.1 years (Table \ref{tab:radio_fluxes}). We thus estimate that the volumetric birth rate for flat spectrum radio transients of similar radio luminosity in similar galaxies is $R =  \left(V_{\textrm{search}} t_{\textrm{span}}f_z\right)^{-1}$ $\approx$ (20 to 600) Gpc$^{-3}$ yr$^{-1}$. 
\\
\\In addition to astrophysical systematics (e.g., less luminous transients that are not detected in VLASS), the rate based on a sample of 1 should be regarded as having a large margin of error due to Poisson noise. Adopting a 95\% confidence interval leads to a range of (1 - 3000) Gpc$^{-3}$ yr$^{-1}$ \citep[][]{Gehrels-Poisson-noise}. For comparison, this corresponds to between 0.001\% and 3\% of the volumetric core collapse supernova (CCSN) rate \citep[][]{Taylor14-CCSN-rate}. The upper limit of this range is comparable to the rate of Ic broadline (Ic-BL) supernovae \citep[$\sim 3\%$ of the CCSN rate;][]{Graham-Schady-16-LGRB-rate}. The rate of low-luminosity GRBs is in the middle of the range \citep[$\sim 0.3\%$ of the CCSN rate;][]{Soderberg06-llGRB060218}, while the beaming-angle corrected rate of classical long GRBs is comparable within uncertainties to lower limit of the range \citep[$\sim 0.005\%$ of the CCSN rate; see][and references therein for a discussion]{Ho2020-Koala}. The rate is also consistent with the estimated birth rate of FRB persistent source candidates ($\sim 500$~Gpc$^{-3}$ yr$^{-1}$) found by \citet[][]{Ofek17-FRB-PRS-search}.

\section{Observations and model fitting}
\label{sec:followup}

We used a combination of follow-up and archival observations to characterize the observational properties of VT 1137-0337 and its surrounding environment. In this section, we describe the observations and associated data processing. We also discuss the model function used to fit the radio and optical follow-up spectra.

\subsection{Radio follow up observations}
\label{subsec:radio}
\begin{figure}
    \includegraphics[width=0.48\textwidth]{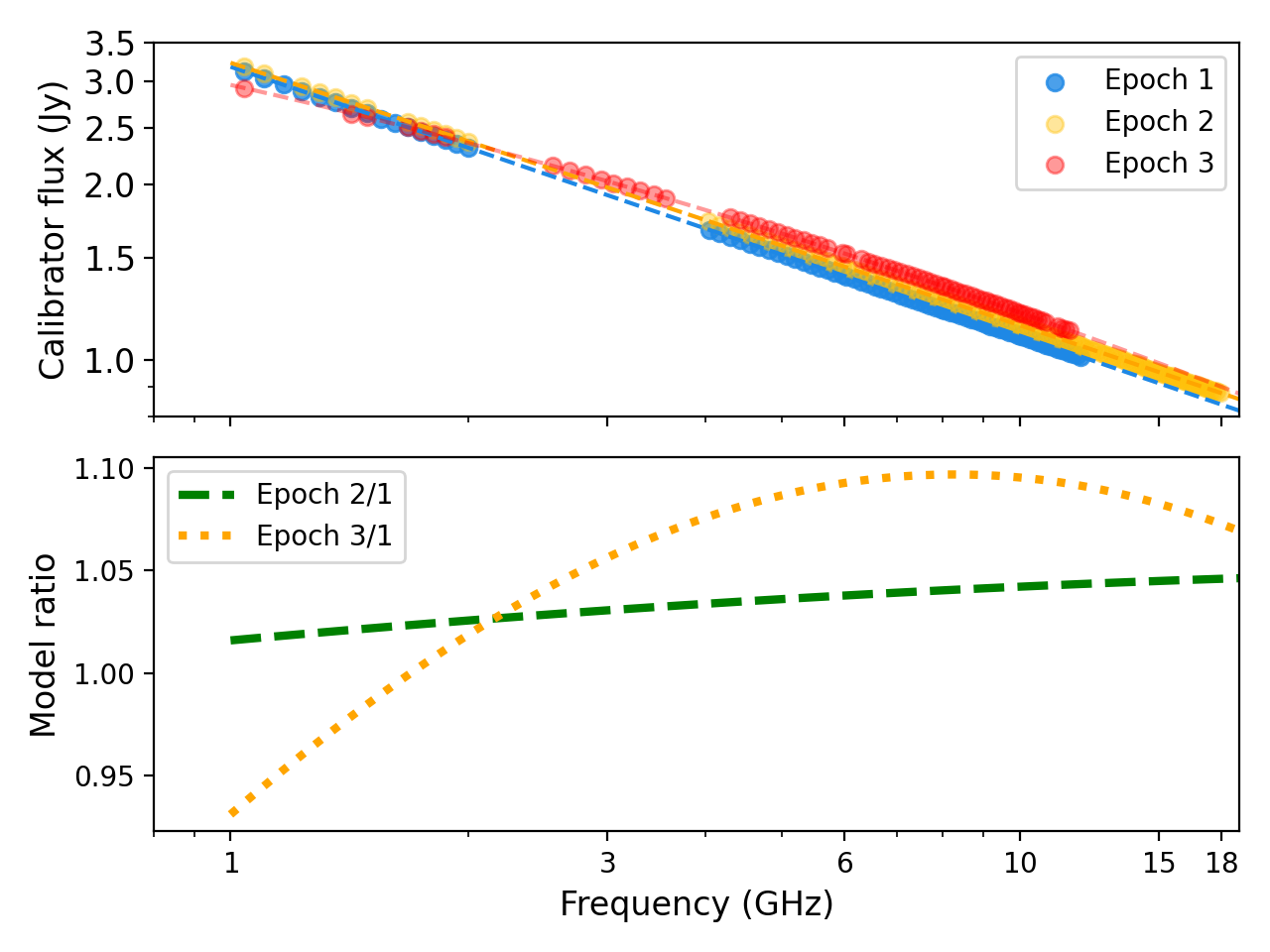}
    \caption{The phase calibrator (J1150-0023) observed in each VLA follow up epoch, plotted as a consistency check for our calibration. \textbf{Top:} The calibrator spectrum after correction with the calibration applied to VT 1137-0337. Data points are measured from images of each VLA spectral window, described in Section \ref{subsec:radio}. Dashed lines show best fit parabolic models for each epoch. \textbf{Bottom:} The ratio of calibrator models, showing $\lesssim 10\%$  variability between epochs at all observed frequencies.}
    \label{fig:pcal-SED}
\end{figure}

Four months after initial discovery in VLASS, we observed VT 1137-0337 with the VLA in its A configuration with the L, S, C, and X bands (1-12 GHz) under program 18A-481 (hereafter referred to as Epoch 1). We used 3C286 as the flux calibrator and J1150-0023 as the phase calibrator. We repeated these broadband radio observations at post-discovery epochs of 1.2 and 4.1 years with the VLA in the B and BnA configurations under programs 19A-422 (Epoch 2) and 21B-393 (Epoch 3). In the second epoch, we also observed in the Ku band (12-18 GHz). To calibrate and image the data, we used standard tasks from the Common Astronomical Software Applications (CASA) package \citep[][]{McMullin07-CASA}. In the first two epochs, we excluded the S band (2-4 GHz) observations which could not be reliably calibrated due to gain compression in the receivers. This is likely due to severe radio frequency interference from geosynchronous satellites in the Clarke Belt during the S band setup scans. For all other observations, we made single-band images of the target using the CASA implementation of \texttt{CLEAN} \citep{Schwab84-Cotton-Schwab-clean}. We additionally made sub-band images from groups of (or individual) independently calibrated 128 MHz spectral windows. In each image, we verified that the target location is not substantially contaminated by image artifacts. We then used the CASA task \texttt{imstat} to measure its flux, which is taken to be the value of the peak pixel. To avoid pixelation noise in this method of photometry, we chose a pixel scale that oversamples the beam size by a conservative margin ($\gtrsim 10\times$) in all images. To estimate the uncertainty, we use the RMS pixel value in a nearby aperture with no sources or image artifacts. The single-band fluxes and uncertainties are listed in Table \ref{tab:radio_fluxes}. The sub-band fluxes are used for model fitting, and are shown in Figure \ref{fig:radio-SED} along with the fluxes from VLASS observations and the FIRST nondetection. 
\\
\\As a check on the calibration in our follow up observations, we made images of the phase calibrator J1150-0023 (a quasar seperated by $\sim$4.7 degrees from the target) in every VLA spectral window, representing frequency chunks of 128 MHz with independent phase, amplitude and bandpass calibrations. In standard VLA observations (including ours), the target flux is established by applying the fluxscale corrections derived for the phase calibrator (which suffers from nearly the same atmospheric and instrumental effects as the target) to the target. Any major issues in the calibration of the target spectrum should therefore show up as irregularities in the calibrated fluxes of the phase calibrator. As shown in the top panel of Figure \ref{fig:pcal-SED}, the calibrated phase calibrator spectrum in each epoch varies smoothly within and across all receiver bands. Since each receiver band contains 2-3 independent hardware basebands, this suggests that there are no substantial systematics due to the instrument. In the first two epochs, the phase calibrator spectrum is well described by a single power law. In Epoch 3, we measure some spectral curvature which is well described in log-space by a parabola. The bottom panel of Figure \ref{fig:pcal-SED} shows that the flux measured in Epoch 2 is offset from the flux in Epoch 1 by $\sim$3\% with marginal frequency structure. Due to the new spectral curvature, the phase calibrator is up to 5\% dimmer and 10\% brighter in Epoch 3 relative to Epoch 1 in our observed frequencies. We take this range to be an upper limit on the systematic fluxscale error introduced by calibration, though we suspect that the true error is smaller since (1) this level of variability is commonly observed in quasars \citep[e.g.,][]{Liodakis17-OVRO-blazar-update-bimodal} and (2) we independently measure the same spectral index for VT 1137-0337 in all 3 epochs (Section \ref{sec:radio-spectral-index-fit}, Figure \ref{fig:radio-SED}). By rerunning the fits described in Section \ref{sec:radio-spectral-index-fit} with fluxes scaled by the phase calibrator model ratio, we find that if the full difference between calibrator epochs were attributed to systematic calibration errors, Epoch 3's best fit spectral index would change from $\alpha = 0.35 \pm 0.02$ to $\alpha = 0.38 \pm 0.02$.\\

\subsection{Modeling the radio spectra}
\label{sec:radio-spectral-index-fit}

\begin{figure*}
    \centering
    \includegraphics[clip=true, trim=0.15cm 0.3cm 0.2cm 0cm, width=\textwidth]{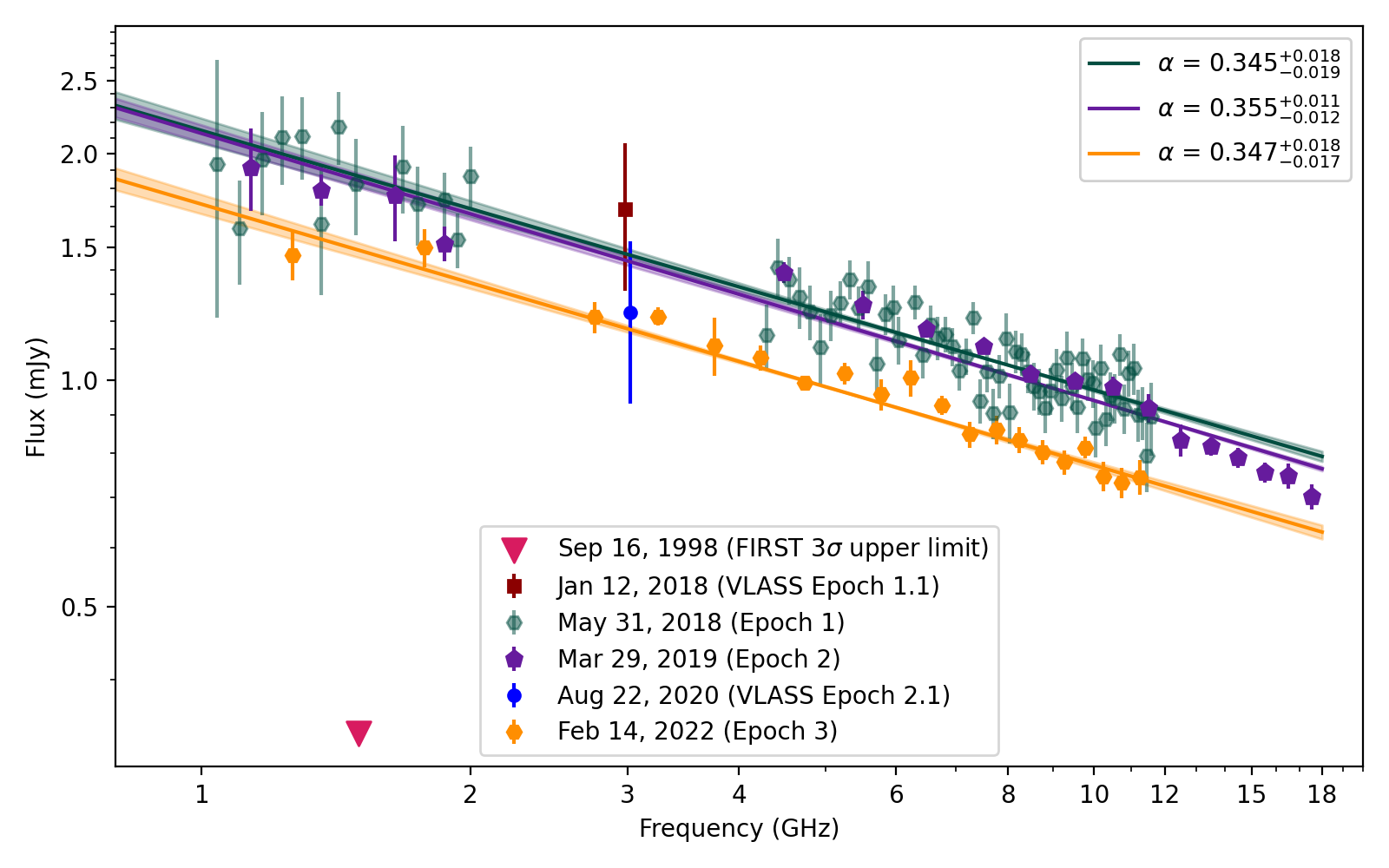}
    \caption{Radio observations of VT 1137-0337 from initial nondetection in FIRST to detections in VLASS and VLA followup epochs, along with single power law fits to followup epochs. To avoid overlap of the plotted errorbars, the frequencies of the VLASS epochs are shown with an arbitrary offset of $\pm$ 0.02~GHz from 3~GHz. The green line shows the median MCMC fit to Epoch 1, observed $\sim$20 years after FIRST: $S_{\nu} = (1.470 \pm 0.026~\textrm{mJy}) \left(\nu / \textrm{3 GHz}\right)^{-0.345 \pm 0.019}$. The purple line shows the median single power law fit to Epoch 2, observed $\sim$1 year after Epoch 1: $S_{\nu} = (1.441 \pm 0.020~\textrm{mJy}) \left(\nu / \textrm{3 GHz}\right)^{-0.355 \pm 0.012}$. The orange line shows the median fit to Epoch 3, observed $\sim$4 years after Epoch 1: $S_{\nu} = (1.171 \pm 0.012~\textrm{mJy}) \left(\nu / \textrm{3 GHz}\right)^{-0.347 \pm 0.018}$.}
    \label{fig:radio-SED}
\end{figure*}

In each epoch, we used the Markov Chain Monte Carlo code \texttt{emcee} \citep{Foreman-Mackey_2013_emcee} to fit the sub-band fluxes with a simple optically-thin synchrotron emission model: 
 
 \begin{equation}
 \label{eqn:spectral-index}
      S_{\nu, mod} = S_{3} \left(\frac{\nu}{3~\textrm{GHz}}\right)^{-\alpha},
 \end{equation}
 
\noindent where $S_{\nu,mod}$ is the model flux at frequency $\nu$, $\alpha$ is the optically thin spectral index, and $S_3$ is the flux at 3 GHz. We used the following log likelihood function which assumes that the uncertainties are Gaussian: 

\begin{equation}
\label{eqn:likelihood}
\ln \mathcal{L} = -\frac{1}{2} \sum_{n} \left[\frac{(S_{\nu,obs,n} - S_{\nu,mod,n})^2}{\sigma_n^2}  + ln (2\pi \sigma_n^2)\right],    
\end{equation}

\noindent where for the $n$-th data point, $S_{\nu,obs,n}$ and $S_{\nu,mod,n}$ are the observed and model fluxes at the data point's frequency $\nu$, and $\sigma_n$ is its uncertainty. In all epochs, we assumed top hat priors, with cutoffs of $0 < \alpha < 1$ and $1~\textrm{mJy} < S_3 < 2~\textrm{mJy}$. For Epoch 1, we found that the data are well described with the parameters $S_{3}$ = $1.470^{+0.026}_{-0.026}$ mJy, and $\alpha$ = $0.345^{+0.018}_{-0.019}$, where the central value and upper and lower uncertainties are taken from the 50th, 86th - 50th and 50th - 14th percentile values of their posterior distributions. For Epoch 2, the best fit power-law model is $S_{3}$ = $1.44^{+0.020}_{-0.020}$ mJy and $\alpha$ = $0.355^{+0.011}_{-0.012}$. For Epoch 3, we find that $S_{3}$ = $1.171^{+0.012}_{-0.012}$ mJy and $\alpha$ = $0.347^{+0.018}_{-0.017}$. The data and best fit models are plotted in Figure \ref{fig:radio-SED}.
\\
\\In Epochs 1 and 2, VT 1137-0337's best fit radio spectra are consistent within 1$\sigma$ uncertainties in each parameter, indicating slow evolution (or a lack thereof) over the $\sim$1 year between epochs. The implied fade rate from the first two epochs is (2.3 $\pm$ 2.8)\% per year. In Epoch 3, its spectral index remains constant but the spectrum's overall amplitude is fainter than that of Epoch 1 by a margin of ($S_{3, \textrm{Epoch 1}} - S_{3, \textrm{Epoch 3}}) / S_{3, \textrm{Epoch 1}}$ = (20 $
\pm$ 2)\%, corresponding to a fade rate of (5.3 $\pm$ 0.4)\% per year. If we attribute the full difference in the phase calibrator's spectrum to systematic calibration errors, the overall level of fading increases to (25 $\pm$ 2)\%, or (6.9 $\pm$ 0.4)\% per year.

\begin{deluxetable*}{lclllllll}
\tablecolumns{8}
\tablecaption{Single-band radio fluxes of VT 1137-0337\label{tab:radio_fluxes}}
\tablehead{
\colhead{Date} & \colhead{$\Delta$t} & \colhead{S$_{\textrm{1.5 GHz}}$} & \colhead{S$_{\textrm{3 GHz}}$} & \colhead{S$_{\textrm{6 GHz}}$} & \colhead{S$_{\textrm{10 GHz}}$} & \colhead{S$_{\textrm{15 GHz}}$} & \colhead{Epoch} \\
\colhead{} & \colhead{Years} & \colhead{mJy} & \colhead{mJy} & \colhead{mJy} & \colhead{mJy} & \colhead{mJy} & \colhead{}
}
\startdata
\hline 
\\ 
Sep 16, 1998 & -19.3 & $<$ 0.34 &  & &  &  & FIRST (3$\sigma$)\\
Jan 12, 2018 & 0 & & 1.69 $\pm$ 0.14  & &  &  & VLASS 1.1\\
May 31, 2018 & 0.38 & 1.86 $\pm$ 0.06 & & 1.21 $\pm$ 0.02 & 1.01 $\pm$ 0.01  &  & Epoch 1\\
Mar 29, 2019 & 1.21 & 1.86 $\pm$ 0.14  &  & 1.23 $\pm$ 0.03 & 0.99 $\pm$ 0.01 & 0.77 $\pm$ 0.01 & Epoch 2\\
Aug 22, 2020 & 2.61 & & 1.23 $\pm$ 0.14 & & & & VLASS 2.1\\
Feb 14, 2022 & 4.09 & 1.41 $\pm$ 0.10 & 1.19 $\pm$ 0.03 & 0.914 $\pm$ 0.02 & 0.75 $\pm$ 0.01 & & Epoch 3
\enddata
\tablecomments{Fluxes from single-band imaging of VT 1137-0337. As in the text, Epochs 1, 2, and 3 refer to follow-up epochs taken with the VLA rather than epochs of VLASS. Uncertainties are estimated using the RMS in a region of the image free of sources. There is an additional $\sim$5\% systematic fluxscale uncertainty in imaging of followup epochs \citep[][]{Perley-Butler-2017}, and a $\sim$20\% systematic error in fluxes derived from VLASS quicklook images.}
\end{deluxetable*}

\subsection{Optical spectra}
\label{subsec:optical}

The host galaxy of VT 1137-0337 was observed as part of the SDSS Legacy Survey on January 20, 2002 ($\sim$3.35 years after nondetection in FIRST and $\sim$16 years before first detection in VLASS). The SDSS spectrum has a spectral resolution of $\Delta \lambda / \lambda$ $\sim$ 2000, and detects numerous emission lines characteristic of star formation. It is sensitive to a 3'' diameter region encompassing the location of the radio transient.
\\
\\To check for any spectral evolution or new spectrally resolved features, we took a follow-up spectrum with the Keck Low Resolution Imaging Spectrograph (LRIS) \citep[][]{Oke1995-LRIS} on Feb 13, 2021 ($\sim$3 years after first detection). We used the 1 arcsecond-wide longslit setup with the 400/3400 grism on the blue arm and the 400/8500 grating on the red arm with a central wavelength of 7830 \AA. These settings correspond to a wavelength coverage of $\sim$3100 - 10,000 \AA~ at a FWHM spectral resolution of $\Delta \lambda / \lambda$ $\sim$ 1000. To acquire the target, we applied a blind offset from a nearby star, resulting in a spectrum of the region around the radio transient of order the slit width (1.0''). We exposed for 20 minutes in both arms, resulting in a typical per-pixel 1$\sigma$ noise of $\sim (1-2) \times$~10$^{-18}$~ erg/s/cm$^2$. We processed the optical spectra using the LRIS data reduction pipeline \texttt{LPIPE} \citep{Perley19-Lpipe}. 

\subsection{Fitting the optical emission features}
\label{sec:fitting-optical-emission-features}

We used \texttt{emcee} to fit the H$\alpha$, H$\beta$, NII 6548/6584\AA, SII 6716/6730\AA, OIII 4363/4959/5007\AA, OII 7320/7330\AA, and OI 6300\AA~emission features in both the LRIS and SDSS spectra. We fit these features in six separate wavelength chunks: 6490-6630\AA~(H$\alpha$/NII), 4750-5200\AA~(OIII 4959+5007/H$\beta$), 4350-4380\AA~(OIII 4363), 6700-6750\AA~(SII 6716/6730), 7300-7350\AA~(OII 7320/7330), and 6270-6350\AA~(OI 6300). We modeled each emission feature as a single-component Gaussian where the amplitude and width are allowed to vary freely (with both constrained to be positive and less than a large value for the purposes of convergence). Within each complex, we marginalized over three additional nuisance parameters: a linear slope and flux offset for the underlying continuum, and a wavelength shift to correct for wavelength calibration errors. We used a uniform prior that loosely constrains the slope and offset to within an order of magnitude of an initial guess from a least-squares fit to the data with the emission lines masked. The wavelength shift was limited to $\pm$1\AA. 
\\
\\The best fit fluxes and their uncertainties listed in Table \ref{tab:optical_line_fluxes}, and the gas properties inferred from these features are discussed in Section \ref{sec:Derived properties from optical features}. No features are spectrally resolved in either observation. We find that despite the $\sim$20 year gap and the difference in spatial scales, the fluxes are similar in all measured lines. The $\sim$5\% higher fluxes in the SDSS spectrum are easily explained by either systematic flux calibration uncertainties, or the larger effective aperture of the SDSS fiber. Thus, on the scale of $\sim$1'', corresponding to $\sim$550~pc at the distance to the host galaxy, we do not detect any significant evolution in the optical spectrum at the location of the transient between 2002 ($\sim$3 years after nondetection in FIRST) and 2021 ($\sim$22 years after nondetection in FIRST and $\sim$3 years after first detection in VLASS). 
\\
\\We additionally checked for emission from the high ionization potential [Fe X] line at 6374\AA, used in a search for AGN in dwarf galaxies by \citet[][]{Molina21b-AGN-sample}. We do not detect this line in either spectrum above the local noise.

\begin{deluxetable}{lcrr}
\centering
\tablecolumns{4}
\tablecaption{Emission line fluxes at the location of VT 1137-0337\label{tab:optical_line_fluxes}}
\tablehead{
\colhead{Line} & \colhead{$\lambda_{\textrm{rest}}$} & \colhead{Flux (SDSS)} & \colhead{Flux (LRIS)} \\
\colhead{} & \colhead{(\AA)} & \colhead{(10$^{-16}$ erg s$^{-1}$ cm$^{-2}$)} & \colhead{(10$^{-16}$ erg s$^{-1}$ cm$^{-2}$)} 
}
\startdata
\hline 
H$\alpha$ & 6562.8  & 387.7$^{+0.4}_{-0.4}$ &  357.6$^{+0.5}_{-0.5}$ \\
H$\beta$ & 4861.4 & 112.6$^{+0.2}_{-0.2}$ & 107.8$^{+0.5}_{-0.5}$ \\
NII & 6548.0 & 5.8$^{+0.5}_{-0.4}$ & 5.0$^{+0.1}_{-0.1}$ \\
NII & 6583.5 & 18.6$^{+0.6}_{-0.6}$ & 17.2$^{+0.2}_{-0.2}$ \\
OIII & 4363.2 & 5.7$^{+0.9}_{-0.8}$ & 5.5$^{+0.3}_{-0.3}$\\
OIII & 4958.9  & 176.7$^{+0.3}_{-0.3}$ & 174.0$^{+0.6}_{-0.5}$\\
OIII & 5006.8 & 519.3$^{+0.7}_{-0.7}$ & 499.8$^{+0.9}_{-0.9}$\\
OI & 6300.3 & 5.2$^{+0.5}_{-0.5 }$ & 4.6$^{+1.9}_{-1.8}$\\
SII & 6716.4 & 28.9$^{+0.8}_{-0.8}$ & 26.6$^{+0.2}_{-0.2}$\\
SII & 6730.8 & 21.9$^{+0.7}_{-0.8}$ & 19.9$^{+0.2}_{-0.2}$\\
OII & 7320.1 & 4.1$^{+0.8}_{-0.8}$ & 3.0$^{+0.2}_{-0.2}$\\
OII & 7330.2 & 2.7$^{+0.5}_{-0.5}$& 3.0$^{+0.2}_{-0.2}$\\
\enddata
\tablecomments{The reported uncertainties are statistical, reflecting the 86th-50th and 50th-16th percentiles of the posterior distributions. There is an additional relative systematic uncertainty between the two spectra due to the different apertures as well as an absolute systematic uncertainty in the spectrophotometric flux calibration in each epoch.}
\end{deluxetable}

\subsection{X-ray upper limit from 2006}
\label{subsec:X-ray}

VT 1137-0337 is located near the edge of an archival XMM-Newton image (observation ID 0305801101; PI Ponman) observed with the European Photon Imaging Camera (EPIC) PN CCD for 21.8 kiloseconds on June 10, 2006. It is not significantly detected in the image. Assuming a galactic HI column of 3 $\times$ 10$^{20}$ cm$^2$ and a power law model with photon index of 1.34 (equal to the radio spectral index), one count corresponds to 3.7 $\times$ 10$^{-12}$ erg s$^{-1}$ cm$^{-2}$ as calculated with the WebPIMMS tool, or a luminosity of 6.5 $\times$ 10$^{42}$ erg/s at the distance to VT 1137-0337. Assuming a Poisson distribution in counts, this nondetection rules out a 0.1 - 15~keV source (or set of sources within a 4.1'' region) active in 2006 with a luminosity of $2 \times 10^{43}$~erg~s$^{-1}$ at the $98\%$ level. We caution, however, that the radio emission first appeared sometime between September 1998 and January 2018 (Table \ref{tab:radio_fluxes}). If the radio emission predates June 2006 and was at the same flux or brighter, then the X-ray / radio luminosity ratio was $r_{X/R} \lesssim$ 10$^{5}$ at that time. If not, this is an upper limit on the quiescent X-ray luminosity of the galaxy. We note that this ratio is not constraining for many astrophysical source classes. In particular, pulsar wind nebulae are generally observed (and expected) to have ratios $r_{X/R} \lesssim 10^{-4}$ \citep[e.g.,][]{Reynolds-Chevalier1984-PWN-lightcurve}. 

\subsection{Nondetection of archival transient counterparts}
\label{subsec:archival}

We searched a variety of archival catalogs for potential transient counterparts to VT 1137-0337. These included the CHIME FRB catalog \citep[][]{CHIME-FRB-catalog-2021}, the Open Supernova Catalog \cite{Guillochon2017-OSC}, the MAXI \citep{Serino-MAXI-GRBs}, INTEGRAL \citep{Rau05-INTEGRAL-GRB-catalog}, and Swift \citep{Lien2016-SWIFT-GRB-catalog} GRB catalogs, and the Transient Name Server\footnote{The Transient Name Server (TNS) is the official IAU repository for reporting transients and can be found at \url{https://www.wis-tns.org/}}. We found no significant matches. We additionally used the ASASSN lightcurve service \citep{Kochanek-ASASSN-lightcurve,Shappee14-ASASSN} to check for optical variability at this location, and found no significant increase in the optical flux beyond the typical ASASSN V band detection limit of $\sim$17 (corresponding to an absolute magnitude of -18.4). This rules out association with superluminous supernovae that exploded after the first ASASSN observation on Jan 29, 2012 \citep{Gal-Yam-SLSNe_review}, but does not rule out the majority of core collapse supernovae \citep[e.g.,][]{Taddia13-IIn-lightcurves,Taddia18-stripped-env-lightcurves,Hicken17-CCSNe-lightcurves}. A similar constraint can be inferred from the lack of an apparent point source in the SDSS image of the host galaxy, which rules out a superluminous supernova in the $\sim 1-3$ years before the observation date (Jan 6, 2006). 

\section{Properties of the transient and its environment}
\label{sec:analysis}

 In this section we explore the physical conditions that can produce the observed features of VT 1137-0337 and its environment, and in Section \ref{sec:comparisons}, we discuss possible astrophysical scenarios that can give rise to these conditions. 

\subsection{Properties of the surrounding starburst}
\label{sec:Derived properties from optical features}

One contextual clue to the origin of VT 1137-0337 comes from the properties of the host galaxy and the local region surrounding the radio transient. We can diagnose the state of the ionized gas near the transient using the optical emission lines discussed in Section \ref{sec:fitting-optical-emission-features}. We use the LRIS flux values due to the tighter spatial area probed around the transient. The quantities presented are averaged over the $\sim$1'' LRIS slit width, corresponding to a linear diameter of $\sim$560~pc. The region directly influenced by the transient is much smaller ($< 6$pc; Section \ref{sec:spectral_breaks}), and may have different properties.
\\
\\The H$\alpha$/H$\beta$ ratio can be used to estimate the extinction due to dust along the line of sight. We measure a ratio of 3.32 $\pm$ 0.02, corresponding to a reddening of E(B-V) $\sim$ 0.13 \citep[][]{Dominguez-2013-extinction}. This reddening corresponds to an extinction of $\sim$0.42 mag for the H$\alpha$ line \citep[][]{Calzetti-2000-extinction}.
\\
\\The [SII] 6717\AA~/6731\AA~ ratio is sensitive to the density of the emitting region, with a small dependence on the metallicity. We measure a ratio of 1.34 $\pm$ 0.02, corresponding to an electron density of $\sim$60  cm$^{-3}$ \citep[][]{Kewley2019-line-ratio-diagnostics}.
\\
\\The faint OIII 4363\AA~line allows us to directly measure the temperature of electrons in the ionized gas producing the emission lines. Since the inferred density is well within the limit where collisional de-excitation is negligible, the temperature can be determined by the ratio of oxygen lines using the relation from \citet[][]{Osterbrock-Ferland-AGN2}:
\begin{equation}
\label{eqn:OIII-temp}
    \frac{j_{4959} + j_{5007}}{j_{4363}} = \frac{7.90~\textrm{exp}(3.29 \times 10^4 / T)}{1 + 4.5 \times 10^{-4}~n_e/T^{1/2}}.
\end{equation}

\begin{figure}
    \centering
    \includegraphics[width=0.5\textwidth]{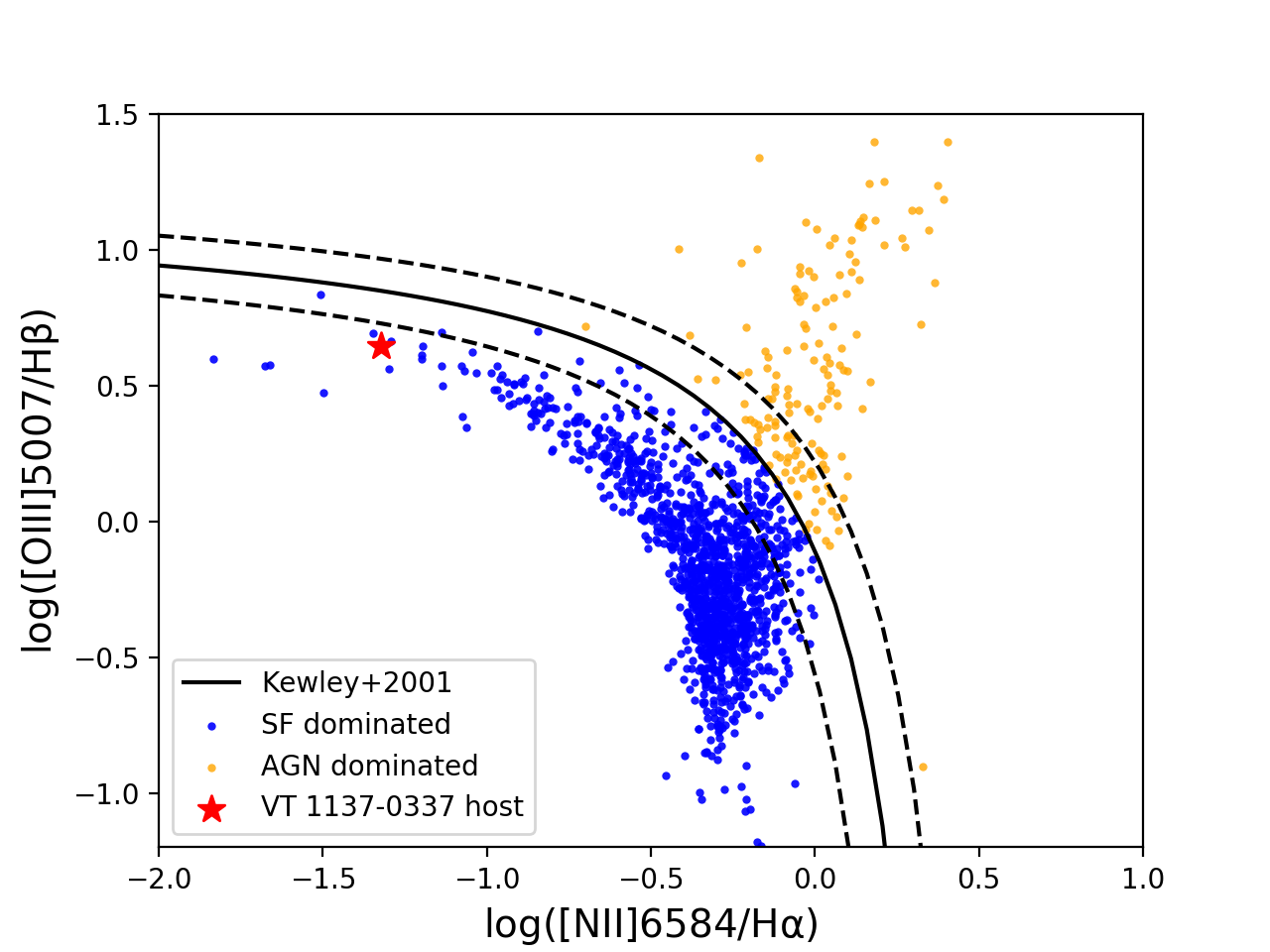}
    \caption{The position of VT 1137-0337's host galaxy (SDSS J113706.18-033737.1}) on a BPT diagram \citep{BPT_diagram-Baldwin-Phillips-Terlevich-1981}. The dividing line between AGN and star formation dominated galaxy spectra from \citet{BPT_diagram-Kewley2001} is shown with $\pm$1$\sigma$ uncertainties (dotted lines). Blue and orange dots show SDSS galaxies classified respectively as star forming or AGN according to this scheme by \citet{Vanderplas2012-astroML}.
    \label{fig:BPT}
\end{figure}

\noindent We measure a line ratio of 122~$\pm$~7, corresponding to a temperature of $\sim$12,000 K.
\\
\\The density and temperature of the surrounding region is consistent with values commonly observed in star forming regions \citep[][]{Draine-physics-of-ISM}. The electron temperature can be used to measure the metallicity of the region to a typical intrinsic scatter of $\sim$0.1 dex \citep[][]{Kewley-Ellison-2008-metallicity-calibrations}. To do so, we adopt the calibration of \citet[][]{Izotov2006-metallicity-calibration} assuming that the total oxygen abundance is the sum of the O$^+$ and O$^{++}$ abundances \citep[][]{Kewley-Ellison-2008-metallicity-calibrations}. We find that 12 + log O$^{+}$/H = 7.68 and 12 + log O$^{++}$/H = 7.99, for a total metallicity of 12 + log O/H = 8.16. This metallicity is 30\% of the solar value \citep[12 + log O/H = 8.69;][]{Asplund-Zsun-review}.
\\
\\To check for signs of AGN activity and to further diagnose the state of the interstellar medium (ISM), we plotted the nucleus of the galaxy on the classical Baldwin, Phillips, and Terlevich diagram \citep[BPT;][]{BPT_diagram-Baldwin-Phillips-Terlevich-1981}. As shown in Figure \ref{fig:BPT}, the source lies on the star forming side, indicating that any AGN activity is not apparent in optical diagnostics. The position of SDSS J113706.18-033737.1 on the left side of the star forming sequence (log OIII 5007/H$\beta$ = 0.66, log NII 6584/H$\alpha$ = -1.32) indicates that the overall star forming complex is irradiated by a hard ionizing photon spectrum \citep[e.g.,][]{Richardson2016-ionization}. At the metallicity inferred above, both line ratios are higher than expected for the highest ionization parameter model presented in Figure 1 of \citep[][]{Kewley2013-ionization}. This suggests that the ionization parameter $q$ exceeds $10^{8.3}$ cm/s, which is substantially harder than the range measured in both local HII regions \citep[][]{Dopita-2000-HII-region-ionization} and star forming galaxies \citep[][]{Moustakas-SINGS-optical-spectroscopy}. This suggests that the transient is surrounded by young, massive, hot stars (e.g. O stars and Wolf-Rayet stars) \citep[][]{Kewley2013-ionization}. 
\\
\\Another indication of a young starburst comes from the high extinction corrected H$\alpha$ luminosity of 9.3 $\times$ 10$^{40}$ erg/s measured in the $\sim$1 arcsecond region around VT 1137-0337. Adopting the calibration of \citet[][]{Murphy2011_SF_calibration}, this corresponds to a star formation rate (SFR) of $\sim$0.5 $M_{\odot}$/yr. If the star formation were to continue at a constant rate, the galaxy \citep[of stellar mass 10$^{8.3} M_{\odot}$;][]{Salim2018-SFR-Mass} would double its stellar mass in the next $\sim$1~Gyr. Nearly all of the star formation is concentrated in the $\sim$ 280 pc radius region surrounding VT 1137-0337, as evidenced by the similar H$\alpha$ fluxes in the LRIS and SDSS spectra which probe radii of $r \sim$ 280~pc and 840~pc respectively. The star formation surface density within a circle of radius $r = 280$~pc is $\Sigma_{\textrm{SFR}} \approx$ 6 M$_{\odot}$ yr$^{-1}$ kpc$^{-2}$. This value is in the high star formation density tail of blue compact star forming galaxies measured by \citet[][]{Perez-Gonzalez-2003-BCDs} and luminous circumnuclear star forming regions \citep[][]{Kormendy-Kennicutt-2004, Kennicutt-Evans-2012}. The surrounding HII region would be among the most H$\alpha$ luminous of the $\sim$18,000 star forming regions compiled by \citet[][]{Bradley2006_HII-region_lum_function}. This H$\alpha$ flux traces ionizing photons from O and B stars, with disproportionate contributions from the youngest, most massive stars. As these stars explode in supernovae over the course of a few Myr, the H$\alpha$ flux is expected to fade rapidly as the gas recombines. The high observed H$\alpha$ luminosity therefore indicates that massive stars were recently formed in the region surrounding VT 1137-0337. In particular, the H$\alpha$ flux of a \citet[][]{Salpeter1955-IMF} initial mass function of stars formed at any given time is expected to fade by $\sim$2 orders of magnitude after $\sim$10~Myr \citep[][]{Leitherer-1999-Starburst99}. If the present-day massive stars are relics of an earlier generation, the star formation rate and surface density at that the time of formation would have been $\sim$2 orders of magnitude more extreme than what we have already observed. This is not likely, so we infer that the massive stars powering the high H$\alpha$ luminosity surrounding VT 1137-0337 must be younger than a few Myr.

\begin{deluxetable}{lll}
\tablecolumns{3}
\tablecaption{The environment around VT 1137-0337\label{tab:derived-properties-optical}}
\tablehead{\colhead{Description} & \colhead{Symbol} & \colhead{Value}}
\startdata
\hline 
Galaxy center RA &  & 11:37:06.18\\
Transient RA & & 11:37:06.19\\
Galaxy center Dec &  & -03:37:37.16\\
Transient Dec & & -03:37:37.29\\
Galaxy stellar mass & M$_*$ & 10$^{8.3} M_{\odot}$\\
Electron temperature & T$_e$ &  12000 K \\
Electron density & n$_e$ & 60 cm$^{-3}$\\
Reddening & E(B-V) & 0.13 mag\\
H$\alpha$ extinction & A(H$\alpha$) & 0.42 mag\\
Metallicity & Z & 0.3 Z$_{\odot}$\\
Star formation rate & SFR & 0.5 M$_{\odot}$ yr$^{-1}$
\enddata
\tablecomments{The host galaxy mass is from SED fitting reported in the GALEX-SDSS-WISE Legacy Catalog \citep[GSWLC;][]{Salim2018-SFR-Mass}. The SFR is from the extinction corrected H$\alpha$ flux, and is consistent with the GSWLC value $0.46 M_{\odot}$/yr. The transient position is fitted from a 10~GHz VLA follow-up observation. All other quantities are derived from a Keck I/LRIS spectrum and are representative of a $\sim$280~pc radius region surrounding VT 1137-0337.}
\end{deluxetable}

\subsection{Flat-spectrum radio emission mechanisms}
\label{sec:flat-spectrum-emission-mechanisms}

Broadband emission at GHz frequencies from extragalactic sources is typically thought to originate from either free-free or synchrotron emission. In Section \ref{sec:free-free-emission} we outline why VT 1137-0337's luminosity and variability timescale are not consistent with having a substantial free-free emission fraction. A full argument is presented in Appendix \ref{app:free-free}. In Section \ref{sec:synchrotron-emission}, we discuss how synchrotron emission is consistent with the observations, though the flat spectrum is unlike typical synchrotron transients in that it cannot be explained with a single shock-accelerated particle distribution. 

\subsubsection{Free-free emission}
\label{sec:free-free-emission}

Free-free emission is a tempting explanation for the radio emission of VT 1137-0337, given that its natural flat spectrum ($\nu^{-0.12}$; \citet[][]{Draine-physics-of-ISM}) can add on to a steeper spectrum component to create a power law similar to what we observe. This combination is commonly observed in extragalactic star forming regions \citep[e.g.,][]{Linden2020-SFRS}. However, with reasonable temperatures and densities, high luminosities of $\sim$10$^{28}$~erg/s/Hz can only be produced via free-free emission from a region $\sim$hundreds of parsecs in size. The required size is not compatible with transient emission appearing on human timescales. In Appendix \ref{app:free-free}, we show that the conditions required to produce VT 1137-0337's luminosity within the upper limit on size set by causality ($R < c$~(20 years) = 6.1~pc) are extreme: a temperature $T \gtrsim 10^{7}$~K and a thermal energy $U \gtrsim 10^{54}$~erg. We conclude that the fraction of VT 1137-0337's observed radio luminosity due to free-free emission is $<<$ 1. We also note that the expected 3~GHz flux due to free-free emission from star formation in the host galaxy is $\sim 50 \mu$Jy \citep[][Figure \ref{fig:galaxy_SED}]{Murphy2011_SF_calibration}, far lower than the $\sim$1.5~mJy we observe.

\subsubsection{Synchrotron emission}
\label{sec:synchrotron-emission}

Having ruled out free-free emission, we are left with synchrotron emission as the likely source of VT 1137-0337. However, its spectral index of $\alpha = 0.35$ (Section \ref{sec:radio-spectral-index-fit}), is much flatter than other synchrotron transients and most synchrotron sources in general and requires further explanation. As we discuss below, the flat spectrum cannot be produced by an ordinary shock and instead requires an alternate model.
\\
\\The simplest model for a power law synchrotron spectrum is a single population of emitting electrons with a power law energy distribution $N(E) \propto E^{-p}$. At frequencies below the synchrotron cooling spectral break, the spectral index $\alpha$ is related to the energy index $p$ by $\alpha = (p-1)/2$. Above the cooling break, the spectral index steepens by 0.5 to $\alpha = p/2$. The single population model implies an energy index $p < 2$, which is too hard to be produced by diffusive shock acceleration (DSA), the mechanism thought to be responsible for most synchrotron transients. In non relativistic shocks, DSA predicts that $p = (r+2)/(r-1)$, where $r$ is the ratio by which the shocked gas is compressed \citep[e.g.,][]{Blandford-Eichler-shocks}. The compression ratio is given by the shock jump conditions: $r = \frac{(\gamma + 1) M^2}{(\gamma -1)M^2 + 2}$, where $M$ is the Mach number (the ratio of the shock speed to the sound speed in the pre-shock gas), and $\gamma$ is the adiabatic index, equal to 5/3 for a monatomic gas. In the limit of a strong shock ($M \gtrsim 10$) in a monatomic gas, $r \approx 4$, and thus $p$ = 2. Nonlinear factors such as the dynamical effect on the shock of the accelerated particles themselves can steepen the energy distribution to $p = 2.5 - 3$ \citep[][]{Jones-Ellison-1991-shock-theory}, a range of values often seen in astrophysical sources. If the shock is relativistic, there are further complications due to e.g. particles being unable to return to the shock front after first crossing. In these conditions, the limiting steepness is estimated to be p $\approx$ 2.3 \citep[][]{Pelletier-2017-relativistic-shocks}. In all cases, ordinary shocks are expected to produce substantially steeper spectra than what we observe. 
\\
\\A variety of astrophysical source classes do display power law spectra that are flatter (over a large bandwidth) than $\alpha = 0.5$, the spectral index corresponding to the $p = 2$ DSA limit. These include pulsar wind nebulae, X-ray binaries in the low/hard state, and a subset of AGN which tend to have jets pointed along our line of sight. Theoretical explanations for their flat spectra often invoke one or more of the following: (1) an alternate particle acceleration mechanism capable of producing an intrinsically hard electron energy distribution, (2) a regular electron energy distribution with a flattened spectrum due to propagation in a non uniform magnetic field, or (3) a flat spectrum constructed from the (discrete or continuous) sum of ordinary self-absorbed synchrotron spectral components. We discuss these mechanisms further in Section \ref{sec:discussion}, where we consider the possible astrophysical sources that may be powering VT 1137-0337.

\subsection{Constraints on the emitting region}
\label{sec:spectral_breaks}

Even with a possibly multi-component synchrotron spectrum where we do not know the value of $p$, we can still derive useful constraints from the overall luminosity of the emission and upper limits on the synchrotron self-absorption and free-free absorption frequencies. In this section, we use these properties to roughly constrain the size and magnetic field of the emitting region, the cooling timescale of the emitting electrons, and the line-of-sight column density of ionized gas. For simplicity, we assume that the emission is from a single hard distribution of electrons with $p = 1.7$. These arguments depend only weakly on $p$. Likewise, if the emission is the sum of components at different optical depth, the arguments should still be applicable to the lowest frequency component observed, which would comprise a large fraction of the total luminosity. In all cases, we stress that the results presented in this section should be regarded as order of magnitude estimates.
\\
\\From the lack of an observed peak in our follow up observations, we infer that the $\tau = 1$ frequencies for synchrotron self-absorption ($\nu_{a}$) and free-free absorption ($\nu_{ff}$) are both $<$ 1 GHz. The upper limit $\nu_a < 1$~GHz allows us to set a lower limit on the size of the nebula producing the synchrotron emission. If we assume energy equipartition between the radiating particles and the magnetic field, the angular diameter of the emitting source is $\theta \gtrsim$ 0.2~mas \citep[][]{Scott-Readhead-1977-equipartition}. At the 115~Mpc angular diameter distance to the source, this corresponds to $D \gtrsim$ 0.1~pc. The lower limit on the source size corresponds to a volume: $V = \frac{4}{3} \pi (D/2)^3 \gtrsim 10^{52}$cm$^3$. From the ~20 year separation between FIRST and Epoch 1, light travel time arguments require that the region responsible for the excess emission is $\lesssim$ 6.1~pc in size.
\\
\\We can use this limit on the source size along with classical energy minimization arguments \citep[][]{Burbidge-Burbidge-1957, Scott-Readhead-1977-equipartition} to estimate the typical strength of the magnetic field $B$ in the emitting region. To do so, we note that the energy stored in the magnetic field is $U_B = V \frac{B^{2}}{8\pi} = \frac{4}{3}\pi R^3 \frac{B^2}{8\pi}$ for a spherical emitting region of radius $R$. Meanwhile, the energy in particles is $U_p = A g(\alpha) L B^{-{3/2}}$ \citep[][]{Scott-Readhead-1977-equipartition}, where $A$ is a constant equal to $1.586 \times 10^{12}$ in cgs units, $L$ is the region's synchrotron luminosity in the range from $\nu_1$ to $\nu_2$ (in units of Hz), $\alpha = -0.35$ is the spectral index in the convention of Equation \ref{eqn:g-alpha}, and $g(\alpha)$ is a quantity from an integral over the electron energy distribution that encapsulates the dependency on frequency range and spectral index:

\begin{equation}
\label{eqn:g-alpha}
    g(\alpha) = \frac{2\alpha + 2}{2\alpha + 1}\left[\frac{\nu_2^{(2\alpha+1)/2} - \nu_1^{(2\alpha+1)/2}}{\nu_2^{\alpha+1} - \nu_1^{\alpha+1}}\right].
\end{equation}

\noindent  The two energies $U_B \propto B^2$ and $U_p \propto B^{-3/2}$ have opposite dependencies on $B$. Thus, there is a value $B_{\textrm{min}}$ that minimizes the total energy $U = U_B + U_p$, which can be found by setting $dU/dB = 0$. The solution is

\begin{equation}
    \label{eqn:equipartition-B}
    B_{\textrm{min}} = \left(\frac{6\pi A g(\alpha) L}{V}\right)^{2/7}.
\end{equation}

\noindent To estimate the value of $B_{\textrm{min}}$ for VT 1137-0337, we integrate over our observed epoch 2 spectrum (from $\nu_1 = 1 \times 10^9$~Hz to $\nu_2 = 1.8 \times 10^{10}$~Hz), resulting in $L \approx 3.2 \times 10^{38}$~erg~s$^{-1}$. Over this frequency range g($\alpha$) = 1.3 $\times$ 10$^{-5}$. For the minimum emitting region size of $R \approx 10^{17}$~cm, we then have $B_{\textrm{min}} \approx 0.04$~G. If $R$ is greater, $B_{\textrm{min}}$ decreases as $R^{-{6/7}}$, to a minimum value of $\sim$6 $\times$ 10$^{-4}$~G at the causality limit of $R = 6.1$~pc.
\\
\\The quantity $B_{\textrm{min}}$ is a reasonable order-of-magnitude estimate for the true magnetic field $B$. If $B >> B_{\textrm{min}}$, then the total energy must increase by $(B/B_{\textrm{min}})^2$. Likewise, if $B << B_{\textrm{min}}$, the total energy must again increase by $(B/B_{\textrm{min}})^{-3/2}$. Perturbations of more than an order of magnitude will thus lead to large increases in the energy. Such an increase may be difficult to explain when considering the energy already required in the emitting region. At the lower limit of $R \approx 10^{17}$~cm, the minimum energy in the magnetic field alone is $U_{B} = \frac{4}{3} \pi R^3 (B_{\textrm{min}}^2 / 8\pi) \approx 4 \times 10^{47}$~erg, comparable to the magnetic energy in an energetic supernova-driven shock \citep[e.g.,][]{Ho19-2018cow}. If $R$ is greater, then the energy scales as $U_B \propto R^{9/7}$, up to $\sim 4 \times 10^{50}$ erg at $R = 6.1$~pc. 
\\
\\One consequence of our magnetic field estimate  is that the spectral break frequency $\nu_c$ associated with synchrotron cooling is not likely to be $<$~1~GHz. The timescale for synchrotron cooling is
\begin{equation}
    \label{eqn:cooling-time}
    t_c \approx 1300 \left(\frac{B}{0.01 G}\right)^{-3/2} \left(\frac{\nu_c}{\textrm{GHz}}\right)^{-1/2} \textrm{years},
\end{equation}

\noindent and the maximum value of $B_{\textrm{min}}$ with our radius constraints is $\sim$0.04~G. This implies that the minimum synchrotron cooling time at 1~GHz is $t_{c,min} \approx 120$ years. At our highest observed frequency of 18~GHz, $t_{c,min} \approx 27$~years. Both timescales are longer than the $\sim$20~year span over which the new emission appeared. If the electrons radiating at those frequencies were accelerated after the FIRST nondetection in 1998, they have not yet had enough time to cool. In the case of a single-component nonthermal electron energy distribution, this implies that our observed spectrum of $S_{\nu} \propto \nu^{-0.35}$ maps to $p = 1.7$ rather than $p = 0.7$. 
\\
\\Finally, the lack of free-free absorption along the line of sight at 1~GHz ($\tau << 1$) implies either a low column density of ionized electrons along the line of sight, or a high temperature for those electrons. We first check that this is the case for a source embedded in the HII region surrounding VT 1137-0337. From the density and temperature in Section \ref{sec:Derived properties from optical features}, and equation \ref{eqn:ff-opacity}, the optical depth to free-free absorption at 1~GHz is $\tau \approx 0.09 (R/\textrm{100 pc})$, where $R$ is the distance that the radio photons must travel through the ionized gas. The effective size of the ionized region is of order the size of a Str\^{o}mgren sphere with the ionizing photon production rate implied by the extinction corrected H$\alpha$ emission $Q_0 \approx 6.9 \times 10^{52}$ s$^{-1}$ \citep[][]{Murphy2011_SF_calibration}, which is $\sim$90~pc \citep[][]{Draine-physics-of-ISM}. Thus, the surrounding HII region should not contribute substantially to the free-free absorption, though if the source were embedded at its center, it may become a factor for observations at frequencies of a few hundred MHz. However, a high density ionized shell at a typical HII region temperature is excluded. From Equation \ref{eqn:ff-opacity}, the limit $\tau << 1$ at $1$~GHz indicates that at a temperature of T = 10$^4$~K, the quantity $\left(n_e/\textrm{cm}^{-3}\right)^2(s/\textrm{pc}) < 3 \times 10^{6}$, where $n_e$ is the average density of ionized gas and $s$ is the thickness of the ionized region. If a hypothetical shell has thickness $s \sim 10^{16}$~cm, the density $n_e < 3 \times 10^{4}$~cm$^{-3}$. If it is more extended, the density constraint becomes more stringent. This excludes the densest winds that have been observed around some supergiant stars \citep[][]{Smith14-mass-loss-review}, though we note that the constraint applies only along the line of sight and dense but asymmetric gas from e.g. binary interaction is allowed \citep[][]{Dong21-VT1210}. \\\\

\section{Astrophysical analogues and possible models}
\label{sec:comparisons}

\begin{figure*}
    \centering
    \includegraphics[clip=true, trim=0cm 0cm 0cm 0cm, width=\textwidth]{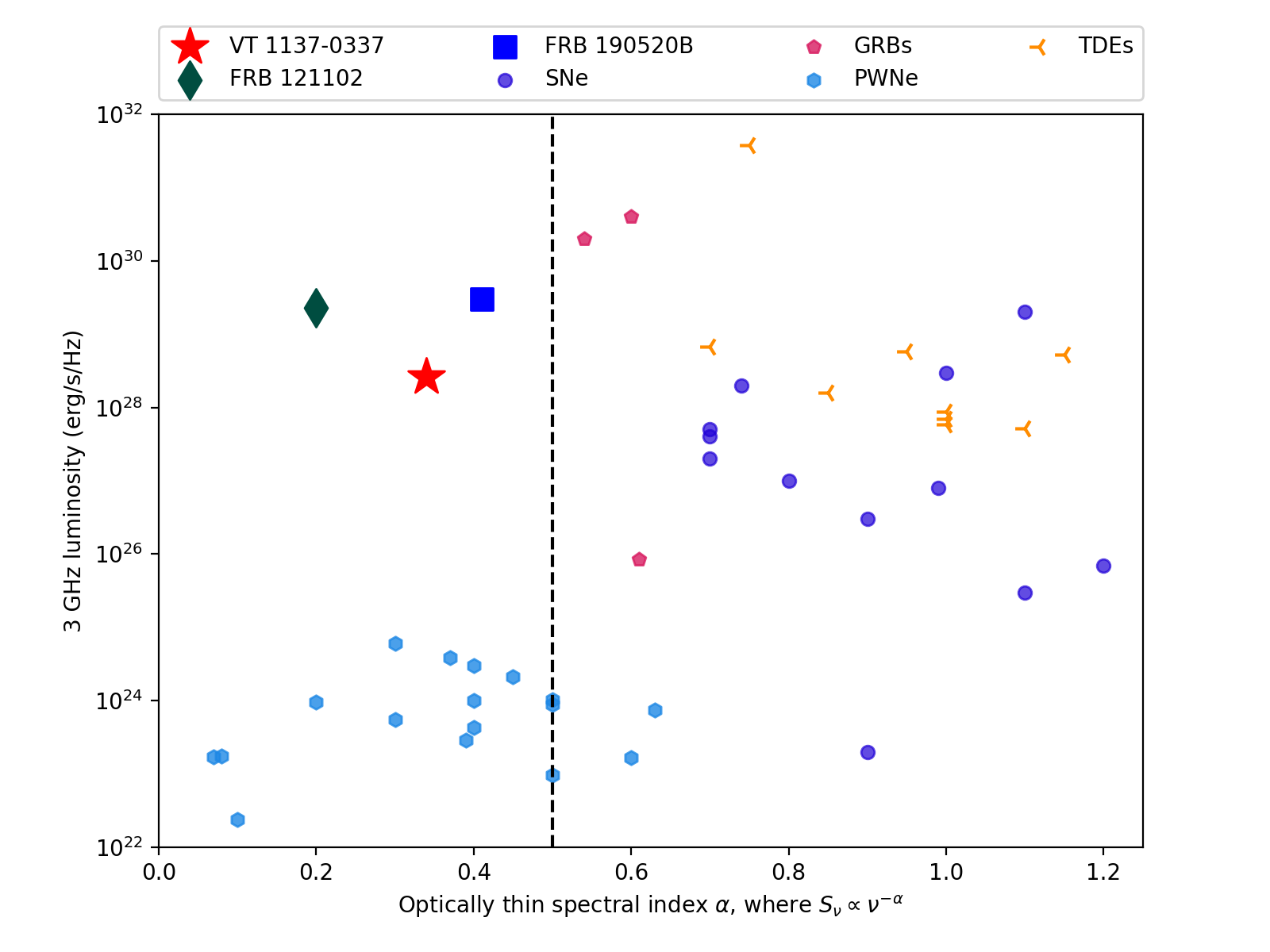}
    \caption{The 3 GHz luminosity and optically thin spectral index of VT 1137-0337 compared against values in the literature from radio observations of supernovae (SNe), gamma ray bursts (GRBs), tidal disruption events (TDEs), pulsar wind nebulae (PWNe), and fast radio burst (FRB) persistent sources. Sources left of the dotted line have harder spectra than the theoretical limit for diffusive shock acceleration: $\alpha = 0.5$. The values plotted and their references are listed in Appendix \ref{app:literature_transients}.}
    \label{fig:lum_vs_alpha}
\end{figure*}

As discussed in Section \ref{sec:followup}, VT 1137-0337 is characterized by the following observational properties:

\begin{itemize}
    \item Located near the nucleus of a dwarf galaxy of stellar mass $\sim$10$^{8.3} M_{\odot}$ at redshift at z = 0.02764
    \item A flat radio spectrum $S_{\nu} \propto \nu^{-0.35}$ with no observed peak between 1 to 15~GHz
    \item A high radio luminosity of L$_{1.5~\textrm{GHz}} \sim 3 \times 10^{28}$ erg/s/Hz, and a $\nu L_{\nu}$ luminosity of 2 $\times$ 10$^{38}$~erg~s$^{-1}$ at our highest observed frequency ($\sim$15~GHz)
    \item Transient-like behavior (an increase of $> 5\times$ in luminosity at 1.5 GHz) on the timescale of $\sim$2 decades
    \item Slow, broadband fading ($\sim$20\%) at 1-12~GHz over $\sim$4 years with a constant spectral index
\end{itemize}

\noindent In Section \ref{sec:analysis}, we used these features to argue that VT 1137-0337 is:

\begin{itemize}
    \item Surrounded by massive stars that have formed in the past few Myr with no indications of AGN activity in the BPT diagram (Section \ref{sec:Derived properties from optical features})
    
    \item Characterized by a synchrotron spectrum that cannot be explained by a single shock under standard magnetic conditions (Section \ref{sec:flat-spectrum-emission-mechanisms}). 
    
    \item Produced by an emitting region of present-day radius $\gtrsim 0.05$~pc and $\lesssim 6$~pc, where the synchrotron cooling time is $\gtrsim$ decades at the GHz frequencies observed. This region contains a magnetic energy that is comparable to or greater than radio-luminous supernovae, and is not presently obscured by a high density ionized shell or wind (Section \ref{sec:spectral_breaks}).

\end{itemize}

In this section, we discuss astrophysical source classes known to produce luminous and variable and/or flat-spectrum radio emission. We assess whether they are consistent with the properties summarized above.

\subsection{Exploding stars: radio SNe and long GRBs}
\label{subsec:SNe}

VT 1137-0337 is located in the midst of a young starburst which, from its star formation rate (SFR) of $\sim$0.5 $M_{\odot}$ / yr, is expected to produce a core collapse supernova (SN) every $\sim$2~centuries. A small fraction of these explosions are expected to reach our observed luminosity of $\sim$10$^{28}$ erg~s$^{-1}$~Hz$^{-1}$, either through interaction with a dense shell in the circumstellar medium (CSM) or by harboring a relativistic outflow such as a gamma ray burst (GRB). The slow evolution and low frequency peak of the radio spectrum can potentially be explained if the explosion were caught at a late epoch ($\sim$years for CSM interaction and $\gtrsim$1 decade for a GRB). However, the flat radio spectrum is unlike any SN or GRB that has been previously observed. Figure \ref{fig:lum_vs_alpha} shows a compilation of radio SNe and GRBs with published optically thin spectral indices. As can be seen in the figure, the spectral indices are $> 0.5$ in each case, consistent with DSA under standard conditions.
\\
\\If VT 1137-0337 is indeed a supernova or GRB, we would likely need to invoke a model such as unusual magnetic conditions in the shock \citep[e.g.,][]{Schlickeiser-1989-shock-acceleration-plasma-beta, Fleishman06-diffusive-sync-rad-GRBs} to explain its flat spectrum. This has not been observed before in the sample of SNe and GRBs discovered in optical and high energy transient surveys. Though we cannot rule out that VT 1137-0337 is drawn from a new class of volumetrically rare flat spectrum stellar explosions discovered at radio wavelengths, the lack of flat spectrum analogs leads us to believe that a SN or GRB is an unlikely explanation. Instead, as discussed in Section \ref{sec:PWNe}, VT 1137-0337 may be associated with the compact remnant of a stellar explosion, rather than the shock.

\begin{figure*}
    \centering
    \includegraphics[clip=true, trim=0cm 0.3cm 0.2cm 0cm, width=\textwidth]{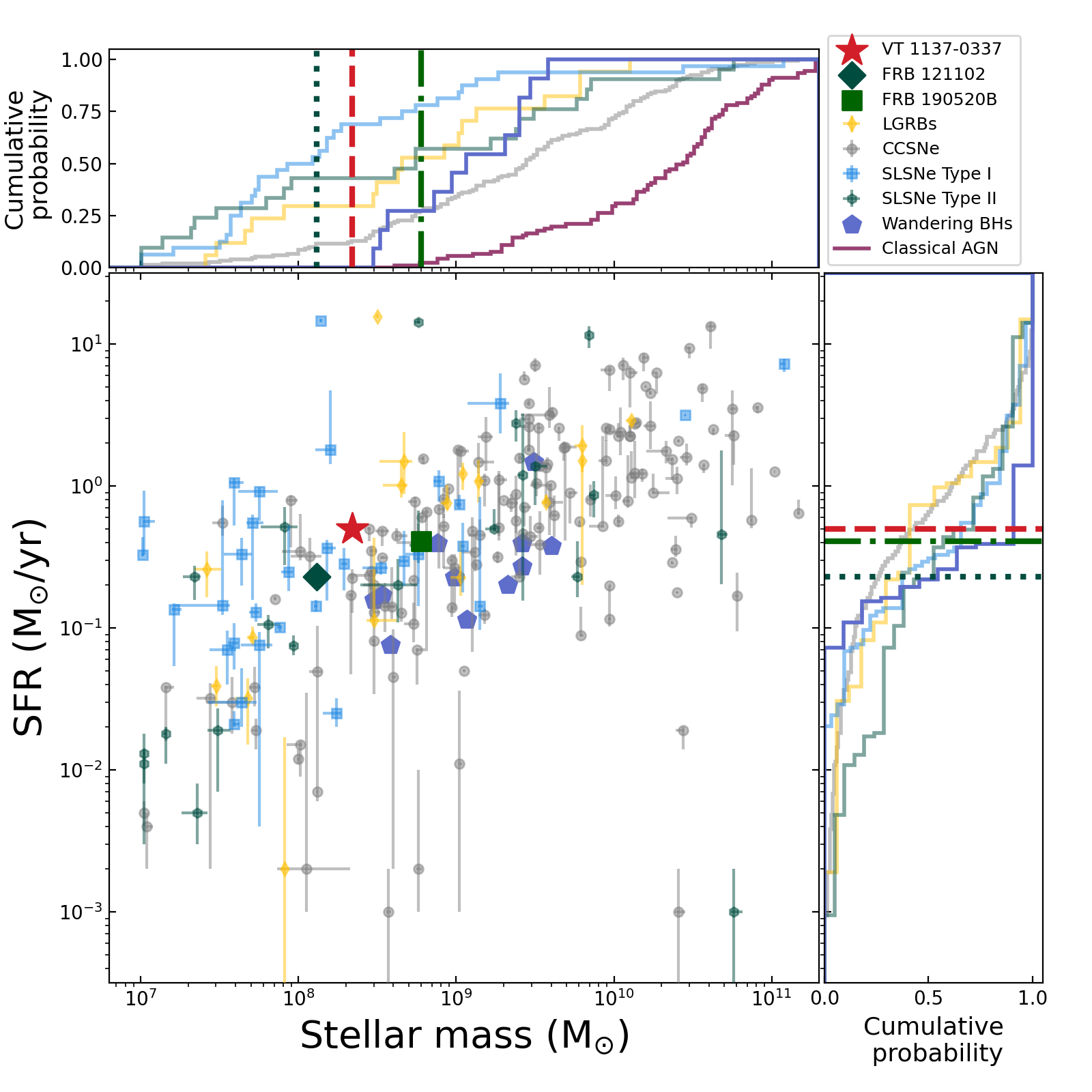}
    \caption{The star formation rate and stellar mass of VT 1137-0337's host galaxy in comparison with the hosts of FRB 121102 \citep[][]{Bassa17-FRB121102-host}, FRB 190520B \citep[][]{Niu21-second-FRB-PRS}, wandering black holes in dwarf galaxies \citep[][their sample A]{Reines2020-Wandering-BHs}, optical/high energy transients compiled by \citet{Taggart2020-SN-hosts}, and broadline AGN host masses from \citet{Reines-Volonteri-2015-BHmass-Mstar-relation}. For all populations with more than one source, the top and right panels show the cumulative distribution in stellar mass and star formation rate respectively. The dotted, dot-dash, and dashed lines show the positions of FRB 121102, FRB 190520B, and VT 1137-0337's host galaxies relative to these histograms.}
    \label{fig:host-galaxies}
\end{figure*}

\subsection{Black hole jets: XRBs, ULXs, TDEs, and AGN}
\label{subsec:AGN}

\begin{figure}
    \centering
    \includegraphics[clip=true, trim=0cm 0cm 0cm 0cm, width=0.45\textwidth]{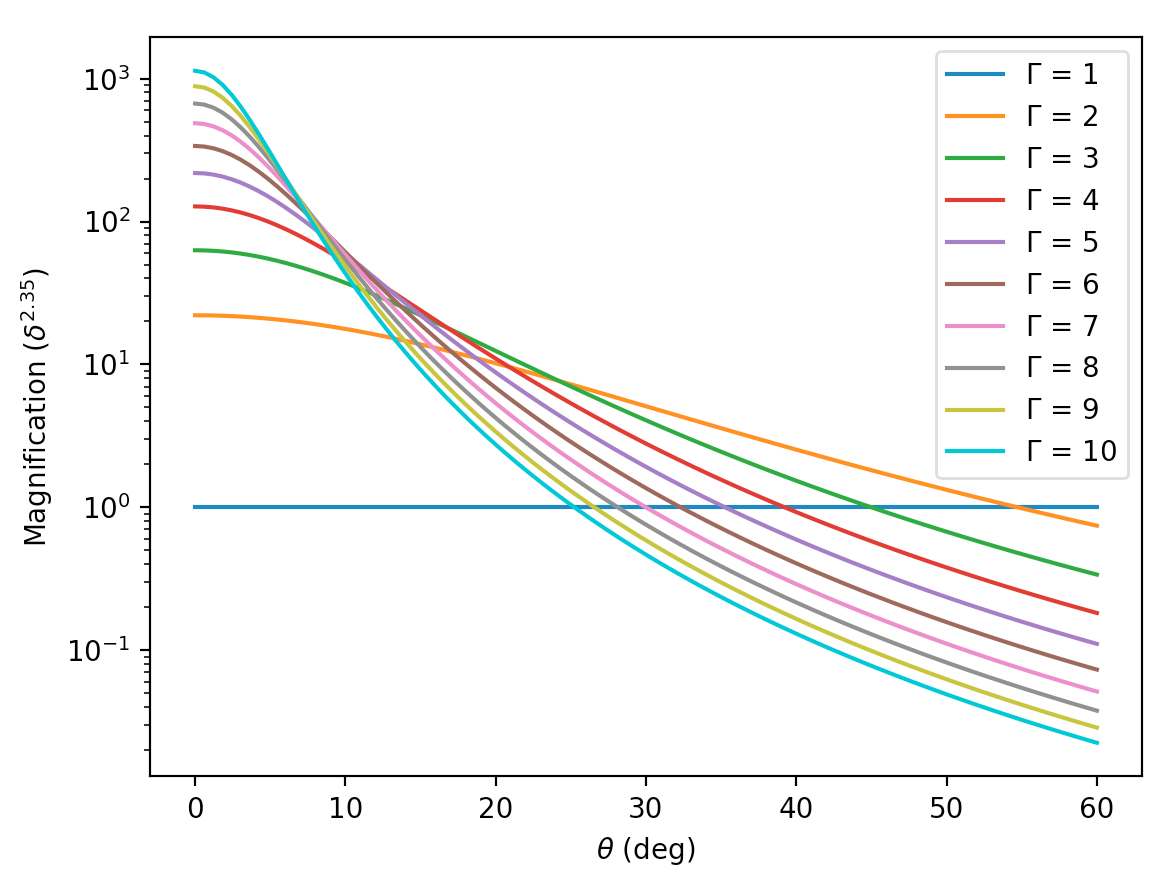}
    \caption{The magnification (or demagnification) due to relativistic beaming for a steady $\alpha = 0.35$ jet with bulk Lorenz factor $\Gamma$ oriented at angle $\theta$ relative to our line of sight.}
    \label{fig:beaming}
\end{figure}

\begin{figure}
    \centering
    \includegraphics[clip=true, trim=0cm 0cm 0cm 0cm, width=0.5\textwidth]{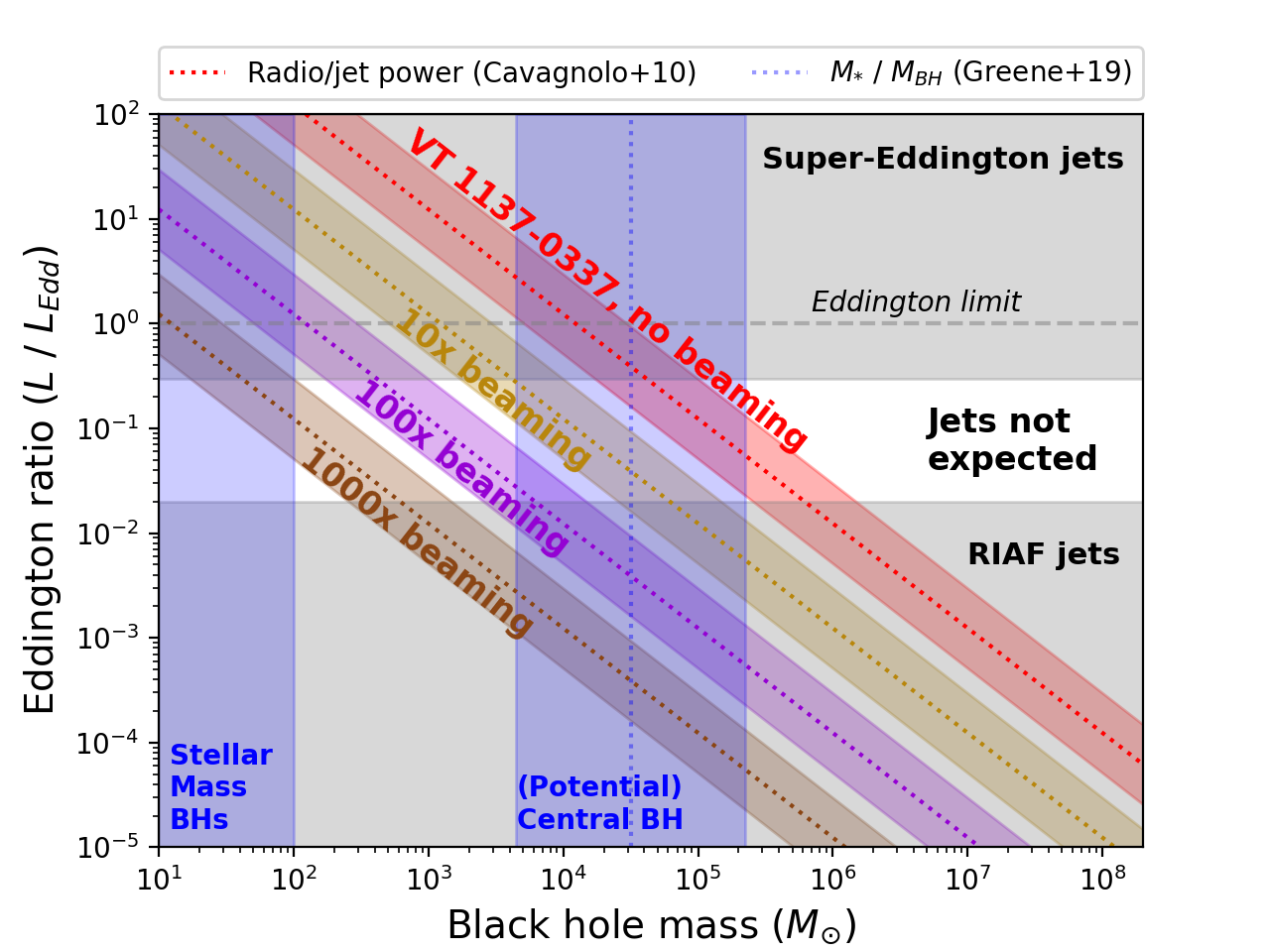}
    \caption{Approximate constraints on the mass and accretion scale required to explain VT 1137-0337 with a BH jet. The dashed lines show the predicted jet power from the radio power - jet power relation of \citet[][]{Cavagnolo10-AGN-jet-power-radio-power} with varying levels of magnification by beaming, with the corresponding shaded regions showing the scatter in the relation. The blue shaded regions show the estimated mass ranges for stellar mass BHs and central massive BHs as predicted by the galaxy mass - black hole mass relation \citep[][]{Green20-IMBH-review}. The gray shaded regions show the accretion regimes (super-Eddington accretion and radiatively inefficient accretion flows) where the inflow is expected to be geometrically thick and thus capable of supporting jets.}
    \label{fig:AGN_constraints}
\end{figure}

\begin{figure*}
    \centering
    \includegraphics[width = \textwidth]{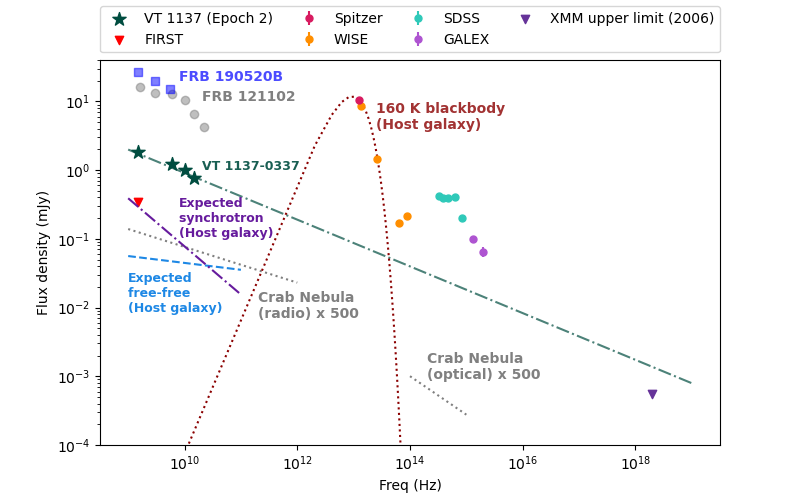}
    \caption{The radio spectrum of VT 1137-0337 superimposed on the spectral energy distribution of its host galaxy. The dotted and dashed green line shows VT 1137-0337's extrapolated spectrum in Epoch 2. The blue dashed and purple dash-dot lines show the expected host galaxy radio emission due to star formation, assuming a typical synchrotron spectral index of 0.7 \citep{Murphy2011_SF_calibration}. The dotted red line shows a blackbody approximation to the host galaxy's infrared dust emission, normalized to the Spitzer 24$\mu$m flux. The 98\% confidence upper limit on the X-ray flux from XMM Newton is shown with the purple triangle. For comparison, the gray circles and blue squares show the spectrum of the persistent counterparts to FRB 121102  \citep[][]{Chatterjee-FRB121102-VLBI-loc} and FRB 190520B \citep[][]{Niu21-second-FRB-PRS}. The gray dotted line shows the integrated radio and optical spectra of the Crab Nebula \citep[][]{Atoyan-Aharonian-Crab-nebula-spectrum, Lyutikov-2019-Crab-synchrotron-spectrum-theory} multiplied by a factor of 500. The FRB persistent source and scaled-up Crab fluxes are normalized to the distance of VT 1137-0337.}
    \label{fig:galaxy_SED}
\end{figure*}

Synchrotron emission from a black hole (BH) jet is a potential explanation for the flat spectrum and variability of VT 1137-0337. Flat radio spectra spanning orders of magnitude in bandwidth have been observed from BHs with a wide range of masses, from stellar mass BHs in X-ray binaries (XRBs) \citep[e.g.,][]{Fender2000-CygX1-flat-spectrum} to supermassive BHs in active galactic nuclei (AGN) \citep[e.g.,][]{Kellerman1981-AGN-review}. The classical model of \citet[][]{Blandford-Konigl-AGN-jets} interprets flat AGN spectra as emission from relativistic electrons that are continuously accelerated by compact conical jets. In this model, the effective area of the emitting region varies with frequency in a way that cancels out the intrinsic optically thick spectral index. This optical depth effect, which can be understood as a continuous superposition of synchrotron components, results in an overall flat spectrum. This general picture is supported by very long baseline interferometry (VLBI) observations of flat spectrum AGN cores, which in some cases directly resolve the roughly conical jets \citep[e.g.,][]{Asada2012-M87-conical}, and in others, show a core shift towards the black hole at high frequencies \citep[e.g.,][]{Plavin19-core-shift}, which are both predictions of the model. A similar model from \citet[][]{Hjellming-Johnston1988-Conical-jets-XRBs}, where the jet has a hollow conical geometry due to jet precession or orbital motion of the binary, is often invoked to explain the flat spectrum radio emission from XRBs. 
\\
\\In addition to producing flat spectra, BH jets are both theoretically expected and observed to be variable. In jets that are pointed within a few degrees of our line of sight, small changes in either the jet angle or jet power can lead to substantial changes in the flux due to relativistic beaming \citep[][]{Blandford-Konigl-AGN-jets, Lister01-beaming}. Stochastic radio variability on timescales of days to months at the $\sim$10 to 40\% level is often observed in blazars and flat spectrum radio quasars, which both have nearly on-axis jets. In some cases, their moving average flux density can change by factors of a few on timescales of years \citep[][]{Liodakis17-OVRO-blazar-update-bimodal}. BHs can also brighten substantially due to new jets being launched. During state changes, XRBs have been observed to brighten by orders of magnitude at radio frequencies while displaying flat radio spectra \citep[][]{Yao20-AT2019wey, Egron21-CygX3-17Jy-flare}. Quasars have also recently been observed to launch new jets, brightening by 2 to $>$25$\times$ on decade timescales, albeit with peaked rather than flat spectra \citep[][]{Nyland2020-VLASS-Quasars}.
\\
\\To assess the viability of a BH jet model in explaining the specific observational properties of VT 1137-0337, we consider BHs in three mass ranges that may exist in its $M_* \approx 10^{8.3} M_{\odot}$ host galaxy: stellar mass BHs ($\lesssim 100 M_{\odot}$), intermediate mass BHs ($10^2 - 10^5 M_{\odot}$), and supermassive BHs ($\gtrsim 10^5 M_{\odot}$). Where possible, we draw analogies to known phenomena from these BH classes in the Milky Way and other galaxies.
\\
\\Stellar mass BHs are expected to exist in large numbers in all galaxies of appreciable mass. While the majority of these BHs are likely isolated and do not emit strongly, a fraction will reside in rapidly accreting systems such as XRBs that could in principle be detected \citep[][]{Wiktorowicz2019-stellar-mass-BH-population-synthesis}. There are, however, a number of orders-of-magnitude mismatches between VT 1137-0337's properties and those of known XRBs. VT 1137-0337's radio luminosity of $L_{\nu} \sim 10^{28}$~erg~s$^{-1}$~Hz$^{-1}$ is $\gtrsim$ 4 orders of magnitude higher than even extreme flares from Galactic XRBs \citep[e.g.,][]{Corbel-1e24-CygX3-flare}. Likewise, its slow fading and lack of spectral index evolution over $\sim$4 years stands in sharp contrast with the large changes in luminosity and spectral shape observed in flaring XRBs over the span of hours to weeks \citep[e.g.,][]{Pietka15-variability-timescale-luminosity, Yao20-AT2019wey}. These issues of scale arise because black holes in this mass range have relatively short dynamical times and are limited in the rate at which they can accrete. In particular, the radio power - jet power relation from \citet[][]{Cavagnolo10-AGN-jet-power-radio-power} (which contains radio sources of similar luminosity to VT 1137-0337 in its calibration set) predicts that the jet power corresponding to the luminosity we observe is $P_{\textrm{jet}} \approx 1.4 \times 10^{42}$ erg s$^{-1}$ with a scatter of $\sim$0.7~dex. This jet power is comparable to the Eddington luminosity of a $M_{\textrm{BH}} = 10^4 M_{\odot}$ BH, given by $L_{\textrm{Edd}} = 1.26 \times 10^{42} \left(M/10^4 M_{\odot}\right)$ erg s$^{-1}$. This suggests that to explain VT 1137-0337, a jet from stellar mass BH would need to suddenly increase in power over $\sim$20~years by a factor of $\sim$5$\times$ to $\sim$100 to 1000 times its Eddington luminosity, and it would need to maintain this new power level for at least the $\sim$4 years over which we have observed the source. 
\\
\\The rate of mass siphoned from a close binary $\dot{M_*}$ can substantially exceed the Eddington accretion rate $\dot{M}_{\textrm{Edd}}$ of the accreting BH. Though the vast majority of this mass is lost through disk winds, the accretion rate $\dot{M}_{\textrm{BH}}$ onto the BH can itself exceed $\dot{M}_{\textrm{Edd}}$ due to photon trapping within the accretion flow. This is thought to occur in most ultra-luminous X-ray sources \citep[ULXs; e.g,][]{Sutton13-ULX-supereddington}. Using 3D general relativistic radiation-MHD simulations, \citet[][]{Sadowski16-superedd-accretion-sims} predict that a rapidly rotating black hole accreting in the super-Eddington regime can launch a jet with a power of a few percent of $\dot{M}_{\textrm{BH}}c^2$ via the \citet[][]{Blandford-Znajek1977} process. In order to explain VT 1137-0337 with such a super-Eddington jet, $\dot{M}_{\textrm{BH}}$ should be $\sim$2 to 3 orders of magnitude higher than Eddington requiring $\dot{M_*}$ to be a few orders of magnitude higher still. 
\\
\\An alternate explanation for the high radio luminosity is relativistic beaming. For a jet with a velocity $\beta = v/c$ oriented at an angle $\theta$ from our line of sight, the Doppler factor is defined as $\delta = (\Gamma - \sqrt{\Gamma^2 - 1}~ \textrm{cos}~\theta)^{-1}$, where $\Gamma = (1-\beta^2)^{-1/2}$ is the bulk Lorenz factor. The jet's luminosity is magnified (or demagnified) by a factor of $\delta^{p_{\textrm{mag}}}$, where the exponent $p_{\textrm{mag}}$ depends on a number of factors including the spectral index of the emission, the structure of the jet, and (if relevant) the lifetime of the emitting blob (see Appendix B of \citet[][]{Urry-Padovani-1995-AGN-unification} for a discussion). For a steady, uniform jet, $p_{\textrm{mag}} = 2 + \alpha$ where $\alpha$ is the spectral index (equal to +0.35 for VT 1137-0337). If the emission arises from a discrete blob of plasma travelling down the jet (i.e. a flare), it will be enhanced by a further factor of $\delta$ due to time dilation, making $p_{\textrm{mag}} = 3 + \alpha$. In either case, beaming can magnify the radio luminosity by a large factor. For a steady jet pointed at $\theta \sim 20^{\circ}$, the magnification is a factor of $\sim$10  depending on $\Gamma$ (see Figure \ref{fig:beaming}). At $\theta \sim 10^{\circ}$, the luminosity is magnified by a factor of $\sim$100, and for $\Gamma \gtrsim 10$ jets pointed at $\theta \approx 0^{\circ}$, beaming can enhance the luminosity by a factor of $\gtrsim 1000$. In Galactic XRBs, $\Gamma$ has been measured to be in the range 1.3 - 3.5 \citep[][]{Saikia19-gammas-for-XRBs}, a range in which the maximum magnification for a steady jet is $\sim$100$\times$ even if it is completely on-axis ($\theta = 0$).
\\
\\Figure \ref{fig:AGN_constraints} shows the Eddington luminosity fraction required to explain VT 1137-0337's inferred jet power for various BH masses and under varying levels of beaming. For stellar mass BHs, highly super-Eddington accretion, strong beaming, or a combination of the two are required to explain the jet power we infer. These scenarios may be difficult to reconcile with the rapid transient behavior between FIRST and VLASS and subsequent slow fading. A large increase in the beaming magnification can occur if the the jet precesses into our line of sight. However, the timescale of this precession should be comparable to the orbital timescale of the binary which, given the high accretion rates, should be $\sim$weeks. Depending on the degree of beaming, the issue of timescales is compounded by time dilation, since the timescale in the jet's rest frame is longer by a factor of $\delta$ than the time we measure on Earth. Alternatively, a sudden increase in the instantaneous jet power may occur if $\dot{M}_{\textrm{BH}}$ increases rapidly. Such increases are observed in XRBs as they transition between the high-soft and low-hard states. However, this increased accretion rate typically happens on a timescale of order the disk dynamical time, which is again much shorter than the $\sim$4~years over which we have observed slow fading. Overall, though it may be possible to explain VT 1137-0337's radio luminosity with the direct emission from a stellar mass BH jet, the characteristic timescales of such systems are not conducive to explaining the broadband lightcurve we observe.

\noindent \\We now consider the possibility of an intermediate mass BH (IMBH) in the mass range $10^{2-5} M_{\odot}$. To date, no examples of IMBHs have been conclusively established, though there are some promising candidates, particularly at masses $\gtrsim $10$^{4} M_{\odot}$ \citep[see][and references therein]{Green20-IMBH-review}. In the lower half of the mass range $\sim 10^{2-3.5} M_{\odot}$, many of the same arguments regarding stellar mass BHs apply. Beaming is still required if the accretion flow is sub or near-Eddington, though for near-Eddington accretion, the required magnification is only $\sim$10$\times$, corresponding to $\theta \lesssim (30^{\circ}~\textrm{to}~15^{\circ})$ for $\Gamma = (2~\textrm{to}~10)$. If the near-Eddington accretion is due to a close binary companion, the orbital timescale will still be of order $\sim$weeks, so the orbital plane would still have to be aligned close to face-on to avoid precessing the jet by more than a few degrees. A flare would need to span  $4\delta$~years $\approx$ 1 decade in its rest frame and would be more akin to a new jet turning on. It is again unclear what could cause the sudden but sustained change in accretion required to power such a flare. Tidal disruption events (TDEs) can launch jets and drive super-Eddington accretion \citep[][]{Zauderer-SwiftJ16}, but the mass fallback rate is expected to decline steeply over time (canonically as $t^{-5/3}$; \citet[][]{Phinney1989-TDE-five-thirds}) and the jet's characteristic fading timescale would be $\sim$months, not $\sim$decades \citep[][]{Tchekhovskoy14-SwiftJ16-gone-MAD}. Likewise, accretion disk instabilities will operate on timescales far too short to explain our observations.
\\
\\The upper half of the IMBH mass range $\sim 10^{3.5 - 5} M_{\odot}$ is more plausible, and is particularly important to consider given the potential presence of a central black hole in VT 1137-0337's $M_* \approx 10^{8.3} M_{\odot}$ host galaxy. Due to selection bias in traditional AGN diagnostics, central black hole detections have historically been limited to galaxies with stellar mass $M_{*} \gtrsim 10^{9.5} M_{\odot}$  \citep[e.g.,][]{Reines-Volonteri-2015-BHmass-Mstar-relation}, where typically $M_{\textrm{BH}} >> 10^{5} M_{\odot}$. However, there is strong evidence of $\sim$10$^{5} M_{\odot}$ BHs in $\sim$10$^{9} M_{\odot}$ galaxies \citep[e.g.,][]{Nyland17-NGC404, Davis20-NGC404-IMBH} and simulations suggest that the BH occupation fraction should be $\gtrsim 50\%$ in $10^{8-9} M_{\odot}$ galaxies \citep[][]{Bellovary19-MBH-multimessenger}. Additionally, new selection methods based on luminous radio emission \citep[][]{Reines2020-Wandering-BHs} and the high ionization potential [FeX] line \citep[][]{Molina21a-single-object-AGN, Molina21b-AGN-sample} have identified promising AGN candidates in dwarf galaxies as low mass as $\sim 10^{7.5} M_{\odot}$. For VT 1137-0337's host galaxy, a slight extrapolation of the most recent stellar mass - BH mass relation \citep[][]{Green20-IMBH-review} predicts a central BH mass of 10$^{4.5 \pm 0.8} M_{\odot}$ when considering all galaxy types, or 10$^{4.7 \pm 0.8} M_{\odot}$ when considering only late-type galaxies. 
\\
\\There are some additional complications that are particularly relevant for jet models in the predicted central BH mass range. First, there is a range of Eddington ratios between $\sim$a few percent and $\sim$30\% for which jets are not observed or theoretically predicted. In this range, the accretion disk is expected to be geometrically thin and incapable of supporting the vertical magnetic fields required to collimate jets \citep[][]{Meier2001-RIAFs}. This jet quenching effect has been observed in XRBs \citep[e.g.,][]{Fender2004-unified-XRB-model}, and has been used to explain the abrupt drop in the X-ray flux from the jetted TDE Swift J1644+57 \citep[][]{Tchekhovskoy14-SwiftJ16-gone-MAD}. The quenched jet zone excludes a substantial part of the BH mass - Eddington ratio parameter space relevant to higher mass IMBHs (Figure \ref{fig:AGN_constraints}). Second, the optical emission lines in the $\sim$280~pc radius region surrounding VT 1137-0337 have intensity ratios that are very far from the region typically associated with AGN (Figure \ref{fig:BPT}), and all of the lines are spectrally unresolved at the $\sim$150~km/s level. This is likely ok for lower mass BHs or those accreting at low Eddington ratios, since sub-Eddington BHs with masses $\lesssim 10^4 M_{\odot}$ are predicted to have weak narrow lines with ratios consistent with star formation, (albeit in a different region of the BPT diagram than VT 1137-0337's host) \citep[][]{Cann19-BPT-limitations-dwarf}. However, for near-Eddington BHs with mass $\gtrsim$10$^5 M_{\odot}$, the expected narrow line luminosity is comparable to what we observe, and their intensity ratios are inconsistent with the ones we measure.
\\
\\At the lower end of the predicted mass range ($\sim$10$^{4} M_{\odot}$), avoiding the quenched jet zone requires either near-Eddington accretion and little to no beaming, or very strong beaming and accretion in the radiatively inefficient accretion flow (RIAF) regime (an Eddington ratio of $\lesssim$ a few percent). If the accretion rate is near Eddington, it is once again difficult to explain the sudden and sustained accretion. If instead the emission is highly beamed, the slow variability is again problematic given time dilation and the expected variability timescale ($\sim$months) for sources in this luminosity range \citep[e.g.,][]{Pietka15-variability-timescale-luminosity}.
\\
\\At the high end of the predicted mass range ($\sim$10$^{5} M_{\odot}$), the two scenarios are demagnification from beaming and near-Eddington accretion, or moderate beaming and accretion in the RIAF regime. As discussed above, the near-Eddington scenario is disfavored by the lack of AGN-like emission ratios in addition to issues with fueling the BH at the right rate on the right timescales. In the RIAF scenario, the power requirement for the jet is more easily satisified, so we investigate it in more detail. If a $\sim$10$^5 M_{\odot}$ BH experienced an increase in its accretion rate to a few percent of Eddington and subsequently launched a flat spectrum jet at an angle where it is magnified by $\sim$10$\times$ by beaming, that could potentially explain the appearance of VT 1137-0337 between FIRST and VLASS. However, in order to explain the slowly fading flat spectrum, the jet would need to maintain a nearly constant luminosity on scales of $\sim$10$^{-3} - 10^{-1}$~pc \citep[the linear scales corresponding to synchrotron self-absorption frequencies of 15 GHz to 1 GHz at the luminosity of VT 1137-0337;][]{Scott-Readhead-1977-equipartition} for an observer-frame timescale of $\sim$4~years. Plasma moving at relativistic velocities in the jet will cycle through these length scales on timescales of $\sim$days to weeks. Thus, the jet would have to maintain a stable power on all scales probed over $\sim$tens to thousands of plasma crossing times. This type of stability is not observed in AGN dominated by their flat spectrum cores. For scale, the 15~GHz length scale for a 1 Jy flat spectrum AGN at a distance of 100~pc is $\sim$0.07~pc, corresponding to a plasma crossing time of $\sim$80 days. Maintaining a constant flux for $\sim$1000 crossing times would be analogous to having that flat-spectrum radio quasar or blazar be stable for $\sim$2 centuries, which is far longer than the typical variability timescale of $\sim$months \citep[][]{Liodakis17-OVRO-blazar-update-bimodal}. Additionally, to avoid spectral index evolution and sharp variability due to changes in beaming, the jet orientation would need to remain stable, excluding models involving a pre-existing jet that precessed into our line of sight. We do not consider any of the above scenarios to be likely.
\\
\\Finally, we consider the possibility of a jet from a supermassive BH. Black holes with mass $>> 10^6 M_{\odot}$ are not generally expected to exist in VT 1137-0337's dwarf host galaxy, though ones in the range 10$^{5-6} M_{\odot}$ are within the scatter of the stellar mass - BH mass relation. Many of the same arguments against $\sim$10$^5 M_{\odot}$ BHs apply in this mass range. While fueling the jet power is no longer an issue, maintaining a low-luminosity flat-spectrum core with only $\sim$20\% variability over $\sim$4 years and no change in the spectral index is once again difficult due to the size scales involved. 
\\
\\In summary, BH jet models at all mass scales face substantial challenges when attempting to explain VT 1137-0337's luminosity, variability properties, and flat spectrum. Stellar mass BHs would require extreme and rapidly changing accretion rates or heavy beaming to produce the required luminosity. Such jets would most likely appear as fast transients (flickering on and off on timescales of $\sim$days) rather than the relatively stable one that we have observed. An IMBH, particularly one towards the higher mass end ($\sim 10^{5} M_{\odot}$), is potentially expected to exist in VT 1137-0337's host galaxy and would be able to accrete at a rate that could power the luminosity we observe. However such a BH cannot explain the flat spectrum and slow variability, since the length scales responsible for emitting at each of our observed frequencies would have been traversed by tens to thousands of separate blobs of relativistic plasma over the course of 4 years. The modest ($\sim 20\%$) and consistent variability  we have observed at all frequencies would require any stochastic jet-power process to be auto-correlated over unrealistic timescales. Supermassive BHs are ruled out by the same argument, and are furthermore not expected to exist in such a low mass galaxy. These arguments, though approximate in nature, show that the specific properties of VT 1137-0337 (particularly its sudden appearance, flat spectrum, and slow fading, which have not been observed for other BHs or BH candidates) are not easily explained by a BH jet. Instead, we believe that there is a more compelling explanation: a young and energetic pulsar wind nebula. We discuss this in the next section.

\subsection{Neutron star nebulae: pulsar winds and magnetar field decay}
\label{sec:PWNe}

The most distinctive feature of VT 1137-0337 is its flat and slowly evolving radio spectrum. Pulsar wind nebulae (PWNe) are one of the only well-observed source classes with this type of spectrum. The vast majority of known PWNe are observed to have spectra flatter than the $\alpha = 0.5$ Fermi acceleration limit \citep[][]{Kaspi-isolated-neutron-stars, Gaensler06-PWN-review, Green19-SNR-catalog} that evolve very slowly on human timescales. As an example, the Crab Nebula has a power law spectrum with no radio-frequency substructure, which fades by only 0.167 $\pm$ 0.015 percent per year \citep[][]{Aller-Reynolds-1985-Crab-fade-rate}. Its spectral index was measured in the 1960s to be $\alpha = 0.299 \pm 0.009$ \citep[][]{Baars1977-flux-density-scale}, and least squares fitting of a compilation of measurements from the 1970s to the early 2000s gives $\alpha = 0.296 \pm 0.006$ \citep[][]{Marcias-Perez2010-Crab-spectral-index-and-fading}. 
\\
\\Many theoretical mechanisms have been proposed to explain the flat spectra of PWNe. Unlike in BH jets, where the time-variable flat spectra are typically explained with optical depth effects that can change rapidly, most PWN theories invoke an intrinsically hard spectrum with an electron energy index $p < 2$. \citet[][]{Atoyan99-flat-spectrum-Crab} proposes that the hard energy distribution is assembled over time from electrons injected at high energy that cool rapidly and pile up at the synchrotron cooling break. As the PWN expands and its magnetic field weakens, the break drifts upwards in energy, freezing in the old low energy electrons and piling up new ones to fill out the higher energy parts of the spectrum. This model ties the spectral index to the evolution of the PWN in a way that reproduces the observed flat spectra for standard energy injection via spindown. It predicts that the current radio-producing electrons in the Crab Nebula were injected at a very early time ($\sim$decades). If this were the case, the Crab Pulsar may have had a spin period of a few milliseconds at birth rather than the $\sim 19~$ms inferred from tracing back its spindown from parameters measured in the present day \citep[e.g.,][]{Kou-Tong15-Crab-initial-period}.
\\
\\There is, however, evidence that radio-emitting electrons in the Crab Nebula are still being accelerated today, given the morphological similarity of radio wisps with much shorter lived optical wisps \citep[][]{Bietenholz2001-Crab-radio-wisps}. In that case, the hard distribution may instead be produced at the acceleration site. \citet[][]{Schlickeiser-1989-shock-acceleration-plasma-beta} propose that $p < 2$ in extremely magnetic dominated shocks, due to second-order Fermi acceleration. By their analytic estimates, reducing $p$ to 1.7 requires that the magnetic energy dominates over the thermal energy by over an order of magnitude (a plasma $\beta$ of $\sim$0.05). Alternatively, simulations of particles accelerated by relativistic magnetic reconnection are consistently able to produce distributions with $p < 2$ with Lorentz factors appropriate for GHz radio emission \citep[e.g.,][]{Kagan-Nakar-Piran-2018-relativistic-reconnection}. This too requires a highly magnetized plasma at the acceleration site. \citet[][]{Werner-2016-power-law-spectra-relativistic-reconnection} found that $p \approx 1.7$ when $\sigma$ (the ratio of the magnetic to relativistic energy density) is $\sim$8. Reconnection-driven acceleration can potentially occur at the termination shock of the highly magnetized and relativistic pulsar wind \citep[][]{Sironi-Spitkovsky11-PWN-flat-spectrum} or within the bulk of the nebula itself \citep[][]{Lyutikov-2019-Crab-synchrotron-spectrum-theory}.
\\
\\Flat spectra can potentially be produced even without a hard electron distribution if the electrons are propagating within a magnetic field that, due to turbulence, is randomly oriented on scales smaller than the particles' orbits. In this scenario, there is a frequency range where $\alpha = (q-1)/2$, where $q$ is the index of the turbulent power spectrum \citep[][]{Fleishman-Bietenholz07-diffusive-sync-rad}. Intriguingly, \citet[][]{Kolmogorov1941-turbulence} turbulence has $q$ = 5/3, which implies a spectral index of $\alpha = 1/3$ similar to our observed value of $\alpha = 0.35$. This mechanism has also been proposed to explain the Crab Nebula's flat radio spectrum \citep[][]{Tanaka-Asano17-Stochastic-acceleration-Crab}. Recent particle in cell simulations have also shown that particles accelerated (typically by magnetic reconnection) within turbulent plasmas have an energy-dependent anisotropy, which for most viewing orientations would also lead to hard spectra $\alpha \sim 0.3$ to 0.5 \citep[][]{Comisso20-PWN-turbulence}
\\
\\At the moment, the theory of PWNe spectra is an evolving field with new ideas being put forward regularly. We therefore remain agnostic to the mechanism and instead point out that VT 1137-0337's flat spectrum and slow fading are \textit{empirically} very similar to what's observed for PWNe. The biggest differences to be explained are that (1) VT 1137-0337 appears as a transient (on a timescale of $\sim$20~years), (2) it is fading at a moderate rate ($\sim$5\% per year, much faster than the Crab's 0.17\% per year), and (3) it is far more radio luminous than any known PWN ($\sim 10^4 \times$ the Crab's luminosity). 
\\
\\All three of these differences are, at least qualitatively, \textit{expected} for a PWN if the central neutron star (NS) is young. VT 1137-0337's starbursting host galaxy is expected to produce young NSs via core collapse supernovae at a relatively high rate. Shortly after birth, a NS will start to spin down, launching a relativistic wind of particles and Poynting flux through the rapid rotation of its magnetic field \citep[][]{Goldreich-Julian-PWNe}. For the most strongly magnetized NSs, there may be a non-negligible or even dominant component of the wind powered by the direct decay of the internal magnetic field, possibly in discrete flares \citep[e.g.,][]{Beloborodov17-magnetar-121102, Margalit18-FRB121102-magnetar}. Regardless of the power source, the wind will deposit its energy in a termination shock, inflating a bubble within the surrounding supernova ejecta and forming a PWN. After some time, the ejecta will expand and dilute to the point where it becomes optically thin to free-free absorption. At that point, the PWN will suddenly appear as a radio transient \citep[][]{Chevalier-Fransson-1992-PWN}. Young PWNe are expected to reach their peak luminosity after just a few years (with the rise time generally occurring while they are still obscured). Those powered by neutron stars with fast periods or high magnetic fields could be orders of magnitude more luminous at early times than the centuries to millenia-old PWNe observed in the Milky Way. Such young nebulae are also expected to fade at their fastest per-year rate \citep[][]{Reynolds-Chevalier1984-PWN-lightcurve}. Together, these qualities make a young, energetic PWN an intriguing potential explanation for VT 1137-0337. In the remainder of this section, we describe some order-of-magnitude properties of PWN models embedded in supernova ejecta, and check whether they are able to explain VT 1137-0337's observed properties.

\subsubsection{Age constraints}
\label{subsubsection:PWN_age}

\begin{figure}
    \includegraphics[width=0.5\textwidth]{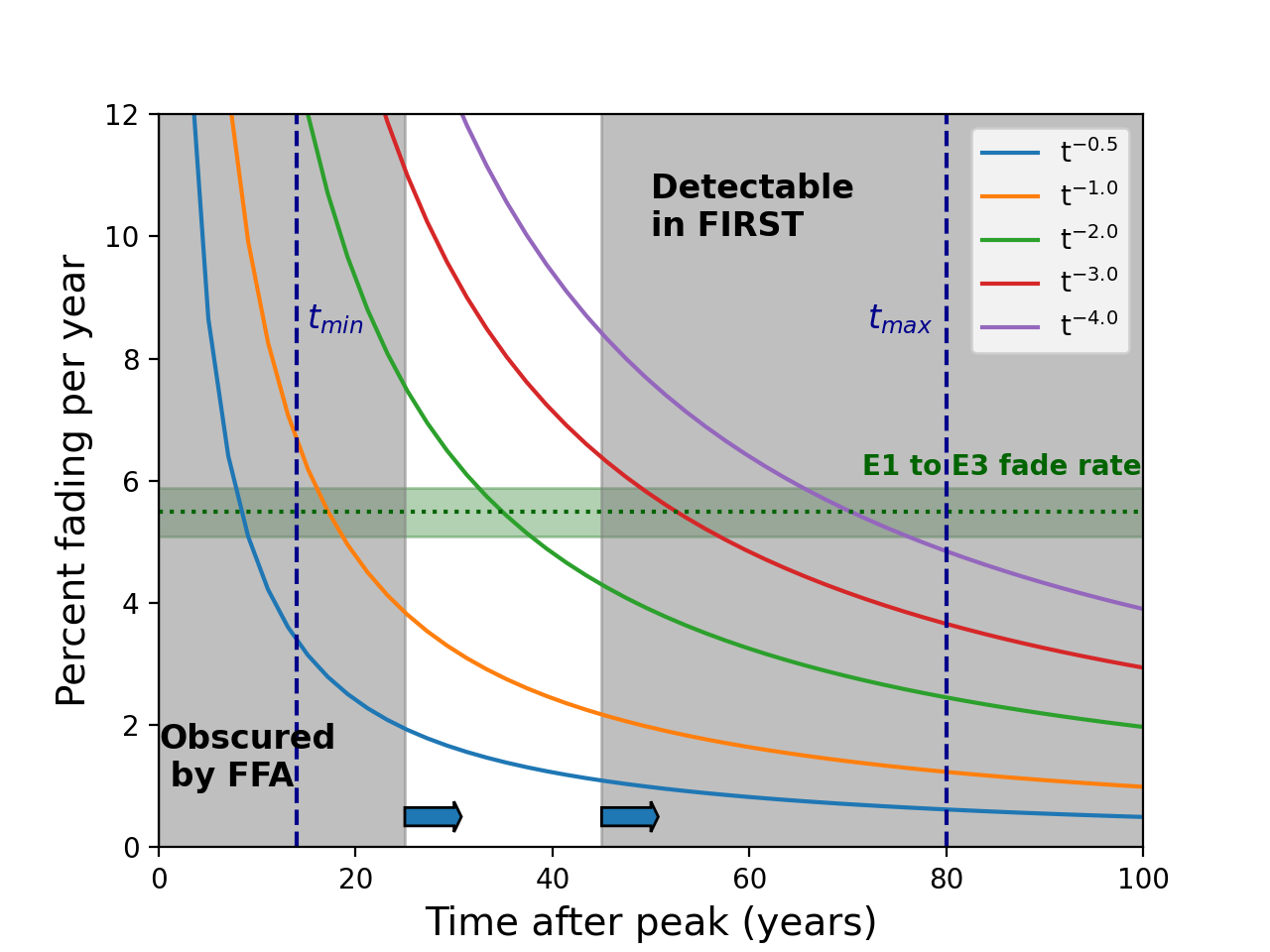}
    \caption{Age constraints for a pulsar wind nebula model of VT 1137-0337. The minimum age of the nebula is $t_{min} \approx 14$~years, while the maximum age is $t_{max} \approx 80$~years. The true age lies within a $\sim$20 year sliding window starting from the age $t_{ff}$ when the ejecta becomes transparent to free-free absorption at 1~GHz (shown in the figure with a fiducial value of $t_{ff}$ = 25~years). The measured per-year fade rate and its uncertainty is shown with the green dotted line and its corresponding shaded region. When combined with an assumed true age, this per-year rate determines the present-day steepness of VT 1137-0337's fading lightcurve, as illustrated by the various power laws plotted as solid lines.}
    \label{fig:fading_constraints}
\end{figure}

We begin by constraining the range of PWN ages ($t_{\textrm{age}}$) consistent with VT 1137-0337's observables, under the assumption that it was born in a supernova. There are thought to be other ways of forming young neutron stars (such as the accretion induced collapse of a white dwarf), but VT 1137-0337's location within an extreme starburst suggests that a supernova would be the most likely channel. The age constraints will be applicable regardless of the nebula's power source. 
\\
\\To start, there is a hard lower limit of $t_{\textrm{age}} > 4$ years in February 2022 based on the duration that we have observed the source. This minimum age is pulled upwards by two requirements: (1) that the supernova ejecta must expand to the point where it is optically thin to free-free absorption, and (2) that the nebula has time to grow to a size $R \gtrsim 10^{17}$~cm, as we inferred in Section \ref{sec:spectral_breaks}. The time for the ejecta to become transparent to free-free absorption is given by \citet[][]{Metzger17-FRB-SLSNe-LGRB}:

\begin{equation}
    t_{ff} \approx 9.85 \left(\frac{f_{\textrm{ion}}}{0.1}\right)^{2/5} T_4^{-3/10} M_{10}^{2/5} v_9^{-1}\nu_{\textrm{GHz}}^{-2/5}~\textrm{years},
\end{equation}
\noindent where $f_{\textrm{ion}}$ is the ionization fraction of the ejecta, $T_4$ is the ejecta temperature in units of 10$^4$K, $M_{10}$ is the ejecta mass in units of 10$M_{\odot}$, $v_9$ is the ejecta velocity in units of 10$^4$ km/s, and $\nu_{\textrm{GHz}}$ is the observing frequency in GHz. Even for a high intensity of ionizing radiation from the nebula, $f_{\textrm{ion}}$ is expected to be $\lesssim$ 0.25 on a timescale of $\sim$decades, with full ionization occurring on a timescale of $\sim$centuries \citep[][]{Metzger17-FRB-SLSNe-LGRB}. For typical supernova kinetic energies and ejecta masses \citep[e.g.,][]{Nicholl15-SLSNe-masses, Taddia15-stripped-envelope-ejecta-mass-energy, Barbarino20-Ic-ejecta-mass, Martinez22-typeII-SN-energy-ejecta_mass}, the quantity $M_{10}^{2/5} v_9^{-1}$ is $\sim 0.5$ for Ic broadline supernovae, $\sim 1$ for superluminous supernovae, $\sim 1.5$ for stripped envelope (type Ib/c) supernovae, and $\sim 3.5$ for type II supernovae. For these various classes, $t_{ff} \approx$ 1 to 4 decades for our lowest observed frequencies of $\sim$1~GHz. 
\\
\\Next, the radius of a PWN expanding within supernova ejecta is given by \citet[][]{Chevalier2004-PWN}:

\begin{equation}
\label{eqn:SN_radius}
    R = 2 \times 10^{17} \dot{E}_{42}^{1/5} E_{51}^{3/10} M_{10}^{-1/2} t_{20}^{6/5}~\textrm{cm},
\end{equation}
where $\dot{E}_{42}$ is the rate of energy injection into the PWN in units of 10$^{42}$~erg~s$^{-1}$, $E_{51}$ is the energy of the supernova in units of 10$^{51}$~erg, and $t_{20}$ is the age normalized to 20 years. For the various supernova types, the quantity $E_{51}^{3/10} M_{10}^{-1/2} \approx 0.9$~to~$3$. For reasonable values of $\dot{E}$ discussed in the sections below, $\dot{E}_{42}^{1/5} \gtrsim 0.25$. The timescale $t_{ff} \gtrsim 10$~years is sufficient to expand the nebula to $\gtrsim 10^{17}$~cm for fiducial values, though less energetic supernovae with higher ejecta masses will require longer to reach this radius. Adding the duration that we have observed the source to $t_{ff}$, we adopt a fiducial lower limit to the nebula's age in 2022 of $t_{\textrm{age}} \gtrsim 14$~years. 
\\
\\We note that at an age of $\gtrsim 10$~years, the shock emission from most (but not all) supernovae should be orders of magnitude fainter than what we can detect. VT 1137-0337 has a radio spectral luminosity of $\sim$10$^{28}$~erg~s~Hz$^{-1}$, which is $\sim$2.5 orders of magnitude more luminous than the median peak of supernova radio lightcurves. These peaks tend to occur at an epoch of $\sim$months to years after explosion \citep[][]{Bietenholz21-SN-luminosity-dist}. At an epoch of $\sim$decades, typical supernovae will be one or more orders of magnitude fainter yet \citep[][]{Chevalier98}. We also note that for typical NS kick velocities, the radius of the nebula should be larger than the distance the NS has travelled, given by $R_{\textrm{kick}} = 1.9 \times 10^{16}~v_{300}~ t_{20}$~cm, where $v_{300}$ is the kick velocity in units of 300 km/s. Even for the largest kick velocities of $\sim$1000~km/s, the NS should still be embedded within the nebula, though the distance travelled and the implied asymmetry of the ejecta would contribute to the nebula partially breaking out at a slightly earlier time. Thus, supernova shocks and neutron star kicks should not be issues for the general picture presented here.
\\
\\An upper limit to the age of the nebula can be inferred from the fact that VT 1137-0337 was detected as a transient. If $t_{ff} \gtrsim 10$~years is the time that it takes for the ejecta to become transparent at 1~GHz, the upper limit to the age in 2018 is $t_{ff} + 20$~years, since if VT 1137-0337 were older than that, it should have been clearly detected in FIRST. Our estimate of $t_{ff} \approx 1 - 4$~decades, therefore suggests that the source is no more than $\sim$ 6 decades old. 
\\
\\A complementary upper limit comes from the $\sim$5.5\% per year fading that we measure between Epoch 2 and Epoch 3. In both spindown and magnetar models, the rate of fading can be approximated with a power law that at early times is generally not steeper than $\sim$t$^{-4}$ \citep[e.g.,][]{Reynolds-Chevalier1984-PWN-lightcurve, Margalit18-FRB121102-magnetar}. Figure \ref{fig:fading_constraints} shows the observed fade rate in comparison with power laws of various steepness. Past an age of $\sim$80~years, even a fade rate of t$^{-4}$ will be slower than the rate that we measure.
\\
\\Altogether, the allowable age range for VT 1137-0337 is a $\sim$20~year window with the lower edge set by $t_{ff}$ (Figure \ref{fig:fading_constraints}). Different values of $t_{\textrm{age}}$ also correspond to different power-law rates of fading when combined with the observed per-year fade rate. In the next section, we use this to constrain the allowable parameter space in the $P-\dot{P}$ diagram for a spindown model.

\subsubsection{Spindown implies a strongly magnetized NS}
\label{subsubsection:spindown}

We now consider a nebula powered by the spindown of a central NS, the mechanism thought to power most Galactic PWNe. In standard spindown models, the rate of energy input to the nebula $\dot{E}$ is assumed to be equal to the rate of rotational energy dissipation (the spindown luminosity $L_{\textrm{sd}}$), which can be expressed as a function of the NS's rotational period $P$ and its time derivative $\dot{P}$:

\begin{equation}
    L_{\textrm{sd}} = 4 \times 10^{41} I_{45} \dot{P}_{-11} P_{10}^{-3}~\textrm{erg s}^{-1}.
\end{equation}
\noindent Here, $I_{45}$ is the NS moment of inertia in units of 10$^{45}$ g cm$^{2}$ (equal to 1.1 for a constant-density 1.4 M$_{\odot}$ sphere of radius $R_{\textrm{NS}} = 10$~km), $\dot{P}_{-11}$ is the period derivative in units of 10$^{-11}$ seconds per second, and $P_{10}$ is the period in units of 10~ms. As the NS's magnetic field rotates, it emits magnetic dipole radiation with a luminosity of

\begin{equation}
    \label{eqn:edot-B-P}
    \dot{E}_{d} = 4 \times 10^{41} B_{d,13}^2 P_{10}^{-4}~\textrm{erg}~\textrm{s}^{-1},
\end{equation}
\noindent where $B_{d,13}$ is the strength of the surface dipole field in units of 10$^{13}$ G. By equating $\dot{E}_{d}$ with $L_{\textrm{sd}}$, we can get an expression for $B_d$ (where we have dropped a constant within a few percent of unity):

\begin{equation}
    \label{eqn:B}
    B_{d,13} = (\dot{P}_{-11} P_{10})^{1/2}.
\end{equation}
\noindent The spindown luminosity evolves over time \citep[][]{Gaensler06-PWN-review}:

\begin{equation}
    \label{eqn:edot-evolution}
    \dot{E}_{d} = \dot{E}_0 \left(1+\frac{t}{\tau_0}\right)^{-\frac{n+1}{n-1}},
\end{equation}
\noindent where $\dot{E}_0$ is the initial value of $\dot{E}_d$, $n$ = 2 to 3 is the braking index, and $\tau_0$ is the initial spindown time:

\begin{equation}
    \label{eqn:tau0}
    \tau_0 = \frac{P_0}{(n-1)\dot{P}_0}.
\end{equation}

\noindent At early times $t << \tau_0$, $\dot{E}_{d}$ will be nearly constant. Later, when $t >> \tau_0$, $\dot{E}_{d}$ will decrease as a power law $\propto t^{-2}$ to $t^{-3}$. By combining Equations \ref{eqn:B} and \ref{eqn:tau0} and assuming $n = 3$, the spindown time can also be recast in terms of the dipole field:

\begin{equation}
    \label{eqn:tau0_B}
    \tau_0 = 15.9~B_{d,13}^{-2}~P_{10}^{2}~\textrm{years}.
\end{equation}

\begin{figure*}
    \centering
    \includegraphics[clip=true, trim=0.15cm 0.3cm 3.1cm 0cm,width=\textwidth]{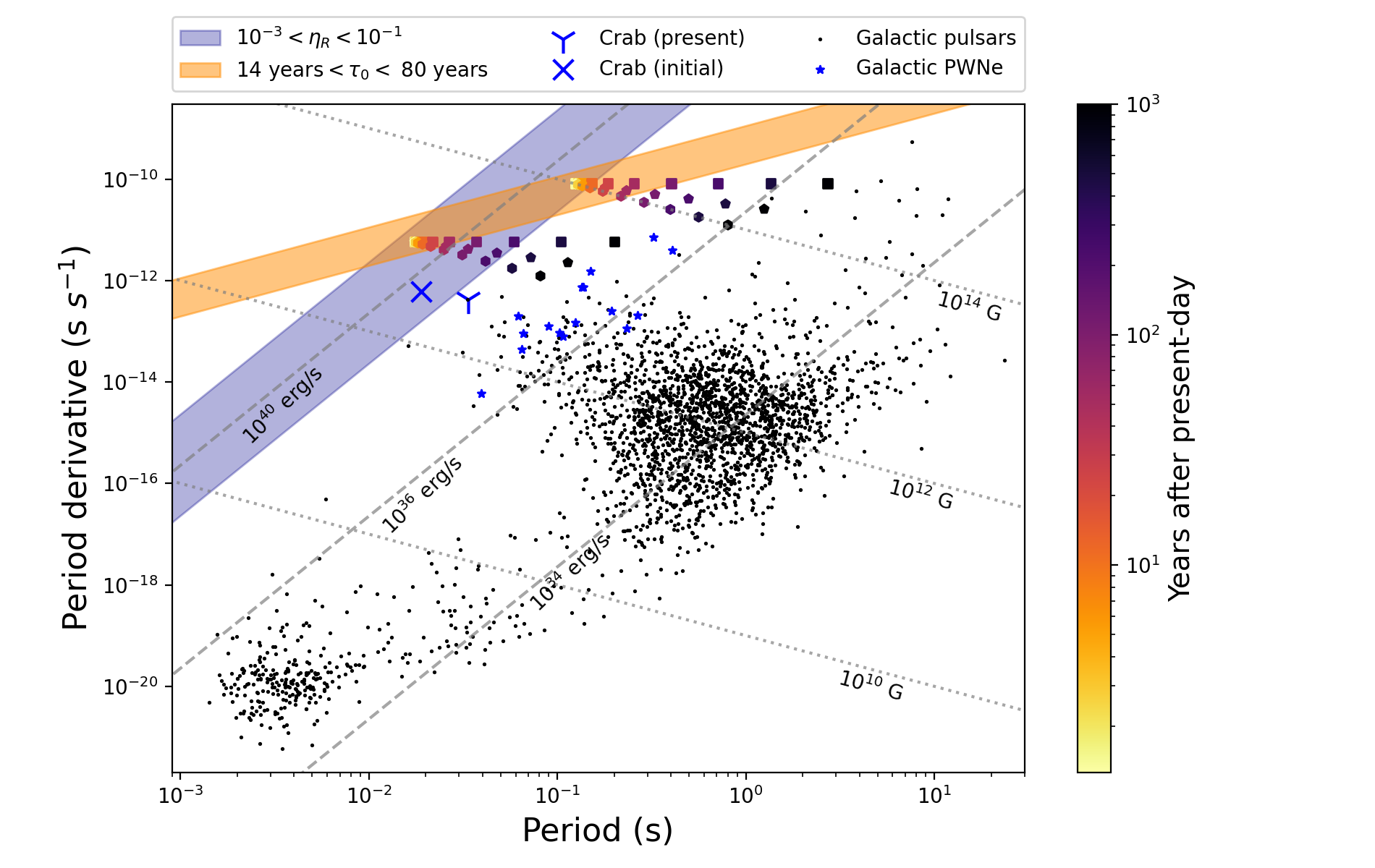}
    \caption{Constraints on a spindown-powered pulsar wind nebula model of VT 1137-0337, plotted in comparison with Galactic pulsars \citep[black points;][]{Manchester05-psrcat}. The dashed and dotted lines show curves of constant spindown luminosity $\dot{E}_d$ and surface dipole field $B_d$ respectively. Pulsars from the \citet[][]{Green19-SNR-catalog} catalog with well-characterized nebulae are shown with blue stars. Using the Y and X symbols, we highlight the present position of the Crab Pulsar \citep[][]{Lyne1993-Crab-braking-index} and its estimated initial position from tracing back its spindown \citep[][]{Kou-Tong15-Crab-initial-period}. In our model, VT 1137-0337's present day position is in the intersection of the orange shaded region (derived from limits on the initial spindown time $\tau_0$) and the blue shaded region (derived from our estimate of $\dot{E}_d$). The colored tracks show the spindown over 1000 years of neutron stars currently located at two different positions in the intersection ($B_d$ = $10^{14}~G$, $\dot{E}_d = 2 \times 10^{39}$~erg~s$^{-1}$) and ($B_d$ = $10^{13}~G$, $\dot{E}_d = 5 \times 10^{40}$~erg~s$^{-1}$). In these tracks, the squares, pentagons, and hexagons represent braking indices of n = 2, 2.5, and 3, respectively.}
    \label{fig:p-pdot}
\end{figure*}

\noindent From this, we can see that the initial spindown time depends strongly on both the initial period and the magnetic field. We will return to this later in this section and the next, but first, we attempt to constrain the values of $\dot{E}_d$ and $\tau_0$ consistent with VT 1137-0337's radio luminosity and lightcurve.
\\
\\In Galactic PWNe, the spindown luminosity can be measured precisely through pulsar timing. However, for our model, the best we can do is to estimate what it should be from the radio luminosity. The conversion between radio and spindown luminosity is often expressed with the fraction $\eta_R \equiv L_R / L_{sd} = L_R / \dot{E}_d$. For well-characterized radio PWNe in the \citet[][]{Green19-SNR-catalog} catalog, the median value of $\eta_R$ is $3 \times 10^{-4}$, with a scatter of $\sim 0.85$~dex, where we have taken properties of the corresponding pulsars from \textit{psrcat} \citep[][]{Manchester05-psrcat}. This is in agreement with results in the literature \citep[e.g.,][]{Frail-Scharringhausen-1997-PWN-radio-survey, Gaensler2000-eta_R_PWNe}, who find $\eta_R \sim 10^{-4}$ for young and energetic pulsars, though some older pulsars with undetected nebulae have values $< 10^{-5}$. 
\\
\\The value of $\eta_R$ is expected to evolve over time as the NS spins down ($\dot{E}_d$ decreases, as described above) and its PWN fades ($L_R$ decreases). The behavior of the radio lightcurve depends on $\dot{E}_d$, the adiabatic cooling of the radio-emitting electrons and the evolution of the magnetic field in the nebula. An analytic model for the lightcurve of a spindown powered PWN expanding within spherically symmetric supernova ejecta undergoing homologous expansion can be found in \citet[][]{Reynolds-Chevalier1984-PWN-lightcurve}. Their model predicts that for $t < \tau_0$, a single injected electron energy distribution of $p = 1.7$, and an observing frequency below the synchrotron cooling break, $L_R$ should fade approximately as a power law of $t^{-0.75}$ to $t^{-1.35}$, depending on the degree to which the PWN has swept up the ejecta. After $\tau_0$, when the spindown luminosity begins to drop, the radio lightcurve fades much more steeply, as $\approx t^{-3.4}$ for $p = 1.7$. Later, after the PWN is compressed by the passage of the reverse shock, the luminosity will undergo a sharp enhancement followed by a slow decline $\approx t^{-1}$. This occurs at a time $\sim 10^4$~years for a thin wind or an ISM-like density, though may occur much sooner if there is a dense shell of circumstellar material around the supernova. The net result is that before the reverse shock sweeps through (both before and after $\tau_0$), $\eta_R$ should very roughly decrease as $\sim t^{-0.75}$ to $t^{-1.5}$. After the reverse shock, $\eta_R$ should increase as $\sim t^{1}$ to $t^{2}$.
\\
\\In Section \ref{sec:spectral_breaks}, we argued that VT 1137-0337 is not surrounded by a dense shell (so the reverse shock is not relevant at our inferred $t_{\textrm{age}}$ of $\sim$decades), and that our radio observations likely lie below the cooling break. We therefore expect that $\eta_R$ should be decreasing with time, and that the value of of $\eta_R$ in Galactic PWNe is likely an underestimate. Extrapolating backwards from the youngest known PWNe (including the Crab Nebula) which have characteristic ages $\sim$1000~years and $\eta_R \sim 10^{-4} - 10^{-3}$, we estimate that a reasonable range of values for a PWN with an age of a few decades is $\eta_R \sim 10^{-1} - 10^{-3}$. For VT 1137-0337's radio luminosity of $L_R \sim 10^{38}$~erg~s$^{-1}$, the corresponding spindown luminosity range is $\dot{E}_d \sim 10^{39}$ to 10$^{41}$~erg~s$^{-1}$. 
\\
\\Using the age constraints in Section \ref{subsubsection:PWN_age}, we can separately constrain the value of the initial spindown time $\tau_0$. As can be seen in Figure \ref{fig:fading_constraints}, if the true age of the nebula $t_{\textrm{age}}$ were close to the minimum age $t_{\textrm{min}} \approx 14$~years, the measured per-year rate of fading corresponds to a rather shallow power law $\approx t^{-1}$. In the framework of \citet[][]{Reynolds-Chevalier1984-PWN-lightcurve}, this would imply that the NS has not yet appreciably spun down, so $\tau_0 \gtrsim t_{\textrm{age}} \gtrsim 14$~years. Alternatively, if $t_{\textrm{age}}$ is closer to the maximum value $t_{\textrm{max}} \approx 80$~years, we would require a steep power law $\sim t^{-4}$ to reach the $\sim$5\% per-year fading we observe. Such a steep power law would imply that the NS has already begun to spin down substantially, and so $80$~years $\gtrsim t_{\textrm{age}} \gtrsim \tau_0$. Intermediate values of $t_{\textrm{age}}$ correspond to power laws of intermediate steepness that would appear when $t_{\textrm{age}} \approx \tau_0$. This implies that for a spindown powered nebula, $\tau_0$ is between $\sim$ 14 to 80 years.
\\
\\As shown in Figure \ref{fig:p-pdot} the constraints on $\dot{E}_d$ and $\tau_0$ combine to form a rather tight region in the classic pulsar parameter space of P and $\dot{P}$. The allowed region corresponds to a dipole field of slightly less than 10$^{13}$~G to a few $\times$ 10$^{14}$~G, roughly an order of magnitude stronger than that of the Crab Pulsar ($\sim 10^{12.5}$~G). Such strongly magnetized neutron stars should spin down rapidly (Equation \ref{eqn:tau0_B}). Within $\sim$100~years, $\dot{E}$ will have decreased by $\sim 10 - 30\times$ and, due to the expected drop in $\eta_R$, the radio luminosity will have decreased by an even larger margin. Within a few thousand years, the central NS should fall within the distribution observed in the Milky Way. If the magnetic field is $\sim 10^{14}$~G, the NS may eventually resemble the population of Galactic magnetars \citep[][]{Kaspi-Beloborodov-magnetar-review-2017}. If instead $B$ is closer to $\sim$10$^{13}$~G, it might eventually resemble a highly magnetized pulsar, similar to the NS at the center of the composite supernova remnant Kes 75 \citep[e.g.,][]{Gavriil08-Kes75, Reynolds18-Kes75-expansion}. 

\subsubsection{Magnetar activity as an alternate power source}
\label{subsubsection:magnetic_energy}

As discussed above, VT 1137-0337's luminosity and lightcurve are consistent with a $\sim$decades old PWN powered by spindown. However, the only part of the argument that depends specifically on spindown is the evolution of $\dot{E}$. In principle, any source of power that can approximately reproduce $\dot{E}$ (and also satisfy one of the mechanisms for producing a flat spectrum) would suffice. A change in the time-dependence of $\dot{E}$ would change the PWN's lightcurve, but as shown in Figure \ref{fig:fading_constraints}, our per-year fading measurement can be reconciled with a broad range of power-law fade rates from $\sim t^{-1}$ to $\sim t^{-4}$ simply by changing $t_{\textrm{age}}$. This leads us to consider a different energy source often used in modeling magnetars: the direct decay of the NS's magnetic field.
\\
\\Galactic magnetars typically have spindown-inferred surface dipole fields of $B \sim 10^{14} - 10^{15}$~G \citep[][]{Kaspi-Beloborodov-magnetar-review-2017}. Their total surface fields may be even stronger if a large fraction is ``hiding" in higher order multipoles \citep[those components fall off more rapidly with radius and make only marginal contributions to the torque at the light cylinder radius where the pulsar wind is launched;][]{Cordes21-dipole-vs-quadrapole-field}. Fields within the NS should be even stronger. Galactic magnetars maintain high blackbody temperatures for much longer than their initial neutrino cooling timescale, suggesting that their internal fields (which heat the NS through their decay) can be as strong as $B_{\textrm{int}} \sim 10^{16}$~G. The strong internal field corresponds to a large energy reservoir $E_{B} \sim (\frac{4\pi}{3} R_{\textrm{NS}}^3) \frac{B_{\textrm{int}}^2}{8\pi} \sim 2 \times 10^{49} B_{\textrm{int},16}^2$~erg. This reservoir is primarily released via ambipolar diffusion on a characteristic timescale of roughly $t_{\textrm{amb}} \sim 400~B_{\textrm{int},16}^{-1.2}~L_{\textrm{km}}^{1.6}$~years, where $B_{\textrm{int}}$ is normalized to 10$^{16}$~G, and $L$ in kilometers is the length scale of magnetic field fluctuations being dissipated in the core \citep[][]{Beloborodov-Li16-magnetar-heating}. The corresponding average luminosity is roughly

\begin{equation}
    \label{eqn:magnetic_Edot}
    \dot{E}_{\textrm{mag}} \sim \frac{E_{B}}{t_{\textrm{amb}}} \sim 10^{39}~B_{\textrm{int},16}^{3.2}~L_{\textrm{km}}^{-1.6}~\textrm{erg}~\textrm{s}^{-1},
\end{equation}

\noindent which scales sharply with the internal magnetic field, and should be able to power VT 1137-0337 (i.e. produce $\dot{E} = 10^{39} - 10^{41}$~erg~s$^{-1}$) provided that $B_{\textrm{int}} \gtrsim 10^{16}$~G.
\\
\\The energy released by ambipolar diffusion should on average decline as a power law with time as the length scale $L$ grows from small scales to the radius of the NS and the internal magnetic field decays \citep[][]{Beloborodov-Li16-magnetar-heating}. Though on short timescales the energy may be released in discrete magnetar flares, these flares should combine to form a variable or intermittent wind which also gets decelerated in a termination shock and transfers its energy to particles in the nebula through wave heating \citep[][]{Beloborodov17-magnetar-121102}. The evolution of a magnetar wind nebula with a power law energy input $\dot{E}_\textrm{mag} = \dot{E}_0 (t/t_0)^{-\alpha_B}$ where the magnetar switches on at a time $t_0$ and injects relativistic electrons through its flares has been modeled semi-analytically by \citet[][]{Margalit18-FRB121102-magnetar}. They find that at an age of $\sim$decades, nearly all of the properties of the persistent radio source associated with the repeating fast radio burst FRB 121102 \citep[including its radio luminosity, which is $\sim 10 \times$ that of VT 1137-0337;][]{Chatterjee-FRB121102-VLBI-loc} can be reproduced with $E_{B} \sim 5 \times 10^{50}$~erg, $\alpha_B$ $\sim$ 1.3, and $t_0 \sim 0.2$~years. By changing the parameters slightly, they are able to scale the luminosity upwards or downwards by an order of magnitude, including to the luminosity of VT 1137-0337. At frequencies below the synchrotron cooling break, their model predicts that the nebula's lightcurve should decrease as $t^{-(\alpha_B^2 + 7\alpha_B - 2)/ 4} \sim t^{-1.5}$ to $t^{-4}$ for $\alpha_B = 1$ to 2. This range of fade rates is consistent with the range inferred for VT 1137-0337 for $t_{\textrm{age}} \sim 20 - 80$~years. 
\\
\\These order-of-magnitude considerations suggest that the decay of a young magnetar's internal field is a plausible alternative to spindown for explaining VT 1137-0337's luminosity and lightcurve. If this is the case, then the field decay should be the nebula's primary power source in the present day, since NSs born with strong enough magnetic fields should have long since spun down. When normalized to a surface dipole field $B_d = 10^{15}$~G (a reasonable value for the strong required internal fields), the initial spindown time (Equation \ref{eqn:tau0_B}) becomes $\tau_0 \sim 14~B_{d,15}^{-2}~P_{10}^2$~hours. For a braking index $n = 3$, the age at which the spindown luminosity drops below the level required to power a radio luminosity $L_R$ is roughly 
\begin{equation}
    \label{eqn:braking-time}
    t_{\textrm{brake}} \sim \tau_0 + (1~\textrm{year})~ B_{d,15}^{-1} L_{38}^{-\frac{1}{2}} \left(\frac{\eta_R}{0.01}\right)^{\frac{1}{2}},
\end{equation}
\noindent where $L_{38}$ = $L_R / 10^{38}$ erg s$^{-1}$. For initial periods $P_0 \lesssim 300$ ms and $B_d = 10^{15}$~G, $t_{\textrm{brake}} \lesssim$ years is much less than $t_{\textrm{age}} \sim$ decades. If the initial period is much longer, the spindown luminosity never reaches the level required to power the nebula (Figure \ref{fig:p-pdot}), and if $n$ is closer to 2, spindown occurs even more rapidly. 
\\
\\At early times, the energy released by a magnetar's spindown could influence the expansion of the nebula and the luminosity of the supernova. If $B_d \sim 10^{15}$~G, a large fraction of the NS's rotational energy will have been released while the ejecta is still coupled to the internal radiation via electron scattering. If the magnetar has an initial period $P_0 \sim$ a few to a few 10s of milliseconds, its rotational energy would be $\gtrsim 10^{50}$~erg, which is enough to power a superluminous supernova when released at this early time \citep[][]{Kasen10-SLSNE-magnetars}. Such rapid rotation may be related to the generation of the strong fields \citep[][]{Raynaud20-millisecond-magnetar-sim}, and due to the age of VT 1137-0337, such a luminous explosion is not ruled out by the lack of an archival supernova detection (Section \ref{subsec:archival}). In this scenario, the radius of the nebula might be larger than expected for its age due to the additional energy imparted to the ejecta and the increase in $\dot{E}$ (Equation \ref{eqn:SN_radius}). If instead $P_0 \gtrsim$ 100 ms, the total rotational energy would be closer to $\sim$10$^{47} - 10^{48}$ erg. In this case, the early spindown luminosity (comparable to $\dot{E}_{\textrm{mag}}$) would have a small effect on the initial expansion of the nebula, and a negligible effect on the supernova. 

\subsection{Comparison with FRB persistent sources}
\label{subsec:FRB-PRSs}

Much of the recent work on luminous magnetar wind nebulae has been done in the context of persistent radio sources (PRSs) associated with repeating fast radio bursts (FRBs) \citep[e.g.,][]{Murase16-FRB-magnetar,Beloborodov17-magnetar-121102,Margalit18-FRB121102-magnetar}, a connection that has recently been strengthened by the detection of a low-luminosity FRB from the Galactic magnetar SGR 1935+2154 \citep[][]{Bochenek20-magnetar-FRB, CHIME-galactic-magnetar}, though see \citet[][]{Sridhar22-XRB-FRB-prs} for an alternative model involving a highly super-Eddington stellar mass black hole. There are a number of observational similarities between VT 1137-0337 and PRSs. The two PRSs identified to date are associated with FRB 121102 \citep[][]{Chatterjee-FRB121102-VLBI-loc} and FRB 190520B \citep[][]{Niu21-second-FRB-PRS}. Both of the PRSs have comparably flat spectral indices at frequencies of a few GHz and radio luminosities $\sim 10\times$ higher than that of VT 1137-0337 (Figure \ref{fig:lum_vs_alpha}). Both PRSs are hosted by starbursting dwarf galaxies of stellar mass $\sim 10^{8} - 10^{9}~M_{\odot}$ and star formation rate $\sim 0.2 - 0.5~M_{\odot} / $yr (Figure \ref{fig:host-galaxies}). These galaxies have specific star formation rates (SSFRs) comparable to the hosts of long GRBs and superluminous supernovae, but there is considerable overlap with the SSFRs of ordinary core collapse supernova hosts \citep[Figure \ref{fig:host-galaxies},][]{Taggart2020-SN-hosts}. Neither PRS has thus far been observed to fade, but the measurement uncertainties for these faint sources ($\sim 250 \mu$Jy at $\sim$1.5~GHz in both cases) do not preclude fading at the level that we have observed. Likewise, neither PRS has been observed to appear as a transient, but the lack of sufficiently deep reference epochs does not exclude this scenario. Some of the best available constraints come from the the NRAO VLA Sky Survey (NVSS), which gives 1.5~GHz 3$\sigma$ upper limits at both PRS locations of $\sim$1.5~mJy roughly two decades before detection \citep[][]{Condon98-NVSS}, suggesting that the average fade rate of these sources on $\sim$2 decade timescales is $\lesssim 10\%$ per year.
\\
\\One feature of the FRB 121102 PRS not present in VT 1137-0337 is the break at $\sim$10~GHz where the spectrum steepens from $\sim \nu^{-0.2}$ to $\sim \nu^{-1.2}$ \citep[][]{Chatterjee-FRB121102-VLBI-loc}. This break is confirmed by \citet[][]{Chen22-FRB121102-scintillation} who find a low variability amplitude at 15 and 22~GHz, suggesting that the break is intrinsic rather than a chromatic artifact of refractive scintillation. A potential break at the same frequency is unconstrained by currently available observations of FRB 190520B, which span 1 - 6 GHz \citep[][]{Niu21-second-FRB-PRS}. The origin of the FRB 121102 break is still somewhat unclear. The most straightforward explanation might be synchrotron cooling, which is generically expected to deplete high-energy electrons radiating at high frequencies where cooling is efficient, piling them up at a lower energy $E_c$ and creating a spectral break at the corresponding cooling frequency $\nu_c$. However, cooling of a single electron population predicts a steepening from $\alpha = (p-1)/2$ for $\nu < \nu_c$ to $p/2$ for $\nu > \nu_c$ for a difference $\Delta \alpha = 0.5$, less than the observed value $\Delta \alpha \sim 1$. One potential solution is if more electrons are injected at early times where the magnetic field is stronger and the cooling frequency lower. As discussed in \citet[][]{Atoyan99-flat-spectrum-Crab}, this would flatten the spectrum at frequencies lower than $\nu_c$, thus alleviating the tension in $\Delta \alpha$. It is likely, however, that other solutions exist.
\\
\\Regardless of the mechanism, if the break were caused by a cooling frequency, the lack of a detected break in VT 1137-0337 is unsurprising given its lower radio luminosity, which suggests an older age, a lower $\dot{E}$, or both. Either of these differences could lead to the cooling break $\nu_c$ being located at a higher frequency than the range we have observed. For electrons with age $t$ and a nebula with field strength $B_n$, the cooling frequency scales as $\nu_c \propto t^{-2}~B_n^{-3}$. In young nebulae, the weakening of the magnetic field (which is primarily driven by adiabatic cooling as the nebula expands) should be the dominant factor, causing the cooling frequency to rise rapidly with time.  In the case of spindown, $B_n \propto t^{-1.3}$ to $t^{-2}$, so $\nu_c \propto t^{1.9}$ to $t^{4}$ \citep[][]{Reynolds-Chevalier1984-PWN-lightcurve}. For a power law magnetic energy injection $\dot{E} \propto (t/t_0)^{-\alpha_B}$ and a nebula radius $R_n \propto \dot{E}^{1/5}~t^{6/5}$ (Equation \ref{eqn:SN_radius}), integrating Equation 11 from \citet[][]{Margalit18-FRB121102-magnetar} yields $B_n \propto t^{-(3\alpha_B + 13)/10} \sim t^{-1.6}$ to $t^{-2}$ for $t >> t_0$ and $\alpha_B = 1$ to 2. This leads to a similar evolution for the cooling break: $\nu_c \propto t^{2.8}$ to $t^{4}$. Even for the shallowest power laws, raising $\nu_c$ from 10~GHz to $\gtrsim 15$~GHz (beyond our observing frequencies) would only require that VT 1137-0337 is $\gtrsim 20\%$ older than FRB 121102. For a fixed nebula age, the value of $B_n$ could also be lowered by reducing $\dot{E}$, though this effect is weaker than that of aging. Roughly, $B_n \propto E_B^{1/2} R_n^{-3/2}$, where $E_B$ is the magnetic energy in the nebula and the nebula radius is a weak function of $\dot{E}$, scaling as $R_n \propto \dot{E}^{1/5}$ (Equation \ref{eqn:SN_radius}). If we neglect the differential adiabatic cooling due to the slightly different expansion rate, $E_B \propto \dot{E}$ for fixed age. Altogether, this leads to $B_n \propto \dot{E}^{0.2}$, implying $\nu_c \propto \dot{E}^{-0.6}$. If we assume that a 10$\times$ smaller radio luminosity corresponds to a 10$\times$ lower $\dot{E}$, this would lead to a $\sim 4\times$ higher $\nu_c$ at the same age. This too would be enough to explain the lack of a cooling break at $\nu < 15$~GHz.
\\
\\Overall, we find the two FRB persistent sources to be intriguing possible analogues for VT 1137-0337, given the theoretical connections to magnetar wind nebulae and the observational similarities. However, a link between the two cannot be firmly established without the detection of associated FRBs. The best publicly available constraints come from the first CHIME data release, which does not report any detections consistent with VT 1137-0337 between July 2018 and July 2019  \citep[][]{CHIME-FRB-catalog-2021}. The lack of detections is not particularly constraining however, since VT 1137-0337 is located close to CHIME's southern declination limit ($-11^{\circ}$) where their observations are likely not as sensitive and since our observations do not constrain the free-free opacity at CHIME's observing frequencies (400 - 800 MHz). From Figure 4 of \citet[][]{CHIME-FRB-catalog-2021}, CHIME's total integration time at VT 1137-0337's location in the reported span was $< 10$ hours, and depending on the exact configuration of their synthesized beams, the actual effective integration time may have been much less. In the most tightly constrained scenario, this implies that VT 1137-0337 is not much more active than FRB 121102, which had one detection by CHIME in a total integration time of 11.3 hours \citep[][]{Josephy19-CHIME-FRB121102}. It remains to be seen whether there are fast timescale bursts detectable from VT 1137-0337. We have started to monitor the source for bursts and will report on our results in a future paper. 

\section{Summary and Conclusions}
\label{sec:discussion}

We have discovered VT 1137-0337, a luminous, flat spectrum radio transient in a dwarf starburst galaxy at redshift z = 0.02764 with mass $\sim 10^{8.3} M_{\odot}$, star formation rate $\sim 0.5 M_{\odot}$~yr$^{-1}$, and metallicity $\sim 0.3 Z_{\odot}$. The transient source became detectable at radio frequencies some time between 1998 and 2018, brightening by $>$550\% at 1.5~GHz. Between 2018 and 2022, it has faded at an average rate of $\sim$5\% per year while maintaining a stable power-law spectrum $S_{\nu} \propto \nu^{-0.35}$ from $\sim 1$ to 15~GHz (Sections \ref{sec:discovery}, \ref{sec:followup}).
\\
\\VT 1137-0337 is surrounded by massive young stars that have formed in the past few Myr, suggesting an association with the massive stars or their remnants. The lack of a self-absorption break in its synchrotron spectrum implies an emitting region larger than $\sim 10^{17}$~cm and causality requires an emitting region smaller than $\sim 6$~pc. Its luminosity implies a magnetic energy similar to the shocks of radio-luminous supernovae.  However, its flat spectrum and slow fading is unique among known radio transients of similar luminosity and cannot be explained by diffusive shock acceleration under standard conditions (Section \ref{sec:analysis}).
\\
\\We considered several possible origins for this transient (Section \ref{sec:comparisons}), summarized by the following:

\begin{itemize}
    \item Shocks from stellar explosions such as supernovae and gamma ray bursts have not been observed to (and are generally not expected to) have a flat spectrum. In the absence of observational analogues or theoretical expectations for such transients, we consider this explanation to be unlikely.
    
    \item Jets launched from black holes can have flat spectra through the continuous superposition of synchrotron self-absorbed components. However, black holes of this radio luminosity are not typically expected to reside in galaxies of this mass, and there are no optical or X-ray indications of accretion onto a black hole. More importantly, the size scale implied by self-absorption is not consistent with the slow and steady fading that we have observed from 1-10~GHz over four years. We do not believe that a black hole jet is a likely explanation, though future observations including radio monitoring will provide important tests for this argument. In particular, stochastic behavior in the radio lightcurve (in excess of expectations for scintillation) or a change in the radio spectral index would reopen the case for an AGN, while continued slow and steady fading would further strengthen our arguments against this scenario.
    
    \item Pulsar wind nebulae are one of the only known source classes that are \textit{expected} to have flat spectra and fade slowly. They will appear as radio transients once the surrounding supernova ejecta becomes optically thin to free-free absorption. Though VT 1137-0337 is much more luminous than pulsar wind nebulae found in the Milky Way, all of its observed properties can be explained with the spin-down of a $\sim$decades old neutron star with a dipole field $B_d \sim 10^{13 - 14}$~G and a period $P \sim 10 - 100$~ms. Alternatively, VT 1137-0337 could be powered by the decay of internal magnetic fields in the core of a magnetar (possibly through discrete flares), provided that the internal field is stronger than $\sim 10^{16}$~G. We find a nebula powered by either spindown or field decay to be a reasonable explanation for VT 1137-0337. 
    
    \item The two known persistent radio sources associated with repeating FRBs have similar radio luminosities, spectral indices and host galaxies to VT 1137-0337.  Though FRB persistent sources have not previously been observed as transients, they have been theorized to be $\sim$decades old magnetar wind nebulae. If that were the case, they should have appeared as transients in the recent past. We consider these sources to be intriguing potential analogs to VT 1137-0337. This scenario can be tested by further monitoring for associated bursts.
\end{itemize}

\noindent Among the scenarios considered, we find that the most compelling explanations for VT 1137-0337 involve highly magnetized neutron stars at an age of $\sim$ 1 to 8 decades. In the case of a spindown-powered pulsar wind nebula, the surface dipole field is constrained by limits on the spindown luminosity and initial spindown time to be approximately $B_d \sim 10^{13 - 14}$~G. In the case of a nebula powered by magnetar field decay, the injected power scales sharply with $B_{\textrm{int}}$, which should be stronger than $\sim 10^{16}$~G in order to power to the nebula. 
\\
\\Our results establish a general picture for VT 1137-0337, but leave open a rich set of questions for further observational and theoretical analysis. We list a few of these questions below:

\begin{itemize}
     \item \textit{Where are the breaks in the spectrum and how do they evolve with time?} Synchrotron spectra should generically turn over at a low frequency due to synchrotron self-absorption or free-free absorption, and should break at a high frequency due to synchrotron cooling. The self-absorption and cooling frequencies depend on the radius and magnetic field of the nebula, which in the case of young pulsar wind nebulae, are expected to evolve with time. Locating and tracking these breaks would provide important constraints on the evolution of the nebula and even nondetections would strengthen the limits we have provided in Section \ref{sec:analysis} \citep[e.g.,][]{Resmi21-low-freq-121102}.
    
    \item \textit{What is the origin of the flat spectrum?} Much theoretical work has been done to explain the flat spectra of pulsar wind nebulae, though observational verification of these theories has proven challenging. Many of the theories are motivated by in-depth studies of the Crab Nebula made possible by multi-wavelength, spatially resolved observations. VT 1137-0337 presents an opportunity to probe a very different region of the pulsar wind nebula parameter space. In particular, theories involving electrons injected at early times \citep[e.g.,][]{Atoyan99-flat-spectrum-Crab} may be constrained by measurements of the nebula's evolution.
    
    \item \textit{Are there associated FRBs or other bursts?} 
    Neutron stars produce a wide range of potentially related bursts, including Crab giant pulses \citep[e.g.,][]{Bera19-Crab-giant-pulses}, X-ray bursts \citep[e.g.,][]{Tavani21-magnetar-burst}, and fast radio bursts \citep[e.g.,][]{Bochenek20-magnetar-FRB}. Most of these burst types are associated with field-decay powered magnetars, though the line between these sources and the most strongly magnetized spindown powered pulsars is somewhat fuzzy \citep[e.g.,][]{Gavriil08-Kes75}. Since VT 1137-0337 has an inferred field in the range of strongly magnetized pulsars to full magnetars, it will be particularly interesting to monitor the source for bursts. In particular, the particles injected by bursts may be important in providing the number of relativistic particles required to produce the high radio luminosity we observe \citep[e.g.,][]{Beloborodov17-magnetar-121102}.
    
    \item \textit{What can VT 1137-0337 teach us about the population of young pulsar wind nebulae as a whole?} In our effective survey volume (Section \ref{subsec:detection-rate}), there should be $\sim 10^2 - 10^3$ young neutron stars born per year, of which at least $\sim 10\%$ should be magnetars \citep[e.g.,][]{Beniamini19-magnetar-birth-rate}. Our detection of only 1 likely pulsar wind nebula in observations spanning 20 years implies that the vast majority of young neutron stars (including magnetars) either produce nebulae that are much less luminous, or do not produce nebulae at all. An even larger gap in rates is inferred for FRB persistent sources by \citet[][]{Law22-FRB-PRS-population}. Understanding the selection factors that lead to luminous pulsar wind nebula will help interpret future searches for extragalactic nebulae and, if FRB persistent sources are truly magnetar wind nebulae, may help reveal why some repeating FRBs have luminous persistent sources \citep[][]{Chatterjee-FRB121102-VLBI-loc, Niu21-second-FRB-PRS} while others do not \citep[e.g.,][]{Kirsten22-FRB-M81}. 
\end{itemize}

\noindent We conclude with a few thoughts on the selection factors leading to radio-luminous nebulae. In both field-decay and spindown powered scenarios, we find that a flux-limited survey such as VLASS may biased towards neutron stars that are particularly strongly magnetized, though detailed verification requires population synthesis, further transient searches, and theoretical modeling beyond the scope of this paper.
\\
\\In Galactic magnetars, the typical timescale of field decay is $\sim$10$^4$ years \citep[][]{Beniamini19-magnetar-birth-rate} rather than the fiducial few $\times$ 10$^2$ years we considered in Section \ref{subsubsection:magnetic_energy}. These longer decay times translate to weaker internal fields $B_{int}$, lower averages rates of energy release $\dot{E}$ and, for fixed values of $\eta_R \equiv L_R / \dot{E}$, lower average radio luminosities $L_R$. Since for our fiducial parameters, $\dot{E}$ is already close to the minimum value for VT 1137-0337 of $\sim 10^{39}$~erg~s$^{-1}$ when $B_{int} = 10^{16}$~G, we may only be sensitive to the highest values of $\dot{E}$, corresponding to strong internal fields. This tracks with the lack of radio-detected magnetar wind nebulae in the Milky Way \citep[the strongest candidate for such a nebula is a faint X-ray source not detected in the radio;][]{Younes16-magnetar-wind-nebula}. Detailed modeling beyond our order of magnitude estimates is required to draw firm conclusions, however.
\\
\\For spindown-powered pulsars, Galactic sources that can be age-dated with associated supernova remnants have initial periods $P_0$ and dipole fields $B_0$ drawn from log-normal distributions centered at $P_0 \sim $100~ms with a scatter of $\sim$0.5~dex and $B_0 \sim 10^{12.4}$~G with a scatter of $\sim$0.4~dex \citep[][]{Igoshev22-pulsar-initial-B-and-P}. By running a Monte Carlo simulation where we assume a volumetrically uniform distance distribution $dN/dr \propto r^3$ and neglect any covariance between $B_0$ and $P_0$, we find that the median peak flux due to spindown of these neutron stars is $L_R \sim 6 \times 10^{-4}~(\eta_R / 10^{-2})$~mJy, roughly 1000 times fainter than we can detect. Due to the sharp scaling of $\dot{E}$ with $P_0$ however, a substantial tail of the distribution ($\sim 8\%$ of nebulae for $\eta_R = 10^{-2}$ or $\sim$3.5\% for $\eta_R = 10^{-3}$) should be brighter at peak than $\sim$1~mJy and would be detectable by our search. The requirement for nebulae to still be luminous at the time of emergence from free-free absorption shrinks this tail by only a small factor due to the slow spindown from ordinary pulsar-like dipole fields (a median value of 10$^{12.7}$~G in the tail). At an age of $\sim$40~years, roughly 80\% in the tail will still be brighter than 1~mJy, dropping to 30\% after 200 years. Altogether, the combined requirements of youth and escape from free-free absorption downselects the pulsar population by a factor of $\sim 100\times$, leaving us with a further factor of $\sim 10-100\times$ to account for our detection rate. 
\\
\\This can be accomplished if VT 1137-0337's unusually strong dipole field is a selected-for feature rather than a fluke. In the bright tail, only $\sim$30\% of neutron stars have dipole fields stronger than 1 $\times$ 10$^{13}$~G, roughly the minimum field strength inferred for VT 1137-0337. The fraction drops to 10\% above 2$\times 10^{13}$~G, and to only 1\% above 5$\times 10^{13}$~G, which is in the middle of the inferred range for spindown. Thus, if a field strength of a few $\times$ 10$^{13}$~G is required to produce a nebula as luminous as VT 1137-0337, that could plausibly explain our detection rate. As with magnetars, further analysis is required to draw a firm conclusion.
\\
\\It is interesting to note that at a dipole field strength of $\sim$10$^{13.5}$~G, the distinction between spindown powered pulsars and field-decay powered magnetars becomes blurred, with each population displaying features of the other \citep[e.g.,][]{Gavriil08-Kes75, Kaspi-Beloborodov-magnetar-review-2017}. If magnetar-like flares are required to populate the radio electron population and a large energy input from spindown (or decay of an unusually strong internal field) is required to power those electrons, the highest radio luminosity tail of young pulsar wind nebulae may be populated by unusually magnetized neutron stars. Future detections of flat spectrum radio transients in surveys such as the remaining epochs of VLASS \citep[][]{Lacy20-VLASS}, ASKAP/VAST \citep[][]{Murphy21-VAST-pilot}, LOTSS \citep[][]{Shimwell17-LOTSS}, or the DSA-2000 \citep[][]{Hallinan19-DSA2000-white-paper} will provide a test of this hypothesis.

\acknowledgments
DZD thanks Brian Metzger for helpful comments and Jim Cordes, Navin Sridhar, Jean Somalwar, Casey Law, Kristina Nyland, Vikram Ravi, and Bryan Gaensler for insightful conversations that helped shape this paper. This material is based upon work supported by the National Science Foundation under Grant No. AST-1654815 and the United States – Israel Binational Science Foundation grant 2018154. This work is based on data taken with the Very Large Array, operated by the National Radio Astronomy Observatory (NRAO). The NRAO is a facility of the National Science Foundation operated under cooperative agreement by Associated Universities, Inc. Some of the data presented here were obtained at the W. M. Keck Observatory, which is operated as a scientific partnership among the California Institute of Technology, the University of California and the National Aeronautics and Space Administration. The Observatory was made possible by the generous financial support of the W. M. Keck Foundation. We wish to recognize and acknowledge the very significant cultural role and reverence that the summit of Maunakea has always had within the indigenous Hawaiian community.  We are very thankful to have the opportunity to conduct observations from this sacred mountain.

\facilities{VLA, Keck}
\software{\texttt{astropy} \citep{Astropy-paper1-2013, Astropy-paper2-2018}, \texttt{numpy} \citep{Van-der-Walt11-numpy,Harris20-numpy}, \texttt{scipy} \citep{Jones01-scipy,Virtanen20-scipy}, \texttt{APLpy} \citep{Robitaille12-aplpy}, \texttt{PyBDSF} \citep{Mohan_Rafferty_2015_pybdsf}, \texttt{CASA} \citep{McMullin07-CASA}, \texttt{LPIPE} \citep{Perley19-Lpipe}, \texttt{emcee} \citep{Foreman-Mackey_2013_emcee}}

\appendix
\section{Free-free emission}
\label{app:free-free}
Here we show that VT 1137-0337's combination of high luminosity and human-timescale variability cannot be explained by free-free emission, as claimed in Section \ref{sec:flat-spectrum-emission-mechanisms}. Let us consider the implications of having a substantial fraction of the observed luminosity be due to free-free emission. For simplicity, we assume that the source of the free-free emission is a sphere of ionized gas with a radius $R$, temperature $T$, and density $n_e$. We denote the frequency where the optical depth to free-free absorption equals 1 as $\nu_{ff}$ and the thermal fraction (i.e. the fraction of the flux at $\nu_{ff}$ due to free-free emission) as $f_{th}$. The flux $S_{\nu_{ff}}$ due to free-free emission at frequency $\nu_{ff}$ can then be expressed in two ways:

\begin{equation}
\label{eqn:spectrum}
    S_{\nu_{ff}} \approx 2~\textrm{mJy}  \left(\frac{\nu_{ff}}{\textrm{GHz}}\right)^{-0.35} \left(\frac{f_{th}}{1.0}\right),
\end{equation}

\noindent from the observed spectrum discussed in Section \ref{sec:radio-spectral-index-fit} and

\begin{equation}
\label{eqn:blackbody}
    S_{\nu_{ff}} \approx \frac{2 \nu_{ff}^2 kT}{c^2} \pi \left(\frac{R}{120~\textrm{Mpc}}\right)^2.
\end{equation}
\noindent since at $\tau = 1$ the thermal free-free emission is close to being optically thick.
\\
\\From the variability timescale and observed spectrum, we have constraints on $R$ and $\nu_{ff}$. Taking the $\Delta t \sim 20$ year span between nondetection at 1.5~GHz in FIRST and and our first follow-up epoch to be an upper limit on the rise time of the transient, we have a maximum length scale for the region responsible for the excess emission of $R < c\Delta t = 6.1$pc. The nondetection in Epoch 1 of any spectral break between $\sim$1-12~GHz and the falling spectrum in that range (Figure \ref{fig:radio-SED}) implies that we are observing the optically thin part of the spectrum and thus that $\nu_{ff} < 1$~GHz. 
\\
\\Combining equations \ref{eqn:spectrum} and \ref{eqn:blackbody}, we have 

\begin{equation}
\label{eqn:temp}
    T = 8 \times 10^{6} \left(\frac{\nu_{ff}}{\textrm{GHz}}\right)^{-2.35} \left(\frac{R}{6.1~\textrm{pc}}\right)^{-2} \left(\frac{f_{th}}{1.0}\right) \textrm{K},
\end{equation}

\noindent which together with the constraints on $R$ and $\nu_{ff}$ indicates a minimum temperature of $\sim$8 $\times 10^6$ f$_{th}$~K. If $f_{th}$ is $\sim$1, the corresponding lower bound is extreme for a region emitting radio free-free emission. Moreover, it is likely a substantial underestimate when considering the range of reasonable radii. For instance, a constant velocity blast wave of 10,000 km/s would reach $R \sim 6 \times 10^{17}$ cm in 20 years, and would thus require a temperature $\gtrsim 8 \times 10^{9}$~K. 
\\
\\We can further highlight the problem of extremes by considering the required density and energy within the emitting region. The density can be constrained from the free-free opacity \citep[][]{Essential-Radio-Astronomy} which, for our constant density sphere of radius $R$, is approximately

\begin{equation}
\label{eqn:ff-opacity}
    \tau \approx 0.082 \left(\frac{T}{K}\right)^{-1.35} \left(\frac{\nu}{\textrm{GHz}}\right)^{-2.1} \left(\frac{n_e}{\textrm{cm}^{-3}}\right)^2 \left(\frac{R}{\textrm{pc}}\right).
\end{equation}

\noindent By definition, $\tau = 1$ at $\nu_{ff}$, so we can combine equations \ref{eqn:temp} and \ref{eqn:ff-opacity} to get:

\begin{equation}
\label{eqn:density}
    n_e = 6.5 \times 10^{4} \left(\frac{\nu_{ff}}{\textrm{GHz}}\right)^{-0.54} \left(\frac{R}{6.1~\textrm{pc}}\right)^{-1.85} \left(\frac{f_{th}}{1.0}\right)^{0.68} \textrm{cm}^{-3}.
\end{equation}

\noindent Again if $f_{th} >> 0$, this density is substantially higher than then $\sim$10$^2$ cm$^{-3}$ inferred for the surrounding star forming region (Section \ref{sec:Derived properties from optical features}), and would thus require an unusual environment or a significant pre-explosion deposition of mass from the source of the transient. It is also likely underestimated by a similar factor to the temperature, given the similar scaling with $R$. These issues come together when considering the thermal energy required for the radiating plasma:

\begin{equation}
\label{eqn:base-energy}
    U = \frac{3}{2} \left(\frac{4}{3} \pi R^3\right) n k T.
\end{equation}

\noindent Folding in equations \ref{eqn:temp} and \ref{eqn:density}, we get

\begin{equation}
\label{eqn:energy}
    U = 3 \times 10^{54} \left(\frac{\nu_{ff}}{\textrm{GHz}}\right)^{-2.9} \left(\frac{R}{6.1~\textrm{pc}}\right)^{-0.85} \left(\frac{f_{th}}{1.0}\right)^{1.68} \textrm{erg}.
\end{equation}

\noindent If there is any substantial thermal fraction, the implied lower limit on the energy is orders of magnitude beyond what can be deposited by e.g. a supernova, tidal disruption event, or black hole flare. For scale, the minimum energy would require a black hole of $\sim$20\% the stellar mass of the galaxy to be emitting at its Eddington luminosity for 20 years. Varying the assumptions (e.g. a nonspherical geometry, or a non constant temperature and density) does not help by much. Considering that the emission appeared within a dwarf galaxy on the timescale of $\sim$20 years, we conclude that free-free emission is not a reasonable explanation for VT 1137-0337.

\section{Literature transients and flat spectrum sources}
\label{app:literature_transients}
\noindent Here we give the luminosities, spectral indices, and references for the points plotted in Figure \ref{fig:lum_vs_alpha}. Due to the evolving nature of supernovae, gamma ray bursts and tidal disruption events, the luminosities for sources in those classes are approximate values, valid at \textit{some} point in the evolution of the transient. We have made an effort to choose luminosities at late times where the transient is optically thin at 3~GHz. Where possible, we use spectral indices that are directly fit from standard peaked SED models. For spectral indices derived from measurements in two bands, we use values from late epochs after the peak has dropped to frequencies below the two bands and the spectral index has converged to its optically thin value.
\\
\\\textbf{Fast radio burst persistent sources:}

\begin{itemize}
    \item \textbf{PRS/FRB 121102:} From a fit of the the 1 - 10 GHz observations in \citet[][]{Chatterjee-FRB121102-VLBI-loc}, we estimate a low frequency spectral index of $\alpha \sim 0.2$. There is evidence of a break at $\sim$10~GHz, with a spectral index of $\alpha \sim 1.2$ toward higher frequencies. The 3~GHz luminosity is $L_{3\textrm{GHz}}$ = 2.3 $\times$ 10$^{29}$ erg s$^{-1}$ Hz$^{-1}$ at the redshift reported in \citet[][]{Tendulkar17-FRB121102-host}. 
    
    \item \textbf{PRS/FRB 190520B:} The spectral index and luminosity come from the values reported in \citet[][]{Niu21-second-FRB-PRS}: $\alpha = $ 0.41 $\pm$ 0.04 and $L_{3\textrm{GHz}}$ = 3 $\times$ 10$^{29}$ erg s$^{-1}$ Hz$^{-1}$.
\end{itemize}

\noindent \textbf{Gamma ray bursts:}
\begin{itemize}
    \item \textbf{GRB 030329:} Our plotted spectral index $\alpha \sim 0.55$ comes from the report in \citet[][]{van-der-Horst07-GRB030329} that $p = 2.1$, though we note that Figure 1 of \citet[][]{Granot-van-der-Horst14-GRB-radio-review} suggests a value closer to $\alpha = 0.7$. This GRB has many observations over a span of years. In the figure, we plot a value of $\sim 2 \times 10^{30}$ erg s$^{-1}$ Hz$^{-1}$ in between the 4.8 and 2.3 GHz temporal peaks.

    \item \textbf{GW 170817:} The spectral index $\alpha \approx 0.6$ comes from the fit in \citet[][]{Mooley18-GW170817}, and the luminosity ($\sim 9 \times 10^{25}$ erg s$^{-1}$ Hz$^{-1}$) comes from their brightest reported 3~GHz data point.
    
    \item \textbf{GRB 980703:} We estimate a spectral index of $\alpha \sim 0.6$ from the late-time data points plotted in Fig 1 of \citet[][]{Granot-van-der-Horst14-GRB-radio-review}. The 5 GHz luminosity at the point where the spectral index plateaus is roughly 4 $\times$ 10$^{29}$~erg~s~Hz$^{-1}$.
    
\end{itemize}

\noindent \textbf{Radio supernovae:}

\begin{itemize}
    
    \item \textbf{SN 1978K:} The spectral index $\alpha = 0.8$ and luminosity $1 \times 10^{27}$ erg s$^{-1}$ Hz$^{-1}$ are from the 1992.5 epoch in Table 5 of \citet[][]{Ryder93-SN1978K}.
    
    \item \textbf{SN 1979C:} The spectral index $\alpha = 0.7$ and luminosity $2 \times 10^{27}$ erg s$^{-1}$ Hz$^{-1}$ are from Table 3 of \citet[][]{Weiler91-SN1979C}.
    
    \item \textbf{SN 1983N:} The plotted optically thin spectral index $\alpha = 1.1$ and luminosity $3 \times 10^{25}$~erg~s~Hz$^{-1}$ are from Table 8 of \citet[][]{Weiler86-radio-SN-review}.
    
    \item \textbf{SN 1984L:} The spectral index $\alpha = 1.2$ and luminosity $7 \times 10^{25}$ erg s$^{-1}$ Hz$^{-1}$ are from the Nov 30. 1984 epoch reported in \citet[][]{Panagia84-SN84L}.
    
    \item \textbf{SN 1986J:} The spectral index is from the report in \citet[][]{Weiler90-SN1986J} that $\alpha = 0.7 \pm 0.1$ from 6 to 2~cm and the luminosity $\sim 5 \times 10^{27}$ erg s$^{-1}$ Hz$^{-1}$ is from Table 5.
     
    \item \textbf{SN 1987A:} The plotted spectral index $\alpha = 0.9$ is from the report in \citet[][]{Manchester02-SN1987A} that the spectral index has changed from 0.97 to 0.88 and the luminosity $\sim 2 \times 10^{23}$~erg~s~Hz$^{-1}$ is estimated from their Figure 1.
    
    \item \textbf{SN 1988Z:} The spectral index $\alpha = 0.74 \pm 0.5$ and luminosity $\sim 2 \times 10^{28}$~erg~s~Hz$^{-1}$ are from Tables 1 and 2 of \citet[][]{vanDyk-Weiler-93-1988Z}.
    
    \item \textbf{SN 1990B:} The spectral index $\alpha = 0.9$ and luminosity $3 \times 10^{26}$ erg s$^{-1}$ Hz$^{-1}$ are from the Apr 26, 1990 epoch reported in \citet[][]{vanDyk93-SN1990B}. 
    
    \item \textbf{SN 1993J:} The spectral index $\alpha \sim -0.99$ is from Table 1 of \citet[][]{vanDyk-Weiler94-SN1993J} and the luminosity $\sim 8 \times 10^{26}$~erg~s~Hz$^{-1}$ is estimated from their Figure 1.
    
    \item \textbf{SN 1998bw:} The plotted spectral index $\alpha = $ 0.7 and luminosity $4 \times 10^{27}$~erg~s~Hz$^{-1}$ come from observations presented in \citet[][]{Li-Chevalier-98bw} at a late epoch where the radio SED is optically thin. 
    
    \item \textbf{SN 2009bb:} The spectral index and luminosity come from \citet[][]{Soderberg-2009bb}, who report $\alpha \sim 1$ and $S_{3\textrm{GHz}} \sim 15$ mJy on day 52, corresponding to $L_{3\textrm{GHz}} \sim$ 3 $\times$ 10$^{28}$ erg s$^{-1}$ Hz$^{-1}$.
    
    \item \textbf{PTF 11qcj:} The spectral index ($\alpha \sim 1.1$) and luminosity ($L_{3\textrm{GHz}} \sim 2 \times 10^{29}$ erg s$^{-1}$ Hz$^{-1}$) come from the most recent broadband radio spectrum (at day 1887) reported in \citet[][]{Palliyaguru-11qcj}.
    
    \item \textbf{VT 1210+4956:} The spectral index $\alpha = 1.04 \pm 0.02$ and 3~GHz luminosity 1.3 $\times$ 10$^{29}$~erg~s~Hz$^{-1}$ come from the May 30, 2018 followup epoch in \citet[][]{Dong21-VT1210}. 
\end{itemize}

\noindent \textbf{Tidal disruption events:}

\begin{itemize}
    \item \textbf{Sw J1644+57:} We use the report in \citet[][]{Eftekhari-SwiftJ1644} that on day 651, the 3.4~GHz flux was 8.78~mJy, corresponding to $L_{\nu} = 4 \times 10^{31}$~erg~s~Hz$^{-1}$. The optically thin slope at this epoch is consistent with $p = 2.5$, corresponding to $\alpha = 0.75$.

    \item \textbf{IGR 12580+0134:} We use the combined jet model from \citet[][]{Perlman17-TDE}, which predicts $p = 2.7$, corresponding to $\alpha = 0.85$. Extrapolating from their 5GHz flux using this spectral index gives a 3~GHz flux of 46~mJy in the 2015 Oct 8 epoch, corresponding to 1.6 $\times$ 10$^{28}$ erg~s~Hz$^{-1}$.
    
    \item \textbf{AT 2019dsg:} The spectral model from \citet[][]{Stein21-neutrino-TDE} predicts $p = 2.9 \pm 0.1$, corresponding to $\alpha = 0.95 \pm 0.05$, consistent with results in \citet[][]{Cendes21-TDE-AT2019dsg}. On day 178, the 3.24~GHz flux was 0.92 $\pm$ 0.08~mJy, corresponding to a luminosity of 5.8 $\times$ 10$^{28}$~erg~s~Hz$^{-1}$.
    
    \item \textbf{Arp 299-B AT1:} We use the spectral model from \citet[][]{Mattila18-jetted-TDE}. Their modeling (Table S8) indicates that p = 3.0 for the jet, corresponding to $\alpha = 1$. From their Table S4, we adopt 3mJy as a representative flux at 3GHz, corresponding to $7 \times 10^{27}$~erg~s~Hz$^{-1}$.
    
    \item \textbf{CNSS J0019+00:} Spectral fitting in \citet[][]{Anderson-CNSS_TDE} indicates that during the 2015 Oct 15 epoch, the 2.9~GHz flux was 7.25~mJy, corresponding to $5.2 \times 10^{28}$~erg~s~Hz$^{-1}$ and the optically thin spectral index was $\alpha = 1.1$.
    
    \item \textbf{ASASSN 14li:} From modeling in \citet[][]{Alexander16-ASASSN-14li}, the transient component in the 2015 June epoch was $\sim 0.9$~mJy, corresponding to $9 \times 10^{27}$~erg~s~Hz$^{-1}$ and the spectral index was $\alpha \sim 1$.
    
    \item \textbf{XMMSL1 J0740-85:} From \citet[][]{Alexander17-TDE-XMMSL1}, the 2.7~GHz emission was 0.87~mJy at peak, corresponding to $6 \times 10^{27}$~erg~s~Hz$^{-1}$. They find a best fit spectral index of $\alpha = 0.7 \pm 0.1$.
    
    \item \textbf{VT J1548+2208:} \citet[][]{Somalwar21-TDE-AGN} observe a 3~GHz flux of $\sim 3$~mJy, corresponding to $\sim 7 \times 10^{28}$~erg~s~Hz$^{-1}$. We use the spectral index of the high frequency component: $\alpha = 1.1$. 
    
    \item \textbf{FIRST J1533+2727:} This TDE was discovered at a late time where the spectrum is optically thin at all observed frequencies. From Table 1 in \citet[][]{Ravi21-reverse-TDE}, we calculate a spectral index of $\alpha = 0.84$. At this late time, the 3~GHz luminosity is $\sim 5 \times 10^{27}$~erg~s~Hz$^{-1}$.

\end{itemize}

\noindent \textbf{Pulsar wind nebulae:}
\noindent \\ The spectral indices and luminosities of plotted PWNe are from the \citet[][]{Green19-SNR-catalog} pulsar wind nebula and supernova remnant catalog. We selected every source classified as a PWN with a measured distance, flux, and spectral index. They are: the Crab Nebula, 3C58, Vela X, G284.3-1.8, G292.2-0.5, G292.0+1.8, G308.8-0.1, G320.4-1.2, G341.2+0.9, G343.1-2.3, G5.4-1.2, G11.2-0.3, G21.5-0.9, G29.7-0.3, G34.7-0.4, G54.1+0.3, and G69.0+2.7.

\bibliography{main}

\begin{thebibliography}{}
\expandafter\ifx\csname natexlab\endcsname\relax\def\natexlab#1{#1}\fi
\providecommand{\url}[1]{\href{#1}{#1}}
\providecommand{\dodoi}[1]{doi:~\href{http://doi.org/#1}{\nolinkurl{#1}}}
\providecommand{\doeprint}[1]{\href{http://ascl.net/#1}{\nolinkurl{http://ascl.net/#1}}}
\providecommand{\doarXiv}[1]{\href{https://arxiv.org/abs/#1}{\nolinkurl{https://arxiv.org/abs/#1}}}

\bibitem[{{Aguado} {et~al.}(2019){Aguado}, {Ahumada}, {Almeida}, {Anderson},
  {Andrews}, {Anguiano}, {Aquino Ort{\'\i}z}, {Arag{\'o}n-Salamanca},
  {Argudo-Fern{\'a}ndez}, {Aubert}, {Avila-Reese}, {Badenes}, {Barboza
  Rembold}, {Barger}, {Barrera-Ballesteros}, {Bates}, {Bautista}, {Beaton},
  {Beers}, {Belfiore}, {Bernardi}, {Bershady}, {Beutler}, {Bird}, {Bizyaev},
  {Blanc}, {Blanton}, {Blomqvist}, {Bolton}, {Boquien}, {Borissova}, {Bovy},
  {Brand t}, {Brinkmann}, {Brownstein}, {Bundy}, {Burgasser}, {Byler}, {Cano
  Diaz}, {Cappellari}, {Carrera}, {Cervantes Sodi}, {Chen}, {Cherinka}, {Choi},
  {Chung}, {Coffey}, {Comerford}, {Comparat}, {Covey}, {da Silva Ilha}, {da
  Costa}, {Dai}, {Damke}, {Darling}, {Davies}, {Dawson}, {de Sainte Agathe},
  {Deconto Machado}, {Del Moro}, {De Lee}, {Diamond-Stanic}, {Dom{\'\i}nguez
  S{\'a}nchez}, {Donor}, {Drory}, {du Mas des Bourboux}, {Duckworth}, {Dwelly},
  {Ebelke}, {Emsellem}, {Escoffier}, {Fern{\'a}ndez-Trincado}, {Feuillet},
  {Fischer}, {Fleming}, {Fraser-McKelvie}, {Freischlad}, {Frinchaboy}, {Fu},
  {Galbany}, {Garcia-Dias}, {Garc{\'\i}a-Hern{\'a}ndez}, {Garma Oehmichen},
  {Geimba Maia}, {Gil-Mar{\'\i}n}, {Grabowski}, {Gu}, {Guo}, {Ha},
  {Harrington}, {Hasselquist}, {Hayes}, {Hearty}, {Hernandez Toledo}, {Hicks},
  {Hogg}, {Holley-Bockelmann}, {Holtzman}, {Hsieh}, {Hunt}, {Hwang},
  {Ibarra-Medel}, {Jimenez Angel}, {Johnson}, {Jones}, {J{\"o}nsson},
  {Kinemuchi}, {Kollmeier}, {Krawczyk}, {Kreckel}, {Kruk}, {Lacerna}, {Lan},
  {Lane}, {Law}, {Lee}, {Li}, {Lian}, {Lin}, {Lin}, {Lintott}, {Long},
  {Longa-Pe{\~n}a}, {Mackereth}, {de la Macorra}, {Majewski}, {Malanushenko},
  {Manchado}, {Maraston}, {Mariappan}, {Marinelli}, {Marques-Chaves},
  {Masseron}, {Masters}, {McDermid}, {Medina Pe{\~n}a}, {Meneses-Goytia},
  {Merloni}, {Merrifield}, {Meszaros}, {Minniti}, {Minsley}, {Muna}, {Myers},
  {Nair}, {Correa do Nascimento}, {Newman}, {Nitschelm}, {Olmstead}, {Oravetz},
  {Oravetz}, {Ortega Minakata}, {Pace}, {Padilla}, {Palicio}, {Pan}, {Pan},
  {Parikh}, {Parker}, {Peirani}, {Penny}, {Percival}, {Perez-Fournon},
  {Peterken}, {Pinsonneault}, {Prakash}, {Raddick}, {Raichoor}, {Riffel},
  {Riffel}, {Rix}, {Robin}, {Roman-Lopes}, {Rose}, {Ross}, {Rossi}, {Rowlands},
  {Rubin}, {S{\'a}nchez}, {S{\'a}nchez-Gallego}, {Sayres}, {Schaefer},
  {Schiavon}, {Schimoia}, {Schlafly}, {Schlegel}, {Schneider}, {Schultheis},
  {Seo}, {Shamsi}, {Shao}, {Shen}, {Shetty}, {Simonian}, {Smethurst}, {Sobeck},
  {Souter}, {Spindler}, {Stark}, {Stassun}, {Steinmetz}, {Storchi-Bergmann},
  {Stringfellow}, {Su{\'a}rez}, {Sun}, {Taghizadeh-Popp}, {Talbot}, {Tayar},
  {Thakar}, {Thomas}, {Tissera}, {Tojeiro}, {Troup}, {Unda-Sanzana},
  {Valenzuela}, {Vargas-Maga{\~n}a}, {V{\'a}zquez-Mata}, {Wake}, {Weaver},
  {Weijmans}, {Westfall}, {Wild}, {Wilson}, {Woods}, {Yan}, {Yang}, {Zamora},
  {Zasowski}, {Zhang}, {Zheng}, {Zheng}, {Zhu}, {Zinn}, \& {Zou}}]{SDSS_DR15}
{Aguado}, D.~S., {Ahumada}, R., {Almeida}, A., {et~al.} 2019, \apjs, 240, 23,
  \dodoi{10.3847/1538-4365/aaf651}

\bibitem[{{Alexander} {et~al.}(2016){Alexander}, {Berger}, {Guillochon},
  {Zauderer}, \& {Williams}}]{Alexander16-ASASSN-14li}
{Alexander}, K.~D., {Berger}, E., {Guillochon}, J., {Zauderer}, B.~A., \&
  {Williams}, P.~K.~G. 2016, \apjl, 819, L25,
  \dodoi{10.3847/2041-8205/819/2/L25}

\bibitem[{{Alexander} {et~al.}(2017){Alexander}, {Wieringa}, {Berger},
  {Saxton}, \& {Komossa}}]{Alexander17-TDE-XMMSL1}
{Alexander}, K.~D., {Wieringa}, M.~H., {Berger}, E., {Saxton}, R.~D., \&
  {Komossa}, S. 2017, \apj, 837, 153, \dodoi{10.3847/1538-4357/aa6192}

\bibitem[{{Aller} \& {Reynolds}(1985)}]{Aller-Reynolds-1985-Crab-fade-rate}
{Aller}, H.~D., \& {Reynolds}, S.~P. 1985, \apjl, 293, L73,
  \dodoi{10.1086/184494}

\bibitem[{{Anderson} {et~al.}(2020){Anderson}, {Mooley}, {Hallinan}, {Dong},
  {Phinney}, {Horesh}, {Bourke}, {Cenko}, {Frail}, {Kulkarni}, \&
  {Myers}}]{Anderson-CNSS_TDE}
{Anderson}, M.~M., {Mooley}, K.~P., {Hallinan}, G., {et~al.} 2020, \apj, 903,
  116, \dodoi{10.3847/1538-4357/abb94b}

\bibitem[{{Asada} \& {Nakamura}(2012)}]{Asada2012-M87-conical}
{Asada}, K., \& {Nakamura}, M. 2012, \apjl, 745, L28,
  \dodoi{10.1088/2041-8205/745/2/L28}

\bibitem[{{Asplund} {et~al.}(2009){Asplund}, {Grevesse}, {Sauval}, \&
  {Scott}}]{Asplund-Zsun-review}
{Asplund}, M., {Grevesse}, N., {Sauval}, A.~J., \& {Scott}, P. 2009, \araa, 47,
  481, \dodoi{10.1146/annurev.astro.46.060407.145222}

\bibitem[{{Astropy Collaboration} {et~al.}(2013){Astropy Collaboration},
  {Robitaille}, {Tollerud}, {Greenfield}, {Droettboom}, {Bray}, {Aldcroft},
  {Davis}, {Ginsburg}, {Price-Whelan}, {Kerzendorf}, {Conley}, {Crighton},
  {Barbary}, {Muna}, {Ferguson}, {Grollier}, {Parikh}, {Nair}, {Unther},
  {Deil}, {Woillez}, {Conseil}, {Kramer}, {Turner}, {Singer}, {Fox}, {Weaver},
  {Zabalza}, {Edwards}, {Azalee Bostroem}, {Burke}, {Casey}, {Crawford},
  {Dencheva}, {Ely}, {Jenness}, {Labrie}, {Lim}, {Pierfederici}, {Pontzen},
  {Ptak}, {Refsdal}, {Servillat}, \& {Streicher}}]{Astropy-paper1-2013}
{Astropy Collaboration}, {Robitaille}, T.~P., {Tollerud}, E.~J., {et~al.} 2013,
  \aap, 558, A33, \dodoi{10.1051/0004-6361/201322068}

\bibitem[{{Astropy Collaboration} {et~al.}(2018){Astropy Collaboration},
  {Price-Whelan}, {Sip{\H{o}}cz}, {G{\"u}nther}, {Lim}, {Crawford}, {Conseil},
  {Shupe}, {Craig}, {Dencheva}, {Ginsburg}, {VanderPlas}, {Bradley},
  {P{\'e}rez-Su{\'a}rez}, {de Val-Borro}, {Aldcroft}, {Cruz}, {Robitaille},
  {Tollerud}, {Ardelean}, {Babej}, {Bach}, {Bachetti}, {Bakanov}, {Bamford},
  {Barentsen}, {Barmby}, {Baumbach}, {Berry}, {Biscani}, {Boquien}, {Bostroem},
  {Bouma}, {Brammer}, {Bray}, {Breytenbach}, {Buddelmeijer}, {Burke},
  {Calderone}, {Cano Rodr{\'\i}guez}, {Cara}, {Cardoso}, {Cheedella}, {Copin},
  {Corrales}, {Crichton}, {D'Avella}, {Deil}, {Depagne}, {Dietrich}, {Donath},
  {Droettboom}, {Earl}, {Erben}, {Fabbro}, {Ferreira}, {Finethy}, {Fox},
  {Garrison}, {Gibbons}, {Goldstein}, {Gommers}, {Greco}, {Greenfield},
  {Groener}, {Grollier}, {Hagen}, {Hirst}, {Homeier}, {Horton}, {Hosseinzadeh},
  {Hu}, {Hunkeler}, {Ivezi{\'c}}, {Jain}, {Jenness}, {Kanarek}, {Kendrew},
  {Kern}, {Kerzendorf}, {Khvalko}, {King}, {Kirkby}, {Kulkarni}, {Kumar},
  {Lee}, {Lenz}, {Littlefair}, {Ma}, {Macleod}, {Mastropietro}, {McCully},
  {Montagnac}, {Morris}, {Mueller}, {Mumford}, {Muna}, {Murphy}, {Nelson},
  {Nguyen}, {Ninan}, {N{\"o}the}, {Ogaz}, {Oh}, {Parejko}, {Parley}, {Pascual},
  {Patil}, {Patil}, {Plunkett}, {Prochaska}, {Rastogi}, {Reddy Janga},
  {Sabater}, {Sakurikar}, {Seifert}, {Sherbert}, {Sherwood-Taylor}, {Shih},
  {Sick}, {Silbiger}, {Singanamalla}, {Singer}, {Sladen}, {Sooley},
  {Sornarajah}, {Streicher}, {Teuben}, {Thomas}, {Tremblay}, {Turner},
  {Terr{\'o}n}, {van Kerkwijk}, {de la Vega}, {Watkins}, {Weaver}, {Whitmore},
  {Woillez}, {Zabalza}, \& {Astropy Contributors}}]{Astropy-paper2-2018}
{Astropy Collaboration}, {Price-Whelan}, A.~M., {Sip{\H{o}}cz}, B.~M., {et~al.}
  2018, \aj, 156, 123, \dodoi{10.3847/1538-3881/aabc4f}

\bibitem[{{Atoyan}(1999)}]{Atoyan99-flat-spectrum-Crab}
{Atoyan}, A.~M. 1999, \aap, 346, L49.
\newblock \doarXiv{astro-ph/9905204}

\bibitem[{{Atoyan} \&
  {Aharonian}(1996)}]{Atoyan-Aharonian-Crab-nebula-spectrum}
{Atoyan}, A.~M., \& {Aharonian}, F.~A. 1996, \mnras, 278, 525,
  \dodoi{10.1093/mnras/278.2.525}

\bibitem[{{Baars} {et~al.}(1977){Baars}, {Genzel}, {Pauliny-Toth}, \&
  {Witzel}}]{Baars1977-flux-density-scale}
{Baars}, J.~W.~M., {Genzel}, R., {Pauliny-Toth}, I.~I.~K., \& {Witzel}, A.
  1977, \aap, 500, 135

\bibitem[{{Baldwin} {et~al.}(1981){Baldwin}, {Phillips}, \&
  {Terlevich}}]{BPT_diagram-Baldwin-Phillips-Terlevich-1981}
{Baldwin}, J.~A., {Phillips}, M.~M., \& {Terlevich}, R. 1981, \pasp, 93, 5,
  \dodoi{10.1086/130766}

\bibitem[{{Barbarino} {et~al.}(2020){Barbarino}, {Sollerman}, {Taddia},
  {Fremling}, {Karamehmetoglu}, {Arcavi}, {Gal-Yam}, {Laher}, {Schulze},
  {Wozniak}, \& {Yan}}]{Barbarino20-Ic-ejecta-mass}
{Barbarino}, C., {Sollerman}, J., {Taddia}, F., {et~al.} 2020, arXiv e-prints,
  arXiv:2010.08392.
\newblock \doarXiv{2010.08392}

\bibitem[{{Bassa} {et~al.}(2017){Bassa}, {Tendulkar}, {Adams}, {Maddox},
  {Bogdanov}, {Bower}, {Burke-Spolaor}, {Butler}, {Chatterjee}, {Cordes},
  {Hessels}, {Kaspi}, {Law}, {Marcote}, {Paragi}, {Ransom}, {Scholz},
  {Spitler}, \& {van Langevelde}}]{Bassa17-FRB121102-host}
{Bassa}, C.~G., {Tendulkar}, S.~P., {Adams}, E.~A.~K., {et~al.} 2017, \apjl,
  843, L8, \dodoi{10.3847/2041-8213/aa7a0c}

\bibitem[{{Becker} {et~al.}(1995){Becker}, {White}, \& {Helfand}}]{Becker95}
{Becker}, R.~H., {White}, R.~L., \& {Helfand}, D.~J. 1995, \apj, 450, 559,
  \dodoi{10.1086/176166}

\bibitem[{{Bellovary} {et~al.}(2019){Bellovary}, {Cleary}, {Munshi}, {Tremmel},
  {Christensen}, {Brooks}, \& {Quinn}}]{Bellovary19-MBH-multimessenger}
{Bellovary}, J.~M., {Cleary}, C.~E., {Munshi}, F., {et~al.} 2019, \mnras, 482,
  2913, \dodoi{10.1093/mnras/sty2842}

\bibitem[{{Beloborodov}(2017)}]{Beloborodov17-magnetar-121102}
{Beloborodov}, A.~M. 2017, \apjl, 843, L26, \dodoi{10.3847/2041-8213/aa78f3}

\bibitem[{{Beloborodov} \& {Li}(2016)}]{Beloborodov-Li16-magnetar-heating}
{Beloborodov}, A.~M., \& {Li}, X. 2016, \apj, 833, 261,
  \dodoi{10.3847/1538-4357/833/2/261}

\bibitem[{{Beniamini} {et~al.}(2019){Beniamini}, {Hotokezaka}, {van der Horst},
  \& {Kouveliotou}}]{Beniamini19-magnetar-birth-rate}
{Beniamini}, P., {Hotokezaka}, K., {van der Horst}, A., \& {Kouveliotou}, C.
  2019, \mnras, 487, 1426, \dodoi{10.1093/mnras/stz1391}

\bibitem[{{Bennett} {et~al.}(2014){Bennett}, {Larson}, {Weiland}, \&
  {Hinshaw}}]{Bennett14-concordence-cosmology}
{Bennett}, C.~L., {Larson}, D., {Weiland}, J.~L., \& {Hinshaw}, G. 2014, \apj,
  794, 135, \dodoi{10.1088/0004-637X/794/2/135}

\bibitem[{{Bera} \& {Chengalur}(2019)}]{Bera19-Crab-giant-pulses}
{Bera}, A., \& {Chengalur}, J.~N. 2019, \mnras, 490, L12,
  \dodoi{10.1093/mnrasl/slz140}

\bibitem[{{Bietenholz} {et~al.}(2021){Bietenholz}, {Bartel}, {Argo}, {Dua},
  {Ryder}, \& {Soderberg}}]{Bietenholz21-SN-luminosity-dist}
{Bietenholz}, M.~F., {Bartel}, N., {Argo}, M., {et~al.} 2021, \apj, 908, 75,
  \dodoi{10.3847/1538-4357/abccd9}

\bibitem[{{Bietenholz} {et~al.}(2001){Bietenholz}, {Frail}, \&
  {Hester}}]{Bietenholz2001-Crab-radio-wisps}
{Bietenholz}, M.~F., {Frail}, D.~A., \& {Hester}, J.~J. 2001, \apj, 560, 254,
  \dodoi{10.1086/322244}

\bibitem[{{Blandford} \& {Eichler}(1987)}]{Blandford-Eichler-shocks}
{Blandford}, R., \& {Eichler}, D. 1987, \physrep, 154, 1,
  \dodoi{10.1016/0370-1573(87)90134-7}

\bibitem[{{Blandford} \& {K{\"o}nigl}(1979)}]{Blandford-Konigl-AGN-jets}
{Blandford}, R.~D., \& {K{\"o}nigl}, A. 1979, \apj, 232, 34,
  \dodoi{10.1086/157262}

\bibitem[{{Blandford} \& {Znajek}(1977)}]{Blandford-Znajek1977}
{Blandford}, R.~D., \& {Znajek}, R.~L. 1977, \mnras, 179, 433,
  \dodoi{10.1093/mnras/179.3.433}

\bibitem[{{Bochenek} {et~al.}(2020){Bochenek}, {Ravi}, {Belov}, {Hallinan},
  {Kocz}, {Kulkarni}, \& {McKenna}}]{Bochenek20-magnetar-FRB}
{Bochenek}, C.~D., {Ravi}, V., {Belov}, K.~V., {et~al.} 2020, \nat, 587, 59,
  \dodoi{10.1038/s41586-020-2872-x}

\bibitem[{{Bradley} {et~al.}(2006){Bradley}, {Knapen}, {Beckman}, \&
  {Folkes}}]{Bradley2006_HII-region_lum_function}
{Bradley}, T.~R., {Knapen}, J.~H., {Beckman}, J.~E., \& {Folkes}, S.~L. 2006,
  \aap, 459, L13, \dodoi{10.1051/0004-6361:20066151}

\bibitem[{{Burbidge} \& {Burbidge}(1957)}]{Burbidge-Burbidge-1957}
{Burbidge}, G.~R., \& {Burbidge}, E.~M. 1957, \apj, 125, 1,
  \dodoi{10.1086/146279}

\bibitem[{{Calzetti} {et~al.}(2000){Calzetti}, {Armus}, {Bohlin}, {Kinney},
  {Koornneef}, \& {Storchi-Bergmann}}]{Calzetti-2000-extinction}
{Calzetti}, D., {Armus}, L., {Bohlin}, R.~C., {et~al.} 2000, \apj, 533, 682,
  \dodoi{10.1086/308692}

\bibitem[{{Cann} {et~al.}(2019){Cann}, {Satyapal}, {Abel}, {Blecha},
  {Mushotzky}, {Reynolds}, \& {Secrest}}]{Cann19-BPT-limitations-dwarf}
{Cann}, J.~M., {Satyapal}, S., {Abel}, N.~P., {et~al.} 2019, \apjl, 870, L2,
  \dodoi{10.3847/2041-8213/aaf88d}

\bibitem[{{Cavagnolo} {et~al.}(2010){Cavagnolo}, {McNamara}, {Nulsen},
  {Carilli}, {Jones}, \& {B{\^\i}rzan}}]{Cavagnolo10-AGN-jet-power-radio-power}
{Cavagnolo}, K.~W., {McNamara}, B.~R., {Nulsen}, P.~E.~J., {et~al.} 2010, \apj,
  720, 1066, \dodoi{10.1088/0004-637X/720/2/1066}

\bibitem[{{Cendes} {et~al.}(2021){Cendes}, {Alexander}, {Berger}, {Eftekhari},
  {Williams}, \& {Chornock}}]{Cendes21-TDE-AT2019dsg}
{Cendes}, Y., {Alexander}, K.~D., {Berger}, E., {et~al.} 2021, \apj, 919, 127,
  \dodoi{10.3847/1538-4357/ac110a}

\bibitem[{{Chatterjee} {et~al.}(2017){Chatterjee}, {Law}, {Wharton},
  {Burke-Spolaor}, {Hessels}, {Bower}, {Cordes}, {Tendulkar}, {Bassa},
  {Demorest}, {Butler}, {Seymour}, {Scholz}, {Abruzzo}, {Bogdanov}, {Kaspi},
  {Keimpema}, {Lazio}, {Marcote}, {McLaughlin}, {Paragi}, {Ransom}, {Rupen},
  {Spitler}, \& {van Langevelde}}]{Chatterjee-FRB121102-VLBI-loc}
{Chatterjee}, S., {Law}, C.~J., {Wharton}, R.~S., {et~al.} 2017, \nat, 541, 58,
  \dodoi{10.1038/nature20797}

\bibitem[{{Chen} {et~al.}(2022){Chen}, {Ravi}, \&
  {Hallinan}}]{Chen22-FRB121102-scintillation}
{Chen}, G., {Ravi}, V., \& {Hallinan}, G.~W. 2022, arXiv e-prints,
  arXiv:2201.00999.
\newblock \doarXiv{2201.00999}

\bibitem[{{Chevalier}(1998)}]{Chevalier98}
{Chevalier}, R.~A. 1998, \apj, 499, 810, \dodoi{10.1086/305676}

\bibitem[{{Chevalier}(2004)}]{Chevalier2004-PWN}
---. 2004, Advances in Space Research, 33, 456,
  \dodoi{10.1016/j.asr.2003.04.018}

\bibitem[{{Chevalier}(2012)}]{Chevalier2012-CE-SNe}
---. 2012, \apjl, 752, L2, \dodoi{10.1088/2041-8205/752/1/L2}

\bibitem[{{Chevalier} \& {Fransson}(1992)}]{Chevalier-Fransson-1992-PWN}
{Chevalier}, R.~A., \& {Fransson}, C. 1992, \apj, 395, 540,
  \dodoi{10.1086/171674}

\bibitem[{{CHIME/FRB Collaboration} {et~al.}(2020){CHIME/FRB Collaboration},
  {Andersen}, {Bandura}, {Bhardwaj}, {Bij}, {Boyce}, {Boyle}, {Brar},
  {Cassanelli}, {Chawla}, {Chen}, {Cliche}, {Cook}, {Cubranic}, {Curtin},
  {Denman}, {Dobbs}, {Dong}, {Fandino}, {Fonseca}, {Gaensler}, {Giri}, {Good},
  {Halpern}, {Hill}, {Hinshaw}, {H{\"o}fer}, {Josephy}, {Kania}, {Kaspi},
  {Landecker}, {Leung}, {Li}, {Lin}, {Masui}, {McKinven}, {Mena-Parra},
  {Merryfield}, {Meyers}, {Michilli}, {Milutinovic}, {Mirhosseini},
  {M{\"u}nchmeyer}, {Naidu}, {Newburgh}, {Ng}, {Patel}, {Pen},
  {Pinsonneault-Marotte}, {Pleunis}, {Quine}, {Rafiei-Ravandi}, {Rahman},
  {Ransom}, {Renard}, {Sanghavi}, {Scholz}, {Shaw}, {Shin}, {Siegel}, {Singh},
  {Smegal}, {Smith}, {Stairs}, {Tan}, {Tendulkar}, {Tretyakov}, {Vanderlinde},
  {Wang}, {Wulf}, \& {Zwaniga}}]{CHIME-galactic-magnetar}
{CHIME/FRB Collaboration}, {Andersen}, B.~C., {Bandura}, K.~M., {et~al.} 2020,
  \nat, 587, 54, \dodoi{10.1038/s41586-020-2863-y}

\bibitem[{{Comisso} {et~al.}(2020){Comisso}, {Sobacchi}, \&
  {Sironi}}]{Comisso20-PWN-turbulence}
{Comisso}, L., {Sobacchi}, E., \& {Sironi}, L. 2020, \apjl, 895, L40,
  \dodoi{10.3847/2041-8213/ab93dc}

\bibitem[{{Condon} {et~al.}(1998){Condon}, {Cotton}, {Greisen}, {Yin},
  {Perley}, {Taylor}, \& {Broderick}}]{Condon98-NVSS}
{Condon}, J.~J., {Cotton}, W.~D., {Greisen}, E.~W., {et~al.} 1998, \aj, 115,
  1693, \dodoi{10.1086/300337}

\bibitem[{{Condon} \& {Ransom}(2016)}]{Essential-Radio-Astronomy}
{Condon}, J.~J., \& {Ransom}, S.~M. 2016, {Essential Radio Astronomy}
  ({Princeton University Press})

\bibitem[{{Cook} {et~al.}(2019){Cook}, {Kasliwal}, {Van Sistine}, {Kaplan},
  {Sutter}, {Kupfer}, {Shupe}, {Laher}, {Masci}, {Dale}, {Sesar}, {Brady},
  {Yan}, {Ofek}, {Reitze}, \& {Kulkarni}}]{Cook19-CLU}
{Cook}, D.~O., {Kasliwal}, M.~M., {Van Sistine}, A., {et~al.} 2019, \apj, 880,
  7, \dodoi{10.3847/1538-4357/ab2131}

\bibitem[{{Corbel} {et~al.}(2012){Corbel}, {Dubus}, {Tomsick}, {Szostek},
  {Corbet}, {Miller-Jones}, {Richards}, {Pooley}, {Trushkin}, {Dubois}, {Hill},
  {Kerr}, {Max-Moerbeck}, {Readhead}, {Bodaghee}, {Tudose}, {Parent}, {Wilms},
  \& {Pottschmidt}}]{Corbel-1e24-CygX3-flare}
{Corbel}, S., {Dubus}, G., {Tomsick}, J.~A., {et~al.} 2012, \mnras, 421, 2947,
  \dodoi{10.1111/j.1365-2966.2012.20517.x}

\bibitem[{{Cordes} \& {Lazio}(2002)}]{Cordes-Lazio02-ne2001}
{Cordes}, J.~M., \& {Lazio}, T.~J.~W. 2002, arXiv e-prints, astro.
\newblock \doarXiv{astro-ph/0207156}

\bibitem[{{Cordes} {et~al.}(2021){Cordes}, {Wasserman}, {Chatterjee}, \&
  {Batra}}]{Cordes21-dipole-vs-quadrapole-field}
{Cordes}, J.~M., {Wasserman}, I., {Chatterjee}, S., \& {Batra}, G. 2021, arXiv
  e-prints, arXiv:2107.12874.
\newblock \doarXiv{2107.12874}

\bibitem[{{Davis} {et~al.}(2020){Davis}, {Nguyen}, {Seth}, {Greene}, {Nyland},
  {Barth}, {Bureau}, {Cappellari}, {den Brok}, {Iguchi}, {Lelli}, {Liu},
  {Neumayer}, {North}, {Onishi}, {Sarzi}, {Smith}, \&
  {Williams}}]{Davis20-NGC404-IMBH}
{Davis}, T.~A., {Nguyen}, D.~D., {Seth}, A.~C., {et~al.} 2020, \mnras, 496,
  4061, \dodoi{10.1093/mnras/staa1567}

\bibitem[{{Dom{\'\i}nguez} {et~al.}(2013){Dom{\'\i}nguez}, {Siana}, {Henry},
  {Scarlata}, {Bedregal}, {Malkan}, {Atek}, {Ross}, {Colbert}, {Teplitz},
  {Rafelski}, {McCarthy}, {Bunker}, {Hathi}, {Dressler}, {Martin}, \&
  {Masters}}]{Dominguez-2013-extinction}
{Dom{\'\i}nguez}, A., {Siana}, B., {Henry}, A.~L., {et~al.} 2013, \apj, 763,
  145, \dodoi{10.1088/0004-637X/763/2/145}

\bibitem[{{Dong} {et~al.}(2021){Dong}, {Hallinan}, {Nakar}, {Ho}, {Hughes},
  {Hotokezaka}, {Myers}, {De}, {Mooley}, {Ravi}, {Horesh}, {Kasliwal}, \&
  {Kulkarni}}]{Dong21-VT1210}
{Dong}, D.~Z., {Hallinan}, G., {Nakar}, E., {et~al.} 2021, Science, 373, 1125,
  \dodoi{10.1126/science.abg6037}

\bibitem[{{Dopita} {et~al.}(2000){Dopita}, {Kewley}, {Heisler}, \&
  {Sutherland}}]{Dopita-2000-HII-region-ionization}
{Dopita}, M.~A., {Kewley}, L.~J., {Heisler}, C.~A., \& {Sutherland}, R.~S.
  2000, \apj, 542, 224, \dodoi{10.1086/309538}

\bibitem[{{Draine}(2011)}]{Draine-physics-of-ISM}
{Draine}, B.~T. 2011, {Physics of the Interstellar and Intergalactic Medium}
  ({Princeton University Press})

\bibitem[{{Eftekhari} {et~al.}(2018){Eftekhari}, {Berger}, {Zauderer},
  {Margutti}, \& {Alexander}}]{Eftekhari-SwiftJ1644}
{Eftekhari}, T., {Berger}, E., {Zauderer}, B.~A., {Margutti}, R., \&
  {Alexander}, K.~D. 2018, \apj, 854, 86, \dodoi{10.3847/1538-4357/aaa8e0}

\bibitem[{{Egron} {et~al.}(2021){Egron}, {Pellizzoni}, {Righini}, {Giroletti},
  {Koljonen}, {Pottschmidt}, {Trushkin}, {Lobina}, {Pilia}, {Wilms}, {Corbel},
  {Grinberg}, {Loru}, {Trois}, {Rodriguez}, {L{\"a}hteenm{\"a}ki},
  {Tornikoski}, {Enestam}, \& {J{\"a}rvel{\"a}}}]{Egron21-CygX3-17Jy-flare}
{Egron}, E., {Pellizzoni}, A., {Righini}, S., {et~al.} 2021, \apj, 906, 10,
  \dodoi{10.3847/1538-4357/abc5b1}

\bibitem[{{Fender} {et~al.}(2004){Fender}, {Belloni}, \&
  {Gallo}}]{Fender2004-unified-XRB-model}
{Fender}, R.~P., {Belloni}, T.~M., \& {Gallo}, E. 2004, \mnras, 355, 1105,
  \dodoi{10.1111/j.1365-2966.2004.08384.x}

\bibitem[{{Fender} {et~al.}(2000){Fender}, {Pooley}, {Durouchoux}, {Tilanus},
  \& {Brocksopp}}]{Fender2000-CygX1-flat-spectrum}
{Fender}, R.~P., {Pooley}, G.~G., {Durouchoux}, P., {Tilanus}, R.~P.~J., \&
  {Brocksopp}, C. 2000, \mnras, 312, 853,
  \dodoi{10.1046/j.1365-8711.2000.03219.x}

\bibitem[{{Fleishman}(2006)}]{Fleishman06-diffusive-sync-rad-GRBs}
{Fleishman}, G.~D. 2006, \apj, 638, 348, \dodoi{10.1086/498732}

\bibitem[{{Fleishman} \&
  {Bietenholz}(2007)}]{Fleishman-Bietenholz07-diffusive-sync-rad}
{Fleishman}, G.~D., \& {Bietenholz}, M.~F. 2007, \mnras, 376, 625,
  \dodoi{10.1111/j.1365-2966.2007.11450.x}

\bibitem[{{Foreman-Mackey} {et~al.}(2013){Foreman-Mackey}, {Hogg}, {Lang}, \&
  {Goodman}}]{Foreman-Mackey_2013_emcee}
{Foreman-Mackey}, D., {Hogg}, D.~W., {Lang}, D., \& {Goodman}, J. 2013, \pasp,
  125, 306, \dodoi{10.1086/670067}

\bibitem[{{Frail} \&
  {Scharringhausen}(1997)}]{Frail-Scharringhausen-1997-PWN-radio-survey}
{Frail}, D.~A., \& {Scharringhausen}, B.~R. 1997, \apj, 480, 364,
  \dodoi{10.1086/303943}

\bibitem[{{Gaensler} \& {Slane}(2006)}]{Gaensler06-PWN-review}
{Gaensler}, B.~M., \& {Slane}, P.~O. 2006, \araa, 44, 17,
  \dodoi{10.1146/annurev.astro.44.051905.092528}

\bibitem[{{Gaensler} {et~al.}(2000){Gaensler}, {Stappers}, {Frail}, {Moffett},
  {Johnston}, \& {Chatterjee}}]{Gaensler2000-eta_R_PWNe}
{Gaensler}, B.~M., {Stappers}, B.~W., {Frail}, D.~A., {et~al.} 2000, \mnras,
  318, 58, \dodoi{10.1046/j.1365-8711.2000.03626.x}

\bibitem[{{Gal-Yam}(2019)}]{Gal-Yam-SLSNe_review}
{Gal-Yam}, A. 2019, \araa, 57, 305, \dodoi{10.1146/annurev-astro-081817-051819}

\bibitem[{{Gavriil} {et~al.}(2008){Gavriil}, {Gonzalez}, {Gotthelf}, {Kaspi},
  {Livingstone}, \& {Woods}}]{Gavriil08-Kes75}
{Gavriil}, F.~P., {Gonzalez}, M.~E., {Gotthelf}, E.~V., {et~al.} 2008, Science,
  319, 1802, \dodoi{10.1126/science.1153465}

\bibitem[{{Gehrels}(1986)}]{Gehrels-Poisson-noise}
{Gehrels}, N. 1986, \apj, 303, 336, \dodoi{10.1086/164079}

\bibitem[{{Goldreich} \& {Julian}(1969)}]{Goldreich-Julian-PWNe}
{Goldreich}, P., \& {Julian}, W.~H. 1969, \apj, 157, 869,
  \dodoi{10.1086/150119}

\bibitem[{{Graham} \& {Schady}(2016)}]{Graham-Schady-16-LGRB-rate}
{Graham}, J.~F., \& {Schady}, P. 2016, \apj, 823, 154,
  \dodoi{10.3847/0004-637X/823/2/154}

\bibitem[{{Granot} \& {van der
  Horst}(2014)}]{Granot-van-der-Horst14-GRB-radio-review}
{Granot}, J., \& {van der Horst}, A.~J. 2014, \pasa, 31, e008,
  \dodoi{10.1017/pasa.2013.44}

\bibitem[{{Green}(2019)}]{Green19-SNR-catalog}
{Green}, D.~A. 2019, Journal of Astrophysics and Astronomy, 40, 36,
  \dodoi{10.1007/s12036-019-9601-6}

\bibitem[{{Greene} {et~al.}(2020){Greene}, {Strader}, \&
  {Ho}}]{Green20-IMBH-review}
{Greene}, J.~E., {Strader}, J., \& {Ho}, L.~C. 2020, \araa, 58, 257,
  \dodoi{10.1146/annurev-astro-032620-021835}

\bibitem[{{Guillochon} {et~al.}(2017){Guillochon}, {Parrent}, {Kelley}, \&
  {Margutti}}]{Guillochon2017-OSC}
{Guillochon}, J., {Parrent}, J., {Kelley}, L.~Z., \& {Margutti}, R. 2017, \apj,
  835, 64, \dodoi{10.3847/1538-4357/835/1/64}

\bibitem[{{Hallinan} {et~al.}(2019){Hallinan}, {Ravi}, {Weinreb}, {Kocz},
  {Huang}, {Woody}, {Lamb}, {D'Addario}, {Catha}, {Law}, {Kulkarni}, {Phinney},
  {Eastwood}, {Bouman}, {McLaughlin}, {Ransom}, {Siemens}, {Cordes}, {Lynch},
  {Kaplan}, {Brazier}, {Bhatnagar}, {Myers}, {Walter}, \&
  {Gaensler}}]{Hallinan19-DSA2000-white-paper}
{Hallinan}, G., {Ravi}, V., {Weinreb}, S., {et~al.} 2019, in Bulletin of the
  American Astronomical Society, Vol.~51, 255.
\newblock \doarXiv{1907.07648}

\bibitem[{Harris {et~al.}(2020)Harris, Millman, van~der Walt, Gommers,
  Virtanen, Cournapeau, Wieser, Taylor, Berg, Smith, Kern, Picus, Hoyer, van
  Kerkwijk, Brett, Haldane, del R{\'{i}}o, Wiebe, Peterson,
  G{\'{e}}rard-Marchant, Sheppard, Reddy, Weckesser, Abbasi, Gohlke, \&
  Oliphant}]{Harris20-numpy}
Harris, C.~R., Millman, K.~J., van~der Walt, S.~J., {et~al.} 2020, Nature, 585,
  357, \dodoi{10.1038/s41586-020-2649-2}

\bibitem[{{Hicken} {et~al.}(2017){Hicken}, {Friedman}, {Blondin}, {Challis},
  {Berlind}, {Calkins}, {Esquerdo}, {Matheson}, {Modjaz}, {Rest}, \&
  {Kirshner}}]{Hicken17-CCSNe-lightcurves}
{Hicken}, M., {Friedman}, A.~S., {Blondin}, S., {et~al.} 2017, \apjs, 233, 6,
  \dodoi{10.3847/1538-4365/aa8ef4}

\bibitem[{{Hjellming} \&
  {Johnston}(1988)}]{Hjellming-Johnston1988-Conical-jets-XRBs}
{Hjellming}, R.~M., \& {Johnston}, K.~J. 1988, \apj, 328, 600,
  \dodoi{10.1086/166318}

\bibitem[{{Ho} {et~al.}(2019){Ho}, {Phinney}, {Ravi}, {Kulkarni}, {Petitpas},
  {Emonts}, {Bhalerao}, {Blundell}, {Cenko}, {Dobie}, {Howie}, {Kamraj},
  {Kasliwal}, {Murphy}, {Perley}, {Sridharan}, \& {Yoon}}]{Ho19-2018cow}
{Ho}, A. Y.~Q., {Phinney}, E.~S., {Ravi}, V., {et~al.} 2019, \apj, 871, 73,
  \dodoi{10.3847/1538-4357/aaf473}

\bibitem[{{Ho} {et~al.}(2020){Ho}, {Perley}, {Kulkarni}, {Dong}, {De},
  {Chandra}, {Andreoni}, {Bellm}, {Burdge}, {Coughlin}, {Dekany}, {Feeney},
  {Frederiks}, {Fremling}, {Golkhou}, {Graham}, {Hale}, {Helou}, {Horesh},
  {Kasliwal}, {Laher}, {Masci}, {Miller}, {Porter}, {Ridnaia}, {Rusholme},
  {Shupe}, {Soumagnac}, \& {Svinkin}}]{Ho2020-Koala}
{Ho}, A. Y.~Q., {Perley}, D.~A., {Kulkarni}, S.~R., {et~al.} 2020, \apj, 895,
  49, \dodoi{10.3847/1538-4357/ab8bcf}

\bibitem[{{Igoshev} {et~al.}(2022){Igoshev}, {Frantsuzova}, {Gourgouliatos},
  {Tsichli}, {Konstantinou}, \& {Popov}}]{Igoshev22-pulsar-initial-B-and-P}
{Igoshev}, A.~P., {Frantsuzova}, A., {Gourgouliatos}, K.~N., {et~al.} 2022,
  arXiv e-prints, arXiv:2205.06823.
\newblock \doarXiv{2205.06823}

\bibitem[{{Izotov} {et~al.}(2006){Izotov}, {Stasi{\'n}ska}, {Meynet}, {Guseva},
  \& {Thuan}}]{Izotov2006-metallicity-calibration}
{Izotov}, Y.~I., {Stasi{\'n}ska}, G., {Meynet}, G., {Guseva}, N.~G., \&
  {Thuan}, T.~X. 2006, \aap, 448, 955, \dodoi{10.1051/0004-6361:20053763}

\bibitem[{Jones {et~al.}(2001)Jones, Oliphant, Peterson,
  {et~al.}}]{Jones01-scipy}
Jones, E., Oliphant, T., Peterson, P., {et~al.} 2001, {SciPy}: Open source
  scientific tools for {Python}.
\newblock \url{http://www.scipy.org/}

\bibitem[{{Jones} \& {Ellison}(1991)}]{Jones-Ellison-1991-shock-theory}
{Jones}, F.~C., \& {Ellison}, D.~C. 1991, \ssr, 58, 259,
  \dodoi{10.1007/BF01206003}

\bibitem[{{Josephy} {et~al.}(2019){Josephy}, {Chawla}, {Fonseca}, {Ng},
  {Patel}, {Pleunis}, {Scholz}, {Andersen}, {Bandura}, {Bhardwaj}, {Boyce},
  {Boyle}, {Brar}, {Cubranic}, {Dobbs}, {Gaensler}, {Gill}, {Giri}, {Good},
  {Halpern}, {Hinshaw}, {Kaspi}, {Landecker}, {Lang}, {Lin}, {Masui},
  {Mckinven}, {Mena-Parra}, {Merryfield}, {Michilli}, {Milutinovic}, {Naidu},
  {Pen}, {Rafiei-Ravandi}, {Rahman}, {Ransom}, {Renard}, {Siegel}, {Smith},
  {Stairs}, {Tendulkar}, {Vanderlinde}, {Yadav}, \&
  {Zwaniga}}]{Josephy19-CHIME-FRB121102}
{Josephy}, A., {Chawla}, P., {Fonseca}, E., {et~al.} 2019, \apjl, 882, L18,
  \dodoi{10.3847/2041-8213/ab2c00}

\bibitem[{{Kagan} {et~al.}(2018){Kagan}, {Nakar}, \&
  {Piran}}]{Kagan-Nakar-Piran-2018-relativistic-reconnection}
{Kagan}, D., {Nakar}, E., \& {Piran}, T. 2018, \mnras, 476, 3902,
  \dodoi{10.1093/mnras/sty452}

\bibitem[{{Kasen} \& {Bildsten}(2010)}]{Kasen10-SLSNE-magnetars}
{Kasen}, D., \& {Bildsten}, L. 2010, \apj, 717, 245,
  \dodoi{10.1088/0004-637X/717/1/245}

\bibitem[{{Kaspi} \&
  {Beloborodov}(2017)}]{Kaspi-Beloborodov-magnetar-review-2017}
{Kaspi}, V.~M., \& {Beloborodov}, A.~M. 2017, \araa, 55, 261,
  \dodoi{10.1146/annurev-astro-081915-023329}

\bibitem[{{Kaspi} {et~al.}(2006){Kaspi}, {Roberts}, \&
  {Harding}}]{Kaspi-isolated-neutron-stars}
{Kaspi}, V.~M., {Roberts}, M. S.~E., \& {Harding}, A.~K. 2006, {Isolated
  neutron stars}, Vol.~39, 279--339

\bibitem[{{Kellermann} \& {Pauliny-Toth}(1981)}]{Kellerman1981-AGN-review}
{Kellermann}, K.~I., \& {Pauliny-Toth}, I.~I.~K. 1981, \araa, 19, 373,
  \dodoi{10.1146/annurev.aa.19.090181.002105}

\bibitem[{{Kennicutt} \& {Evans}(2012)}]{Kennicutt-Evans-2012}
{Kennicutt}, R.~C., \& {Evans}, N.~J. 2012, \araa, 50, 531,
  \dodoi{10.1146/annurev-astro-081811-125610}

\bibitem[{{Kewley} {et~al.}(2013){Kewley}, {Dopita}, {Leitherer}, {Dav{\'e}},
  {Yuan}, {Allen}, {Groves}, \& {Sutherland}}]{Kewley2013-ionization}
{Kewley}, L.~J., {Dopita}, M.~A., {Leitherer}, C., {et~al.} 2013, \apj, 774,
  100, \dodoi{10.1088/0004-637X/774/2/100}

\bibitem[{{Kewley} {et~al.}(2001){Kewley}, {Dopita}, {Sutherland}, {Heisler},
  \& {Trevena}}]{BPT_diagram-Kewley2001}
{Kewley}, L.~J., {Dopita}, M.~A., {Sutherland}, R.~S., {Heisler}, C.~A., \&
  {Trevena}, J. 2001, \apj, 556, 121, \dodoi{10.1086/321545}

\bibitem[{{Kewley} \&
  {Ellison}(2008)}]{Kewley-Ellison-2008-metallicity-calibrations}
{Kewley}, L.~J., \& {Ellison}, S.~L. 2008, \apj, 681, 1183,
  \dodoi{10.1086/587500}

\bibitem[{{Kewley} {et~al.}(2019){Kewley}, {Nicholls}, {Sutherland}, {Rigby},
  {Acharya}, {Dopita}, \& {Bayliss}}]{Kewley2019-line-ratio-diagnostics}
{Kewley}, L.~J., {Nicholls}, D.~C., {Sutherland}, R., {et~al.} 2019, \apj, 880,
  16, \dodoi{10.3847/1538-4357/ab16ed}

\bibitem[{{Kirsten} {et~al.}(2022){Kirsten}, {Marcote}, {Nimmo}, {Hessels},
  {Bhardwaj}, {Tendulkar}, {Keimpema}, {Yang}, {Snelders}, {Scholz},
  {Pearlman}, {Law}, {Peters}, {Giroletti}, {Paragi}, {Bassa}, {Hewitt},
  {Bach}, {Bezrukovs}, {Burgay}, {Buttaccio}, {Conway}, {Corongiu}, {Feiler},
  {Forss{\'e}n}, {Gawro{\'n}ski}, {Karuppusamy}, {Kharinov}, {Lindqvist},
  {Maccaferri}, {Melnikov}, {Ould-Boukattine}, {Possenti}, {Surcis}, {Wang},
  {Yuan}, {Aggarwal}, {Anna-Thomas}, {Bower}, {Blaauw}, {Burke-Spolaor},
  {Cassanelli}, {Clarke}, {Fonseca}, {Gaensler}, {Gopinath}, {Kaspi}, {Kassim},
  {Lazio}, {Leung}, {Li}, {Lin}, {Masui}, {Mckinven}, {Michilli}, {Mikhailov},
  {Ng}, {Orbidans}, {Pen}, {Petroff}, {Rahman}, {Ransom}, {Shin}, {Smith},
  {Stairs}, \& {Vlemmings}}]{Kirsten22-FRB-M81}
{Kirsten}, F., {Marcote}, B., {Nimmo}, K., {et~al.} 2022, \nat, 602, 585,
  \dodoi{10.1038/s41586-021-04354-w}

\bibitem[{{Kochanek} {et~al.}(2017){Kochanek}, {Shappee}, {Stanek}, {Holoien},
  {Thompson}, {Prieto}, {Dong}, {Shields}, {Will}, {Britt}, {Perzanowski}, \&
  {Pojma{\'n}ski}}]{Kochanek-ASASSN-lightcurve}
{Kochanek}, C.~S., {Shappee}, B.~J., {Stanek}, K.~Z., {et~al.} 2017, \pasp,
  129, 104502, \dodoi{10.1088/1538-3873/aa80d9}

\bibitem[{{Kolmogorov}(1941)}]{Kolmogorov1941-turbulence}
{Kolmogorov}, A. 1941, Akademiia Nauk SSSR Doklady, 30, 301

\bibitem[{{Kormendy} \& {Kennicutt}(2004)}]{Kormendy-Kennicutt-2004}
{Kormendy}, J., \& {Kennicutt}, Robert~C., J. 2004, \araa, 42, 603,
  \dodoi{10.1146/annurev.astro.42.053102.134024}

\bibitem[{{Kou} \& {Tong}(2015)}]{Kou-Tong15-Crab-initial-period}
{Kou}, F.~F., \& {Tong}, H. 2015, \mnras, 450, 1990,
  \dodoi{10.1093/mnras/stv734}

\bibitem[{{Kulkarni} {et~al.}(2018){Kulkarni}, {Perley}, \&
  {Miller}}]{Kulkarni18-galaxy-catalog-completeness}
{Kulkarni}, S.~R., {Perley}, D.~A., \& {Miller}, A.~A. 2018, \apj, 860, 22,
  \dodoi{10.3847/1538-4357/aabf85}

\bibitem[{{Lacy} {et~al.}(2020){Lacy}, {Baum}, {Chandler}, {Chatterjee},
  {Clarke}, {Deustua}, {English}, {Farnes}, {Gaensler}, {Gugliucci},
  {Hallinan}, {Kent}, {Kimball}, {Law}, {Lazio}, {Marvil}, {Mao}, {Medlin},
  {Mooley}, {Murphy}, {Myers}, {Osten}, {Richards}, {Rosolowsky}, {Rudnick},
  {Schinzel}, {Sivakoff}, {Sjouwerman}, {Taylor}, {White}, {Wrobel},
  {Andernach}, {Beasley}, {Berger}, {Bhatnager}, {Birkinshaw}, {Bower},
  {Brandt}, {Brown}, {Burke-Spolaor}, {Butler}, {Comerford}, {Demorest}, {Fu},
  {Giacintucci}, {Golap}, {G{\"u}th}, {Hales}, {Hiriart}, {Hodge}, {Horesh},
  {Ivezi{\'c}}, {Jarvis}, {Kamble}, {Kassim}, {Liu}, {Loinard}, {Lyons},
  {Masters}, {Mezcua}, {Moellenbrock}, {Mroczkowski}, {Nyland}, {O'Dea},
  {O'Sullivan}, {Peters}, {Radford}, {Rao}, {Robnett}, {Salcido}, {Shen},
  {Sobotka}, {Witz}, {Vaccari}, {van Weeren}, {Vargas}, {Williams}, \&
  {Yoon}}]{Lacy20-VLASS}
{Lacy}, M., {Baum}, S.~A., {Chandler}, C.~J., {et~al.} 2020, \pasp, 132,
  035001, \dodoi{10.1088/1538-3873/ab63eb}

\bibitem[{{Law} {et~al.}(2022){Law}, {Connor}, \&
  {Aggarwal}}]{Law22-FRB-PRS-population}
{Law}, C.~J., {Connor}, L., \& {Aggarwal}, K. 2022, \apj, 927, 55,
  \dodoi{10.3847/1538-4357/ac4c42}

\bibitem[{{Law} {et~al.}(2018){Law}, {Gaensler}, {Metzger}, {Ofek}, \&
  {Sironi}}]{Law18-FIRSTJ1419}
{Law}, C.~J., {Gaensler}, B.~M., {Metzger}, B.~D., {Ofek}, E.~O., \& {Sironi},
  L. 2018, \apjl, 866, L22, \dodoi{10.3847/2041-8213/aae5f3}

\bibitem[{{Leitherer} {et~al.}(1999){Leitherer}, {Schaerer}, {Goldader},
  {Delgado}, {Robert}, {Kune}, {de Mello}, {Devost}, \&
  {Heckman}}]{Leitherer-1999-Starburst99}
{Leitherer}, C., {Schaerer}, D., {Goldader}, J.~D., {et~al.} 1999, \apjs, 123,
  3, \dodoi{10.1086/313233}

\bibitem[{{Li} \& {Chevalier}(1999)}]{Li-Chevalier-98bw}
{Li}, Z.-Y., \& {Chevalier}, R.~A. 1999, \apj, 526, 716, \dodoi{10.1086/308031}

\bibitem[{{Lien} {et~al.}(2016){Lien}, {Sakamoto}, {Barthelmy}, {Baumgartner},
  {Cannizzo}, {Chen}, {Collins}, {Cummings}, {Gehrels}, {Krimm}, {Markwardt},
  {Palmer}, {Stamatikos}, {Troja}, \& {Ukwatta}}]{Lien2016-SWIFT-GRB-catalog}
{Lien}, A., {Sakamoto}, T., {Barthelmy}, S.~D., {et~al.} 2016, \apj, 829, 7,
  \dodoi{10.3847/0004-637X/829/1/7}

\bibitem[{{Linden} {et~al.}(2020){Linden}, {Murphy}, {Dong}, {Momjian},
  {Kennicutt}, {Meier}, {Schinnerer}, \& {Turner}}]{Linden2020-SFRS}
{Linden}, S.~T., {Murphy}, E.~J., {Dong}, D., {et~al.} 2020, \apjs, 248, 25,
  \dodoi{10.3847/1538-4365/ab8a4d}

\bibitem[{{Liodakis} {et~al.}(2017){Liodakis}, {Pavlidou}, {Hovatta},
  {Max-Moerbeck}, {Pearson}, {Richards}, \&
  {Readhead}}]{Liodakis17-OVRO-blazar-update-bimodal}
{Liodakis}, I., {Pavlidou}, V., {Hovatta}, T., {et~al.} 2017, \mnras, 467,
  4565, \dodoi{10.1093/mnras/stx432}

\bibitem[{{Lister}(2001)}]{Lister01-beaming}
{Lister}, M.~L. 2001, \apj, 561, 676, \dodoi{10.1086/323528}

\bibitem[{{Lyne} {et~al.}(1993){Lyne}, {Pritchard}, \& {Graham
  Smith}}]{Lyne1993-Crab-braking-index}
{Lyne}, A.~G., {Pritchard}, R.~S., \& {Graham Smith}, F. 1993, \mnras, 265,
  1003, \dodoi{10.1093/mnras/265.4.1003}

\bibitem[{{Lyutikov} {et~al.}(2019){Lyutikov}, {Temim}, {Komissarov}, {Slane},
  {Sironi}, \& {Comisso}}]{Lyutikov-2019-Crab-synchrotron-spectrum-theory}
{Lyutikov}, M., {Temim}, T., {Komissarov}, S., {et~al.} 2019, \mnras, 489,
  2403, \dodoi{10.1093/mnras/stz2023}

\bibitem[{{Mac{\'\i}as-P{\'e}rez} {et~al.}(2010){Mac{\'\i}as-P{\'e}rez},
  {Mayet}, {Aumont}, \&
  {D{\'e}sert}}]{Marcias-Perez2010-Crab-spectral-index-and-fading}
{Mac{\'\i}as-P{\'e}rez}, J.~F., {Mayet}, F., {Aumont}, J., \& {D{\'e}sert},
  F.~X. 2010, \apj, 711, 417, \dodoi{10.1088/0004-637X/711/1/417}

\bibitem[{{Manchester} {et~al.}(2002){Manchester}, {Gaensler}, {Wheaton},
  {Staveley-Smith}, {Tzioumis}, {Bizunok}, {Kesteven}, \&
  {Reynolds}}]{Manchester02-SN1987A}
{Manchester}, R.~N., {Gaensler}, B.~M., {Wheaton}, V.~C., {et~al.} 2002, \pasa,
  19, 207, \dodoi{10.1071/AS01042}

\bibitem[{{Manchester} {et~al.}(2005){Manchester}, {Hobbs}, {Teoh}, \&
  {Hobbs}}]{Manchester05-psrcat}
{Manchester}, R.~N., {Hobbs}, G.~B., {Teoh}, A., \& {Hobbs}, M. 2005, \aj, 129,
  1993, \dodoi{10.1086/428488}

\bibitem[{{Margalit} \& {Metzger}(2018)}]{Margalit18-FRB121102-magnetar}
{Margalit}, B., \& {Metzger}, B.~D. 2018, \apjl, 868, L4,
  \dodoi{10.3847/2041-8213/aaedad}

\bibitem[{{Martinez} {et~al.}(2022){Martinez}, {Bersten}, {Anderson}, {Hamuy},
  {Gonz{\'a}lez-Gait{\'a}n}, {F{\"o}rster}, {Orellana}, {Stritzinger},
  {Phillips}, {Guti{\'e}rrez}, {Burns}, {Contreras}, {de Jaeger}, {Ertini},
  {Folatelli}, {Galbany}, {Hoeflich}, {Hsiao}, {Morrell}, {Pessi}, \&
  {Suntzeff}}]{Martinez22-typeII-SN-energy-ejecta_mass}
{Martinez}, L., {Bersten}, M.~C., {Anderson}, J.~P., {et~al.} 2022, \aap, 660,
  A41, \dodoi{10.1051/0004-6361/202142076}

\bibitem[{{Mattila} {et~al.}(2018){Mattila}, {P{\'e}rez-Torres}, {Efstathiou},
  {Mimica}, {Fraser}, {Kankare}, {Alberdi}, {Aloy}, {Heikkil{\"a}}, {Jonker},
  {Lundqvist}, {Mart{\'\i}-Vidal}, {Meikle}, {Romero-Ca{\~n}izales}, {Smartt},
  {Tsygankov}, {Varenius}, {Alonso-Herrero}, {Bondi}, {Fransson},
  {Herrero-Illana}, {Kangas}, {Kotak}, {Ram{\'\i}rez-Olivencia},
  {V{\"a}is{\"a}nen}, {Beswick}, {Clements}, {Greimel}, {Harmanen},
  {Kotilainen}, {Nandra}, {Reynolds}, {Ryder}, {Walton}, {Wiik}, \&
  {{\"O}stlin}}]{Mattila18-jetted-TDE}
{Mattila}, S., {P{\'e}rez-Torres}, M., {Efstathiou}, A., {et~al.} 2018,
  Science, 361, 482, \dodoi{10.1126/science.aao4669}

\bibitem[{{McMullin} {et~al.}(2007){McMullin}, {Waters}, {Schiebel}, {Young},
  \& {Golap}}]{McMullin07-CASA}
{McMullin}, J.~P., {Waters}, B., {Schiebel}, D., {Young}, W., \& {Golap}, K.
  2007, Astronomical Society of the Pacific Conference Series, Vol. 376, {CASA
  Architecture and Applications}, ed. R.~A. {Shaw}, F.~{Hill}, \& D.~J. {Bell},
  127

\bibitem[{{Meier}(2001)}]{Meier2001-RIAFs}
{Meier}, D.~L. 2001, \apjl, 548, L9, \dodoi{10.1086/318921}

\bibitem[{{Metzger} {et~al.}(2017){Metzger}, {Berger}, \&
  {Margalit}}]{Metzger17-FRB-SLSNe-LGRB}
{Metzger}, B.~D., {Berger}, E., \& {Margalit}, B. 2017, \apj, 841, 14,
  \dodoi{10.3847/1538-4357/aa633d}

\bibitem[{{Metzger} {et~al.}(2015){Metzger}, {Williams}, \&
  {Berger}}]{Metzger15-VLASS-predictions}
{Metzger}, B.~D., {Williams}, P.~K.~G., \& {Berger}, E. 2015, \apj, 806, 224,
  \dodoi{10.1088/0004-637X/806/2/224}

\bibitem[{{Mohan} \& {Rafferty}(2015)}]{Mohan_Rafferty_2015_pybdsf}
{Mohan}, N., \& {Rafferty}, D. 2015, {PyBDSF: Python Blob Detection and Source
  Finder}, Astrophysics Source Code Library.
\newblock \doeprint{1502.007}

\bibitem[{{Molina} {et~al.}(2021{\natexlab{a}}){Molina}, {Reines}, {Greene},
  {Darling}, \& {Condon}}]{Molina21a-single-object-AGN}
{Molina}, M., {Reines}, A.~E., {Greene}, J.~E., {Darling}, J., \& {Condon},
  J.~J. 2021{\natexlab{a}}, \apj, 910, 5, \dodoi{10.3847/1538-4357/abe120}

\bibitem[{{Molina} {et~al.}(2021{\natexlab{b}}){Molina}, {Reines}, {Latimer},
  {Baldassare}, \& {Salehirad}}]{Molina21b-AGN-sample}
{Molina}, M., {Reines}, A.~E., {Latimer}, C.~J., {Baldassare}, V., \&
  {Salehirad}, S. 2021{\natexlab{b}}, \apj, 922, 155,
  \dodoi{10.3847/1538-4357/ac1ffa}

\bibitem[{{Mooley} {et~al.}(2019){Mooley}, {Myers}, {Frail}, {Hallinan},
  {Butler}, {Kimball}, \& {Golap}}]{Mooley19-CNSS2-OTF-methodology}
{Mooley}, K.~P., {Myers}, S.~T., {Frail}, D.~A., {et~al.} 2019, \apj, 870, 25,
  \dodoi{10.3847/1538-4357/aaef7c}

\bibitem[{{Mooley} {et~al.}(2018){Mooley}, {Nakar}, {Hotokezaka}, {Hallinan},
  {Corsi}, {Frail}, {Horesh}, {Murphy}, {Lenc}, {Kaplan}, {de}, {Dobie},
  {Chandra}, {Deller}, {Gottlieb}, {Kasliwal}, {Kulkarni}, {Myers}, {Nissanke},
  {Piran}, {Lynch}, {Bhalerao}, {Bourke}, {Bannister}, \&
  {Singer}}]{Mooley18-GW170817}
{Mooley}, K.~P., {Nakar}, E., {Hotokezaka}, K., {et~al.} 2018, \nat, 554, 207,
  \dodoi{10.1038/nature25452}

\bibitem[{{Moustakas} {et~al.}(2010){Moustakas}, {Kennicutt}, {Tremonti},
  {Dale}, {Smith}, \& {Calzetti}}]{Moustakas-SINGS-optical-spectroscopy}
{Moustakas}, J., {Kennicutt}, Robert~C., J., {Tremonti}, C.~A., {et~al.} 2010,
  \apjs, 190, 233, \dodoi{10.1088/0067-0049/190/2/233}

\bibitem[{{Murase} {et~al.}(2016){Murase}, {Kashiyama}, \&
  {M{\'e}sz{\'a}ros}}]{Murase16-FRB-magnetar}
{Murase}, K., {Kashiyama}, K., \& {M{\'e}sz{\'a}ros}, P. 2016, \mnras, 461,
  1498, \dodoi{10.1093/mnras/stw1328}

\bibitem[{{Murphy} {et~al.}(2011){Murphy}, {Condon}, {Schinnerer}, {Kennicutt},
  {Calzetti}, {Armus}, {Helou}, {Turner}, {Aniano}, {Beir{\~a}o}, {Bolatto},
  {Brandl}, {Croxall}, {Dale}, {Donovan Meyer}, {Draine}, {Engelbracht},
  {Hunt}, {Hao}, {Koda}, {Roussel}, {Skibba}, \&
  {Smith}}]{Murphy2011_SF_calibration}
{Murphy}, E.~J., {Condon}, J.~J., {Schinnerer}, E., {et~al.} 2011, \apj, 737,
  67, \dodoi{10.1088/0004-637X/737/2/67}

\bibitem[{{Murphy} {et~al.}(2021){Murphy}, {Kaplan}, {Stewart}, {O'Brien},
  {Lenc}, {Pintaldi}, {Pritchard}, {Dobie}, {Fox}, {Leung}, {An}, {Bell},
  {Broderick}, {Chatterjee}, {Dai}, {d'Antonio}, {Doyle}, {Gaensler}, {Heald},
  {Horesh}, {Jones}, {McConnell}, {Moss}, {Raja}, {Ramsay}, {Ryder}, {Sadler},
  {Sivakoff}, {Wang}, {Wang}, {Wheatland}, {Whiting}, {Allison}, {Anderson},
  {Ball}, {Bannister}, {Bock}, {Bolton}, {Bunton}, {Chekkala}, {Chippendale},
  {Cooray}, {Gupta}, {Hayman}, {Jeganathan}, {Koribalski}, {Lee-Waddell},
  {Mahony}, {Marvil}, {McClure-Griffiths}, {Mirtschin}, {Ng}, {Pearce},
  {Phillips}, \& {Voronkov}}]{Murphy21-VAST-pilot}
{Murphy}, T., {Kaplan}, D.~L., {Stewart}, A.~J., {et~al.} 2021, \pasa, 38,
  e054, \dodoi{10.1017/pasa.2021.44}

\bibitem[{{Nicholl} {et~al.}(2015){Nicholl}, {Smartt}, {Jerkstrand}, {Inserra},
  {Sim}, {Chen}, {Benetti}, {Fraser}, {Gal-Yam}, {Kankare}, {Maguire}, {Smith},
  {Sullivan}, {Valenti}, {Young}, {Baltay}, {Bauer}, {Baumont}, {Bersier},
  {Botticella}, {Childress}, {Dennefeld}, {Della Valle}, {Elias-Rosa},
  {Feindt}, {Galbany}, {Hadjiyska}, {Le Guillou}, {Leloudas}, {Mazzali},
  {McKinnon}, {Polshaw}, {Rabinowitz}, {Rostami}, {Scalzo}, {Schmidt},
  {Schulze}, {Sollerman}, {Taddia}, \& {Yuan}}]{Nicholl15-SLSNe-masses}
{Nicholl}, M., {Smartt}, S.~J., {Jerkstrand}, A., {et~al.} 2015, \mnras, 452,
  3869, \dodoi{10.1093/mnras/stv1522}

\bibitem[{{Niu} {et~al.}(2021){Niu}, {Aggarwal}, {Li}, {Zhang}, {Chatterjee},
  {Tsai}, {Yu}, {Law}, {Burke-Spolaor}, {Cordes}, {Zhang}, {Ocker}, {Yao},
  {Wang}, {Feng}, {Niino}, {Bochenek}, {Cruces}, {Connor}, {Jiang}, {Dai},
  {Luo}, {Li}, {Miao}, {Niu}, {Anna-Thomas}, {Sydnor}, {Stern}, {Wang}, {Yuan},
  {Yue}, {Zhou}, {Yan}, {Zhu}, \& {Zhang}}]{Niu21-second-FRB-PRS}
{Niu}, C.~H., {Aggarwal}, K., {Li}, D., {et~al.} 2021, arXiv e-prints,
  arXiv:2110.07418.
\newblock \doarXiv{2110.07418}

\bibitem[{{Nyland} {et~al.}(2017){Nyland}, {Davis}, {Nguyen}, {Seth}, {Wrobel},
  {Kamble}, {Lacy}, {Alatalo}, {Karovska}, {Maksym}, {Mukherjee}, \&
  {Young}}]{Nyland17-NGC404}
{Nyland}, K., {Davis}, T.~A., {Nguyen}, D.~D., {et~al.} 2017, \apj, 845, 50,
  \dodoi{10.3847/1538-4357/aa7ecf}

\bibitem[{{Nyland} {et~al.}(2020){Nyland}, {Dong}, {Patil}, {Lacy}, {van
  Velzen}, {Kimball}, {Sarbadhicary}, {Hallinan}, {Baldassare}, {Clarke},
  {Goulding}, {Greene}, {Hughes}, {Kassim}, {Kunert-Bajraszewska}, {Maccarone},
  {Mooley}, {Mukherjee}, {Peters}, {Petrov}, {Polisensky}, {Rujopakarn},
  {Whittle}, \& {Vaccari}}]{Nyland2020-VLASS-Quasars}
{Nyland}, K., {Dong}, D.~Z., {Patil}, P., {et~al.} 2020, \apj, 905, 74,
  \dodoi{10.3847/1538-4357/abc341}

\bibitem[{{Ofek}(2017)}]{Ofek17-FRB-PRS-search}
{Ofek}, E.~O. 2017, \apj, 846, 44, \dodoi{10.3847/1538-4357/aa8310}

\bibitem[{{Oke} {et~al.}(1995){Oke}, {Cohen}, {Carr}, {Cromer}, {Dingizian},
  {Harris}, {Labrecque}, {Lucinio}, {Schaal}, {Epps}, \&
  {Miller}}]{Oke1995-LRIS}
{Oke}, J.~B., {Cohen}, J.~G., {Carr}, M., {et~al.} 1995, \pasp, 107, 375,
  \dodoi{10.1086/133562}

\bibitem[{{Osterbrock} \& {Ferland}(2006)}]{Osterbrock-Ferland-AGN2}
{Osterbrock}, D.~E., \& {Ferland}, G.~J. 2006, {Astrophysics of gaseous nebulae
  and active galactic nuclei} ({University Science Books})

\bibitem[{{Palliyaguru} {et~al.}(2019){Palliyaguru}, {Corsi}, {Frail},
  {Vink{\'o}}, {Wheeler}, {Gal-Yam}, {Cenko}, {Kulkarni}, \&
  {Kasliwal}}]{Palliyaguru-11qcj}
{Palliyaguru}, N.~T., {Corsi}, A., {Frail}, D.~A., {et~al.} 2019, \apj, 872,
  201, \dodoi{10.3847/1538-4357/aaf64d}

\bibitem[{{Panagia} {et~al.}(1986){Panagia}, {Sramek}, \&
  {Weiler}}]{Panagia84-SN84L}
{Panagia}, N., {Sramek}, R.~A., \& {Weiler}, K.~W. 1986, \apjl, 300, L55,
  \dodoi{10.1086/184602}

\bibitem[{{Pelletier} {et~al.}(2017){Pelletier}, {Bykov}, {Ellison}, \&
  {Lemoine}}]{Pelletier-2017-relativistic-shocks}
{Pelletier}, G., {Bykov}, A., {Ellison}, D., \& {Lemoine}, M. 2017, \ssr, 207,
  319, \dodoi{10.1007/s11214-017-0364-6}

\bibitem[{{P{\'e}rez-Gonz{\'a}lez} {et~al.}(2003){P{\'e}rez-Gonz{\'a}lez},
  {Zamorano}, {Gallego}, {Arag{\'o}n-Salamanca}, \& {Gil de
  Paz}}]{Perez-Gonzalez-2003-BCDs}
{P{\'e}rez-Gonz{\'a}lez}, P.~G., {Zamorano}, J., {Gallego}, J.,
  {Arag{\'o}n-Salamanca}, A., \& {Gil de Paz}, A. 2003, \apj, 591, 827,
  \dodoi{10.1086/375364}

\bibitem[{{Perley}(2019)}]{Perley19-Lpipe}
{Perley}, D.~A. 2019, \pasp, 131, 084503, \dodoi{10.1088/1538-3873/ab215d}

\bibitem[{{Perley} \& {Butler}(2017)}]{Perley-Butler-2017}
{Perley}, R.~A., \& {Butler}, B.~J. 2017, \apjs, 230, 7,
  \dodoi{10.3847/1538-4365/aa6df9}

\bibitem[{{Perley} {et~al.}(2011){Perley}, {Chandler}, {Butler}, \&
  {Wrobel}}]{Perley11-EVLA}
{Perley}, R.~A., {Chandler}, C.~J., {Butler}, B.~J., \& {Wrobel}, J.~M. 2011,
  \apjl, 739, L1, \dodoi{10.1088/2041-8205/739/1/L1}

\bibitem[{{Perlman} {et~al.}(2017){Perlman}, {Meyer}, {Wang}, {Yuan},
  {Henriksen}, {Irwin}, {Krause}, {Wiegert}, {Murphy}, {Heald}, \&
  {Dettmar}}]{Perlman17-TDE}
{Perlman}, E.~S., {Meyer}, E.~T., {Wang}, Q.~D., {et~al.} 2017, \apj, 842, 126,
  \dodoi{10.3847/1538-4357/aa71b1}

\bibitem[{{Phinney}(1989)}]{Phinney1989-TDE-five-thirds}
{Phinney}, E.~S. 1989, in The Center of the Galaxy, ed. M.~{Morris}, Vol. 136,
  543

\bibitem[{{Pietka} {et~al.}(2015){Pietka}, {Fender}, \&
  {Keane}}]{Pietka15-variability-timescale-luminosity}
{Pietka}, M., {Fender}, R.~P., \& {Keane}, E.~F. 2015, \mnras, 446, 3687,
  \dodoi{10.1093/mnras/stu2335}

\bibitem[{{Plavin} {et~al.}(2019){Plavin}, {Kovalev}, {Pushkarev}, \&
  {Lobanov}}]{Plavin19-core-shift}
{Plavin}, A.~V., {Kovalev}, Y.~Y., {Pushkarev}, A.~B., \& {Lobanov}, A.~P.
  2019, \mnras, 485, 1822, \dodoi{10.1093/mnras/stz504}

\bibitem[{{Rau} {et~al.}(2005){Rau}, {Kienlin}, {Hurley}, \&
  {Lichti}}]{Rau05-INTEGRAL-GRB-catalog}
{Rau}, A., {Kienlin}, A.~V., {Hurley}, K., \& {Lichti}, G.~G. 2005, \aap, 438,
  1175, \dodoi{10.1051/0004-6361:20053159}

\bibitem[{{Ravi} {et~al.}(2021){Ravi}, {Dykaar}, {Codd}, {Zaccagnini}, {Dong},
  {Drout}, {Gaensler}, {Hallinan}, \& {Law}}]{Ravi21-reverse-TDE}
{Ravi}, V., {Dykaar}, H., {Codd}, J., {et~al.} 2021, arXiv e-prints,
  arXiv:2102.05795.
\newblock \doarXiv{2102.05795}

\bibitem[{{Raynaud} {et~al.}(2020){Raynaud}, {Guilet}, {Janka}, \&
  {Gastine}}]{Raynaud20-millisecond-magnetar-sim}
{Raynaud}, R., {Guilet}, J., {Janka}, H.-T., \& {Gastine}, T. 2020, Science
  Advances, 6, eaay2732, \dodoi{10.1126/sciadv.aay2732}

\bibitem[{{Reines} {et~al.}(2020){Reines}, {Condon}, {Darling}, \&
  {Greene}}]{Reines2020-Wandering-BHs}
{Reines}, A.~E., {Condon}, J.~J., {Darling}, J., \& {Greene}, J.~E. 2020, \apj,
  888, 36, \dodoi{10.3847/1538-4357/ab4999}

\bibitem[{{Reines} \&
  {Volonteri}(2015)}]{Reines-Volonteri-2015-BHmass-Mstar-relation}
{Reines}, A.~E., \& {Volonteri}, M. 2015, \apj, 813, 82,
  \dodoi{10.1088/0004-637X/813/2/82}

\bibitem[{{Resmi} {et~al.}(2021){Resmi}, {Vink}, \&
  {Ishwara-Chandra}}]{Resmi21-low-freq-121102}
{Resmi}, L., {Vink}, J., \& {Ishwara-Chandra}, C.~H. 2021, \aap, 655, A102,
  \dodoi{10.1051/0004-6361/202039771}

\bibitem[{{Reynolds} {et~al.}(2018){Reynolds}, {Borkowski}, \&
  {Gwynne}}]{Reynolds18-Kes75-expansion}
{Reynolds}, S.~P., {Borkowski}, K.~J., \& {Gwynne}, P.~H. 2018, \apj, 856, 133,
  \dodoi{10.3847/1538-4357/aab3d3}

\bibitem[{{Reynolds} \&
  {Chevalier}(1984)}]{Reynolds-Chevalier1984-PWN-lightcurve}
{Reynolds}, S.~P., \& {Chevalier}, R.~A. 1984, \apj, 278, 630,
  \dodoi{10.1086/161831}

\bibitem[{{Richardson} {et~al.}(2016){Richardson}, {Allen}, {Baldwin},
  {Hewett}, {Ferland}, {Crider}, \& {Meskhidze}}]{Richardson2016-ionization}
{Richardson}, C.~T., {Allen}, J.~T., {Baldwin}, J.~A., {et~al.} 2016, \mnras,
  458, 988, \dodoi{10.1093/mnras/stw100}

\bibitem[{{Robitaille} \& {Bressert}(2012)}]{Robitaille12-aplpy}
{Robitaille}, T., \& {Bressert}, E. 2012, {APLpy: Astronomical Plotting Library
  in Python}, Astrophysics Source Code Library, record ascl:1208.017.
\newblock \doeprint{1208.017}

\bibitem[{{Ryder} {et~al.}(1993){Ryder}, {Staveley-Smith}, {Dopita}, {Petre},
  {Colbert}, {Malin}, \& {Schlegel}}]{Ryder93-SN1978K}
{Ryder}, S., {Staveley-Smith}, L., {Dopita}, M., {et~al.} 1993, \apj, 416, 167,
  \dodoi{10.1086/173223}

\bibitem[{{Saikia} {et~al.}(2019){Saikia}, {Russell}, {Bramich},
  {Miller-Jones}, {Baglio}, \& {Degenaar}}]{Saikia19-gammas-for-XRBs}
{Saikia}, P., {Russell}, D.~M., {Bramich}, D.~M., {et~al.} 2019, \apj, 887, 21,
  \dodoi{10.3847/1538-4357/ab4a09}

\bibitem[{{Salim} {et~al.}(2018){Salim}, {Boquien}, \&
  {Lee}}]{Salim2018-SFR-Mass}
{Salim}, S., {Boquien}, M., \& {Lee}, J.~C. 2018, \apj, 859, 11,
  \dodoi{10.3847/1538-4357/aabf3c}

\bibitem[{{Salpeter}(1955)}]{Salpeter1955-IMF}
{Salpeter}, E.~E. 1955, \apj, 121, 161, \dodoi{10.1086/145971}

\bibitem[{{Schlickeiser} \&
  {Fuerst}(1989)}]{Schlickeiser-1989-shock-acceleration-plasma-beta}
{Schlickeiser}, R., \& {Fuerst}, E. 1989, \aap, 219, 192

\bibitem[{{Schwab}(1984)}]{Schwab84-Cotton-Schwab-clean}
{Schwab}, F.~R. 1984, \aj, 89, 1076, \dodoi{10.1086/113605}

\bibitem[{{Scott} \& {Readhead}(1977)}]{Scott-Readhead-1977-equipartition}
{Scott}, M.~A., \& {Readhead}, A.~C.~S. 1977, \mnras, 180, 539,
  \dodoi{10.1093/mnras/180.4.539}

\bibitem[{{Serino} {et~al.}(2014){Serino}, {Sakamoto}, {Kawai}, {Yoshida},
  {Ohno}, {Ogawa}, {Nishimura}, {Fukushima}, {Higa}, {Ishikawa}, {Ishikawa},
  {Kawamuro}, {Kimura}, {Matsuoka}, {Mihara}, {Morii}, {Nakagawa}, {Nakahira},
  {Nakajima}, {Nakano}, {Negoro}, {Onodera}, {Sasaki}, {Shidatsu}, {Sugimoto},
  {Sugizaki}, {Suwa}, {Suzuki}, {Tachibana}, {Takagi}, {Toizumi}, {Tomida},
  {Tsuboi}, {Tsunemi}, {Ueda}, {Ueno}, {Usui}, {Yamada}, {Yamamoto}, {Yamaoka},
  {Yamauchi}, {Yoshidome}, \& {Yoshii}}]{Serino-MAXI-GRBs}
{Serino}, M., {Sakamoto}, T., {Kawai}, N., {et~al.} 2014, \pasj, 66, 87,
  \dodoi{10.1093/pasj/psu063}

\bibitem[{{Shappee} {et~al.}(2014){Shappee}, {Prieto}, {Grupe}, {Kochanek},
  {Stanek}, {De Rosa}, {Mathur}, {Zu}, {Peterson}, {Pogge}, {Komossa}, {Im},
  {Jencson}, {Holoien}, {Basu}, {Beacom}, {Szczygie{\l}}, {Brimacombe},
  {Adams}, {Campillay}, {Choi}, {Contreras}, {Dietrich}, {Dubberley},
  {Elphick}, {Foale}, {Giustini}, {Gonzalez}, {Hawkins}, {Howell}, {Hsiao},
  {Koss}, {Leighly}, {Morrell}, {Mudd}, {Mullins}, {Nugent}, {Parrent},
  {Phillips}, {Pojmanski}, {Rosing}, {Ross}, {Sand}, {Terndrup}, {Valenti},
  {Walker}, \& {Yoon}}]{Shappee14-ASASSN}
{Shappee}, B.~J., {Prieto}, J.~L., {Grupe}, D., {et~al.} 2014, \apj, 788, 48,
  \dodoi{10.1088/0004-637X/788/1/48}

\bibitem[{{Shimwell} {et~al.}(2017){Shimwell}, {R{\"o}ttgering}, {Best},
  {Williams}, {Dijkema}, {de Gasperin}, {Hardcastle}, {Heald}, {Hoang},
  {Horneffer}, {Intema}, {Mahony}, {Mandal}, {Mechev}, {Morabito}, {Oonk},
  {Rafferty}, {Retana-Montenegro}, {Sabater}, {Tasse}, {van Weeren},
  {Br{\"u}ggen}, {Brunetti}, {Chy{\.z}y}, {Conway}, {Haverkorn}, {Jackson},
  {Jarvis}, {McKean}, {Miley}, {Morganti}, {White}, {Wise}, {van Bemmel},
  {Beck}, {Brienza}, {Bonafede}, {Calistro Rivera}, {Cassano}, {Clarke},
  {Cseh}, {Deller}, {Drabent}, {van Driel}, {Engels}, {Falcke}, {Ferrari},
  {Fr{\"o}hlich}, {Garrett}, {Harwood}, {Heesen}, {Hoeft}, {Horellou},
  {Israel}, {Kapi{\'n}ska}, {Kunert-Bajraszewska}, {McKay}, {Mohan},
  {Orr{\'u}}, {Pizzo}, {Prandoni}, {Schwarz}, {Shulevski}, {Sipior}, {Smith},
  {Sridhar}, {Steinmetz}, {Stroe}, {Varenius}, {van der Werf}, {Zensus}, \&
  {Zwart}}]{Shimwell17-LOTSS}
{Shimwell}, T.~W., {R{\"o}ttgering}, H.~J.~A., {Best}, P.~N., {et~al.} 2017,
  \aap, 598, A104, \dodoi{10.1051/0004-6361/201629313}

\bibitem[{{Sironi} \&
  {Spitkovsky}(2011)}]{Sironi-Spitkovsky11-PWN-flat-spectrum}
{Sironi}, L., \& {Spitkovsky}, A. 2011, \apj, 741, 39,
  \dodoi{10.1088/0004-637X/741/1/39}

\bibitem[{{S{\k{a}}dowski} \&
  {Narayan}(2016)}]{Sadowski16-superedd-accretion-sims}
{S{\k{a}}dowski}, A., \& {Narayan}, R. 2016, \mnras, 456, 3929,
  \dodoi{10.1093/mnras/stv2941}

\bibitem[{{Smith}(2014)}]{Smith14-mass-loss-review}
{Smith}, N. 2014, \araa, 52, 487, \dodoi{10.1146/annurev-astro-081913-040025}

\bibitem[{{Soderberg} {et~al.}(2006){Soderberg}, {Kulkarni}, {Nakar}, {Berger},
  {Cameron}, {Fox}, {Frail}, {Gal-Yam}, {Sari}, {Cenko}, {Kasliwal},
  {Chevalier}, {Piran}, {Price}, {Schmidt}, {Pooley}, {Moon}, {Penprase},
  {Ofek}, {Rau}, {Gehrels}, {Nousek}, {Burrows}, {Persson}, \&
  {McCarthy}}]{Soderberg06-llGRB060218}
{Soderberg}, A.~M., {Kulkarni}, S.~R., {Nakar}, E., {et~al.} 2006, \nat, 442,
  1014, \dodoi{10.1038/nature05087}

\bibitem[{{Soderberg} {et~al.}(2010){Soderberg}, {Chakraborti}, {Pignata},
  {Chevalier}, {Chandra}, {Ray}, {Wieringa}, {Copete}, {Chaplin},
  {Connaughton}, {Barthelmy}, {Bietenholz}, {Chugai}, {Stritzinger}, {Hamuy},
  {Fransson}, {Fox}, {Levesque}, {Grindlay}, {Challis}, {Foley}, {Kirshner},
  {Milne}, \& {Torres}}]{Soderberg-2009bb}
{Soderberg}, A.~M., {Chakraborti}, S., {Pignata}, G., {et~al.} 2010, \nat, 463,
  513, \dodoi{10.1038/nature08714}

\bibitem[{{Somalwar} {et~al.}(2021){Somalwar}, {Ravi}, {Dong}, {Graham},
  {Hallinan}, {Law}, {Lu}, \& {Myers}}]{Somalwar21-TDE-AGN}
{Somalwar}, J.~J., {Ravi}, V., {Dong}, D., {et~al.} 2021, arXiv e-prints,
  arXiv:2108.12431.
\newblock \doarXiv{2108.12431}

\bibitem[{{Sridhar} \& {Metzger}(2022)}]{Sridhar22-XRB-FRB-prs}
{Sridhar}, N., \& {Metzger}, B.~D. 2022, arXiv e-prints, arXiv:2206.10486.
\newblock \doarXiv{2206.10486}

\bibitem[{{Stein} {et~al.}(2021){Stein}, {Velzen}, {Kowalski}, {Franckowiak},
  {Gezari}, {Miller-Jones}, {Frederick}, {Sfaradi}, {Bietenholz}, {Horesh},
  {Fender}, {Garrappa}, {Ahumada}, {Andreoni}, {Belicki}, {Bellm},
  {B{\"o}ttcher}, {Brinnel}, {Burruss}, {Cenko}, {Coughlin}, {Cunningham},
  {Drake}, {Farrar}, {Feeney}, {Foley}, {Gal-Yam}, {Golkhou}, {Goobar},
  {Graham}, {Hammerstein}, {Helou}, {Hung}, {Kasliwal}, {Kilpatrick}, {Kong},
  {Kupfer}, {Laher}, {Mahabal}, {Masci}, {Necker}, {Nordin}, {Perley},
  {Rigault}, {Reusch}, {Rodriguez}, {Rojas-Bravo}, {Rusholme}, {Shupe},
  {Singer}, {Sollerman}, {Soumagnac}, {Stern}, {Taggart}, {van Santen}, {Ward},
  {Woudt}, \& {Yao}}]{Stein21-neutrino-TDE}
{Stein}, R., {Velzen}, S.~v., {Kowalski}, M., {et~al.} 2021, Nature Astronomy,
  5, 510, \dodoi{10.1038/s41550-020-01295-8}

\bibitem[{{Sutton} {et~al.}(2013){Sutton}, {Roberts}, \&
  {Middleton}}]{Sutton13-ULX-supereddington}
{Sutton}, A.~D., {Roberts}, T.~P., \& {Middleton}, M.~J. 2013, \mnras, 435,
  1758, \dodoi{10.1093/mnras/stt1419}

\bibitem[{{Taddia} {et~al.}(2013){Taddia}, {Stritzinger}, {Sollerman},
  {Phillips}, {Anderson}, {Boldt}, {Campillay}, {Castell{\'o}n}, {Contreras},
  {Folatelli}, {Hamuy}, {Heinrich-Josties}, {Krzeminski}, {Morrell}, {Burns},
  {Freedman}, {Madore}, {Persson}, \& {Suntzeff}}]{Taddia13-IIn-lightcurves}
{Taddia}, F., {Stritzinger}, M.~D., {Sollerman}, J., {et~al.} 2013, \aap, 555,
  A10, \dodoi{10.1051/0004-6361/201321180}

\bibitem[{{Taddia} {et~al.}(2015){Taddia}, {Sollerman}, {Leloudas},
  {Stritzinger}, {Valenti}, {Galbany}, {Kessler}, {Schneider}, \&
  {Wheeler}}]{Taddia15-stripped-envelope-ejecta-mass-energy}
{Taddia}, F., {Sollerman}, J., {Leloudas}, G., {et~al.} 2015, \aap, 574, A60,
  \dodoi{10.1051/0004-6361/201423915}

\bibitem[{{Taddia} {et~al.}(2018){Taddia}, {Stritzinger}, {Bersten}, {Baron},
  {Burns}, {Contreras}, {Holmbo}, {Hsiao}, {Morrell}, {Phillips}, {Sollerman},
  \& {Suntzeff}}]{Taddia18-stripped-env-lightcurves}
{Taddia}, F., {Stritzinger}, M.~D., {Bersten}, M., {et~al.} 2018, \aap, 609,
  A136, \dodoi{10.1051/0004-6361/201730844}

\bibitem[{{Taggart} \& {Perley}(2019)}]{Taggart2020-SN-hosts}
{Taggart}, K., \& {Perley}, D. 2019, arXiv e-prints, arXiv:1911.09112.
\newblock \doarXiv{1911.09112}

\bibitem[{{Tanaka} \&
  {Asano}(2017)}]{Tanaka-Asano17-Stochastic-acceleration-Crab}
{Tanaka}, S.~J., \& {Asano}, K. 2017, \apj, 841, 78,
  \dodoi{10.3847/1538-4357/aa6f13}

\bibitem[{{Tavani} {et~al.}(2021){Tavani}, {Casentini}, {Ursi}, {Verrecchia},
  {Addis}, {Antonelli}, {Argan}, {Barbiellini}, {Baroncelli}, {Bernardi},
  {Bianchi}, {Bulgarelli}, {Caraveo}, {Cardillo}, {Cattaneo}, {Chen}, {Costa},
  {Del Monte}, {Di Cocco}, {Di Persio}, {Donnarumma}, {Evangelista}, {Feroci},
  {Ferrari}, {Fioretti}, {Fuschino}, {Galli}, {Gianotti}, {Giuliani},
  {Labanti}, {Lazzarotto}, {Lipari}, {Longo}, {Lucarelli}, {Magro},
  {Marisaldi}, {Mereghetti}, {Morelli}, {Morselli}, {Naldi}, {Pacciani},
  {Parmiggiani}, {Paoletti}, {Pellizzoni}, {Perri}, {Perotti}, {Piano},
  {Picozza}, {Pilia}, {Pittori}, {Puccetti}, {Pupillo}, {Rapisarda},
  {Rappoldi}, {Rubini}, {Setti}, {Soffitta}, {Trifoglio}, {Trois},
  {Vercellone}, {Vittorini}, {Giommi}, \& {D'Amico}}]{Tavani21-magnetar-burst}
{Tavani}, M., {Casentini}, C., {Ursi}, A., {et~al.} 2021, Nature Astronomy, 5,
  401, \dodoi{10.1038/s41550-020-01276-x}

\bibitem[{{Taylor} {et~al.}(2014){Taylor}, {Cinabro}, {Dilday}, {Galbany},
  {Gupta}, {Kessler}, {Marriner}, {Nichol}, {Richmond}, {Schneider}, \&
  {Sollerman}}]{Taylor14-CCSN-rate}
{Taylor}, M., {Cinabro}, D., {Dilday}, B., {et~al.} 2014, \apj, 792, 135,
  \dodoi{10.1088/0004-637X/792/2/135}

\bibitem[{{Tchekhovskoy} {et~al.}(2014){Tchekhovskoy}, {Metzger}, {Giannios},
  \& {Kelley}}]{Tchekhovskoy14-SwiftJ16-gone-MAD}
{Tchekhovskoy}, A., {Metzger}, B.~D., {Giannios}, D., \& {Kelley}, L.~Z. 2014,
  \mnras, 437, 2744, \dodoi{10.1093/mnras/stt2085}

\bibitem[{{Tendulkar} {et~al.}(2017){Tendulkar}, {Bassa}, {Cordes}, {Bower},
  {Law}, {Chatterjee}, {Adams}, {Bogdanov}, {Burke-Spolaor}, {Butler},
  {Demorest}, {Hessels}, {Kaspi}, {Lazio}, {Maddox}, {Marcote}, {McLaughlin},
  {Paragi}, {Ransom}, {Scholz}, {Seymour}, {Spitler}, {van Langevelde}, \&
  {Wharton}}]{Tendulkar17-FRB121102-host}
{Tendulkar}, S.~P., {Bassa}, C.~G., {Cordes}, J.~M., {et~al.} 2017, \apjl, 834,
  L7, \dodoi{10.3847/2041-8213/834/2/L7}

\bibitem[{{The CHIME/FRB Collaboration} {et~al.}(2021){The CHIME/FRB
  Collaboration}, {:}, {Amiri}, {Andersen}, {Bandura}, {Berger}, {Bhardwaj},
  {Boyce}, {Boyle}, {Brar}, {Breitman}, {Cassanelli}, {Chawla}, {Chen},
  {Cliche}, {Cook}, {Cubranic}, {Curtin}, {Deng}, {Dobbs}, {Fengqiu}, {Dong},
  {Eadie}, {Fandino}, {Fonseca}, {Gaensler}, {Giri}, {Good}, {Halpern}, {Hill},
  {Hinshaw}, {Josephy}, {Kaczmarek}, {Kader}, {Kania}, {Kaspi}, {Landecker},
  {Lang}, {Leung}, {Li}, {Lin}, {Masui}, {Mckinven}, {Mena-Parra},
  {Merryfield}, {Meyers}, {Michilli}, {Milutinovic}, {Mirhosseini},
  {M{\"u}nchmeyer}, {Naidu}, {Newburgh}, {Ng}, {Patel}, {Pen}, {Petroff},
  {Pinsonneault-Marotte}, {Pleunis}, {Rafiei-Ravandi}, {Rahman}, {Ransom},
  {Renard}, {Sanghavi}, {Scholz}, {Shaw}, {Shin}, {Siegel}, {Sikora}, {Singh},
  {Smith}, {Stairs}, {Tan}, {Tendulkar}, {Vanderlinde}, {Wang}, {Wulf}, \&
  {Zwaniga}}]{CHIME-FRB-catalog-2021}
{The CHIME/FRB Collaboration}, {:}, {Amiri}, M., {et~al.} 2021, arXiv e-prints,
  arXiv:2106.04352.
\newblock \doarXiv{2106.04352}

\bibitem[{{Urry} \& {Padovani}(1995)}]{Urry-Padovani-1995-AGN-unification}
{Urry}, C.~M., \& {Padovani}, P. 1995, \pasp, 107, 803, \dodoi{10.1086/133630}

\bibitem[{{van der Horst} {et~al.}(2007){van der Horst}, {Kamble}, {Wijers},
  {Resmi}, {Bhattacharya}, {Rol}, {Strom}, {Kouveliotou}, {Oosterloo}, \&
  {Ishwara-Chandra}}]{van-der-Horst07-GRB030329}
{van der Horst}, A.~J., {Kamble}, A., {Wijers}, R.~A.~M.~J., {et~al.} 2007,
  Philosophical Transactions of the Royal Society of London Series A, 365,
  1241, \dodoi{10.1098/rsta.2006.1993}

\bibitem[{{van der Walt} {et~al.}(2011){van der Walt}, {Colbert}, \&
  {Varoquaux}}]{Van-der-Walt11-numpy}
{van der Walt}, S., {Colbert}, S.~C., \& {Varoquaux}, G. 2011, Computing in
  Science and Engineering, 13, 22, \dodoi{10.1109/MCSE.2011.37}

\bibitem[{{van Dyk} {et~al.}(1993{\natexlab{a}}){van Dyk}, {Sramek}, {Weiler},
  \& {Panagia}}]{vanDyk93-SN1990B}
{van Dyk}, S.~D., {Sramek}, R.~A., {Weiler}, K.~W., \& {Panagia}, N.
  1993{\natexlab{a}}, \apj, 409, 162, \dodoi{10.1086/172652}

\bibitem[{{van Dyk} {et~al.}(1993{\natexlab{b}}){van Dyk}, {Weiler}, {Sramek},
  \& {Panagia}}]{vanDyk-Weiler-93-1988Z}
{van Dyk}, S.~D., {Weiler}, K.~W., {Sramek}, R.~A., \& {Panagia}, N.
  1993{\natexlab{b}}, \apjl, 419, L69, \dodoi{10.1086/187139}

\bibitem[{{van Dyk} {et~al.}(1994){van Dyk}, {Weiler}, {Sramek}, {Rupen}, \&
  {Panagia}}]{vanDyk-Weiler94-SN1993J}
{van Dyk}, S.~D., {Weiler}, K.~W., {Sramek}, R.~A., {Rupen}, M.~P., \&
  {Panagia}, N. 1994, \apjl, 432, L115, \dodoi{10.1086/187525}

\bibitem[{{Vanderplas} {et~al.}(2012){Vanderplas}, {Connolly}, {Ivezi{\'c}}, \&
  {Gray}}]{Vanderplas2012-astroML}
{Vanderplas}, J., {Connolly}, A., {Ivezi{\'c}}, {\v Z}., \& {Gray}, A. 2012, in
  Conference on Intelligent Data Understanding (CIDU), 47 --54,
  \dodoi{10.1109/CIDU.2012.6382200}

\bibitem[{Virtanen {et~al.}(2020)Virtanen, Gommers, Oliphant, Haberland, Reddy,
  Cournapeau, Burovski, Peterson, Weckesser, Bright, {van der Walt}, Brett,
  Wilson, Millman, Mayorov, Nelson, Jones, Kern, Larson, Carey, Polat, Feng,
  Moore, {VanderPlas}, Laxalde, Perktold, Cimrman, Henriksen, Quintero, Harris,
  Archibald, Ribeiro, Pedregosa, {van Mulbregt}, \& {SciPy 1.0
  Contributors}}]{Virtanen20-scipy}
Virtanen, P., Gommers, R., Oliphant, T.~E., {et~al.} 2020, Nature Methods, 17,
  261, \dodoi{10.1038/s41592-019-0686-2}

\bibitem[{{Walker}(1998)}]{Walker1998-scintillation}
{Walker}, M.~A. 1998, \mnras, 294, 307,
  \dodoi{10.1046/j.1365-8711.1998.01238.x}

\bibitem[{{Weiler} {et~al.}(1990){Weiler}, {Panagia}, \&
  {Sramek}}]{Weiler90-SN1986J}
{Weiler}, K.~W., {Panagia}, N., \& {Sramek}, R.~A. 1990, \apj, 364, 611,
  \dodoi{10.1086/169444}

\bibitem[{{Weiler} {et~al.}(1986){Weiler}, {Sramek}, {Panagia}, {van der
  Hulst}, \& {Salvati}}]{Weiler86-radio-SN-review}
{Weiler}, K.~W., {Sramek}, R.~A., {Panagia}, N., {van der Hulst}, J.~M., \&
  {Salvati}, M. 1986, \apj, 301, 790, \dodoi{10.1086/163944}

\bibitem[{{Weiler} {et~al.}(1991){Weiler}, {van Dyk}, {Panagia}, {Sramek}, \&
  {Discenna}}]{Weiler91-SN1979C}
{Weiler}, K.~W., {van Dyk}, S.~D., {Panagia}, N., {Sramek}, R.~A., \&
  {Discenna}, J.~L. 1991, \apj, 380, 161, \dodoi{10.1086/170571}

\bibitem[{{Werner} {et~al.}(2016){Werner}, {Uzdensky}, {Cerutti}, {Nalewajko},
  \& {Begelman}}]{Werner-2016-power-law-spectra-relativistic-reconnection}
{Werner}, G.~R., {Uzdensky}, D.~A., {Cerutti}, B., {Nalewajko}, K., \&
  {Begelman}, M.~C. 2016, \apjl, 816, L8, \dodoi{10.3847/2041-8205/816/1/L8}

\bibitem[{{Wiktorowicz} {et~al.}(2019){Wiktorowicz}, {Wyrzykowski},
  {Chruslinska}, {Klencki}, {Rybicki}, \&
  {Belczynski}}]{Wiktorowicz2019-stellar-mass-BH-population-synthesis}
{Wiktorowicz}, G., {Wyrzykowski}, {\L}., {Chruslinska}, M., {et~al.} 2019,
  \apj, 885, 1, \dodoi{10.3847/1538-4357/ab45e6}

\bibitem[{{Wright}(2006)}]{Wright06-cosmology-calc}
{Wright}, E.~L. 2006, \pasp, 118, 1711, \dodoi{10.1086/510102}

\bibitem[{{Yao} {et~al.}(2020){Yao}, {Kulkarni}, {Burdge}, {Caiazzo}, {De},
  {Dong}, {Fremling}, {Kasliwal}, {Kupfer}, {van Roestel}, {Sollerman},
  {Bagdasaryan}, {Bellm}, {Cenko}, {Drake}, {Duev}, {Graham}, {Kaye}, {Masci},
  {Miranda}, {Prince}, {Riddle}, {Rusholme}, \& {Soumagnac}}]{Yao20-AT2019wey}
{Yao}, Y., {Kulkarni}, S.~R., {Burdge}, K.~B., {et~al.} 2020, arXiv e-prints,
  arXiv:2012.00169.
\newblock \doarXiv{2012.00169}

\bibitem[{{Younes} {et~al.}(2016){Younes}, {Kouveliotou}, {Kargaltsev}, {Gill},
  {Granot}, {Watts}, {Gelfand}, {Baring}, {Harding}, {Pavlov}, {van der Horst},
  {Huppenkothen}, {G{\"o}{\u{g}}{\"u}{\c{s}}}, {Lin}, \&
  {Roberts}}]{Younes16-magnetar-wind-nebula}
{Younes}, G., {Kouveliotou}, C., {Kargaltsev}, O., {et~al.} 2016, \apj, 824,
  138, \dodoi{10.3847/0004-637X/824/2/138}

\bibitem[{{Zauderer} {et~al.}(2011){Zauderer}, {Berger}, {Soderberg}, {Loeb},
  {Narayan}, {Frail}, {Petitpas}, {Brunthaler}, {Chornock}, {Carpenter},
  {Pooley}, {Mooley}, {Kulkarni}, {Margutti}, {Fox}, {Nakar}, {Patel},
  {Volgenau}, {Culverhouse}, {Bietenholz}, {Rupen}, {Max-Moerbeck}, {Readhead},
  {Richards}, {Shepherd}, {Storm}, \& {Hull}}]{Zauderer-SwiftJ16}
{Zauderer}, B.~A., {Berger}, E., {Soderberg}, A.~M., {et~al.} 2011, \nat, 476,
  425, \dodoi{10.1038/nature10366}

\end{thebibliography}
\bibliographystyle{aasjournal}

\end{document}